\documentclass[aps,prb,twocolumn,superscriptaddress]{revtex4-2}
\usepackage{amsmath,amssymb,graphicx,color,hyperref}
\usepackage{bm}
\usepackage{amsfonts}
\usepackage{bbold}
\newcommand{\fn}{{\,\mathfrak{n}\,}}
\newcommand{\fr}{\mathfrak{r}}
\def\CC{\mathbb{C}}
\newcommand{\bc}{\mathbf{c}}
\newcommand{\bt}{\mathbf{t}}
\newcommand{\bJ}{\mathbf{J}}
\newcommand{\bK}{\mathbf{K}}

\newcommand{\be}{\begin{equation}}
	\newcommand{\ee}{\end{equation}}
\newcommand{\bea}{\begin{eqnarray}}
	\newcommand{\eea}{\end{eqnarray}}

\newcommand{\bi}{\begin{itemize}}
\newcommand{\ei}{\end{itemize}}

\newcommand{\bce}{\begin{center}}
\newcommand{\ece}{\end{center}}

\begin{document}

	\title{Transport in spinful non-Hermitian gauge lattices}
\author{H. Ghaemi-Dizicheh}
\affiliation{College of Undergraduate Studies, The Chicago School, Chicago, Illinois, USA}
\affiliation{Department of Physics and Astronomy, University of Texas Rio Grande Valley, Edinburg, Texas 78539, USA}

\author{S. Dehdashti}
\affiliation{Emmy-Noether Group Theoretical Quantum Systems Design, Technical University of Munich, Munich, Germany}

\author{J. N\"{o}tzel}
\affiliation{Emmy-Noether Group Theoretical Quantum Systems Design, Technical University of Munich, Munich, Germany}
\begin{abstract}
We extend the generalized transfer-matrix method introduced in
[H. Ghaemi-Dizicheh, Phys.\ Rev.\ B \textbf{107}, 125155 (2023)]
to one-dimensional spinful non-Hermitian tight-binding lattices with
matrix-valued Abelian and non-Abelian gauge couplings. We derive a
minimal channel-resolved transfer matrix and identify its eigenvalues
with the generalized Bloch factors. The same transfer matrix determines
the periodic- and open-boundary spectra, generalized Brillouin zone,
skin localization, and finite-sample scattering conditions. In the
Abelian limit, the model separates into fixed spin sectors, yielding
exact spectral results and closed finite-length transport expressions
in terms of Chebyshev polynomials. These results reveal bipolar
spin-dependent skin localization and distinguish the reciprocal hopping
scale, which governs poles, zeros, and reflectionlessness, from the
directed-amplification factor controlling transmission asymmetry,
transparency, and invisibility. For non-Abelian gauge fields,
noncommuting link matrices coherently mix the spin sectors, such that
four transfer eigenmodes jointly determine the non-Bloch spectrum and
transport eigenchannels. We show that this mixing produces
spin-hybridized lasing and coherent-perfect-absorption channels,
Zeeman-tunable lasing thresholds, and direction-dependent
reflectionless transport accompanied by pronounced spin conversion.
The resulting boundary-condition criteria describe lasing, coherent
perfect absorption, reflectionlessness, transparency, and invisibility,
while identifying spin polarization and conversion as experimentally
accessible signatures that distinguish Abelian from non-Abelian
non-Hermitian transport.
\end{abstract}

	\maketitle
\section{Introduction}
\label{sec:introduction}

Non-Hermitian Hamiltonians provide effective descriptions of open and
nonconservative systems that exchange probability, energy, or wave
intensity with their surroundings
\cite{Moiseyev2011,ElGanainy2018,Ashida2020,Ding2022,Roccati2022,Ghaemi-Dizicheh:23,Ghaemi-Dizicheh_2024,Ghaemidizicheh:25,Rodriguez-Lara:26}.
In quantum settings, they commonly arise after eliminating
environmental degrees of freedom or conditioning the dynamics on
selected measurement outcomes \cite{rodriguez2026grid}. Classical-wave platforms provide
particularly flexible realizations, with non-Hermiticity engineered
through gain and loss, radiation, feedback, and nonreciprocal
couplings in photonic, acoustic, mechanical, and electrical systems
\cite{Huang2024Acoustic,Xiao2025Photonics}.
These mechanisms produce complex spectra, exceptional points,
nonorthogonal eigenstates, point-gap topology, and boundary
sensitivity without direct Hermitian counterparts
\cite{Bergholtz2021,OkumaSato2023}.

A prominent manifestation of this boundary sensitivity is the
non-Hermitian skin effect, in which an extensive number of bulk states
accumulate exponentially near a boundary
\cite{HatanoNelson1996,YaoWang2018,Okuma2020,LinTaiLiLee2023}.
The accompanying spectral rearrangement between periodic and open
boundary conditions invalidates conventional Bloch theory in many
nonreciprocal lattices. Non-Bloch band theory and the generalized
Brillouin zone restore a bulk description by allowing the Bloch factor
to acquire a nonunit modulus that accounts for spatial growth or decay
\cite{Yokomizo2019,Longhi2019}.

The transfer matrix offers a natural real-space framework for this
boundary-sensitive physics because propagation and boundary
conditions remain explicit throughout the calculation. In
one-dimensional scattering, it provides direct conditions for
reflection, transmission, spectral singularities, coherent perfect
absorption, and invisibility
\cite{Mostafazadeh2009,Chong2010,Lin2011,Longhi2015,
	Mostafazadeh2020}. Generalized transfer-matrix constructions extend
this approach to multichannel tight-binding systems and to
rank-deficient intercell hopping, for which a conventional transfer
matrix cannot be obtained through direct inversion
\cite{Dwivedi2016,ghaemiPRA2021, ghaemiPRB2023}. In non-Hermitian lattices, the transfer
eigenvalues additionally diagnose the breakdown of conventional
bulk-boundary correspondence, generalized Bloch propagation, and skin
localization \cite{KunstDwivedi2019,ChenZhang2025}. The same formalism
can therefore describe both the closed-chain spectrum and the
scattering response of a finite sample.

Internal degrees of freedom substantially enrich this connection.
Spin, pseudospin, polarization, orbital, and sublattice structure turn
scalar hopping amplitudes into matrices and permit synthetic gauge
fields to rotate the internal state during propagation. Artificial
Abelian and non-Abelian gauge potentials have been developed
extensively in cold-atom and photonic settings, where they emulate
magnetic fields, spin-orbit interactions, and matrix-valued parallel
transport
\cite{Osterloh2005,Ruseckas2005,Goldman2014,
	Ozawa2019}. Commuting gauge links can be diagonalized in a common
internal-state basis and reduce the propagation problem to independent
channels. Noncommuting links instead mix these channels, so their
propagation and scattering eigenstates need not coincide with any
fixed spin quantization axis.\\
The combination of matrix-valued gauge fields and non-Hermiticity has
recently enabled spectral braiding, spin-selective skin localization,
and tunable bipolar and unidirectional accumulation. This interplay is
particularly evident in generalized Hatano--Nelson lattices, where
synthetic non-Abelian gauge fields generate Hopf-link complex-energy
braiding together with tunable bipolar skin modes
\cite{Pang2024,Zhao2025NonAbelian}. More recently, reciprocal
non-Abelian couplings have also been shown to produce
time-reversal-symmetry-protected $\mathbb{Z}_2$ skin effects with
pseudospin-resolved boundary localization and boundary-controlled
conversion from bidirectional to unidirectional accumulation
\cite{Tang2026Z2}. Spin-dependent Abelian and matrix-valued gauge couplings have produced
spin-polarized skin effects, bipolar boundary localization, and
magnetic control of the accumulation direction
\cite{Sanahal2025}. Recent
electrical-circuit realizations have further demonstrated
non-Abelian spectral braiding and spin-hybridized skin modes
\cite{ChenNonAbelian2026,WuNonAbelian2026}. Related proposals and
mechanical implementations have explored time-reversal-protected skin
effects and directly measurable non-Abelian gauge observables
\cite{Tang2026Z2,Velkovsky2026non}.

These advances have concentrated primarily on closed-system spectra,
generalized Bloch bands, exceptional structures, and the spatial
localization of eigenmodes. The transport consequences of
non-Abelian gauge structure remain comparatively less developed.
In this study, we show that by looking at transport effects one can investigate information that the complex spectrum alone does
not determine: it identifies which incident spinor is absorbed,
reflected, transmitted, or converted into an outgoing lasing channel.
Moreover, in a multichannel problem, the singularity of a
transfer-matrix block may select only one spinor rather than impose a
condition on the complete spin space. It is therefore necessary to
distinguish transport eigenchannel conditions from stronger matrix
conditions that apply to every channel.

Previous transfer-matrix studies classified the compatibility of
lasing, coherent perfect absorption, reflectionlessness,
transparency, and invisibility in scalar nonreciprocal chains and
extended this classification to reduced multichannel transfer
matrices \cite{ghaemiPRA2021, ghaemiPRB2023}. However, a unified treatment
connecting generalized Bloch factors, spinful skin modes, and the
finite-length transport eigenchannels of noncommuting gauge lattices
has remained lacking. In particular, it has not been clear how the
same transfer modes responsible for the non-Bloch spectrum determine
the spin polarization of scattering poles, zeros, reflectionless
states, and invisible channels.

In this work, we establish this connection for a one-dimensional
spinful non-Hermitian tight-binding lattice with independent
matrix-valued hopping operators in the two directions. We begin with
a general finite-range multichannel lattice and construct a reduced
transfer matrix by identifying the active hopping subspaces through a
singular-value decomposition. Inactive intracell amplitudes are
eliminated through a Schur complement, allowing the propagation
dimension to be determined by the number of active channels rather
than the full supercell Hilbert space. Specializing this construction
to the spinful gauge chain gives a four-by-four transfer matrix that
preserves the ordering of noncommuting link operations.

For the closed chain, the transfer eigenvalues are identified with
the generalized Bloch factors, consistent with transfer-matrix
formulations of non-Bloch band theory
\cite{Yokomizo2019,KunstDwivedi2019}. This provides the periodic- and
open-boundary spectra, finite-chain quantization, generalized
Brillouin zones, and the spatial growth exponents underlying the skin
effect within a single framework. In the Abelian limit, the model
separates into two spin sectors, yielding exact finite-chain spectra,
spin-resolved generalized Brillouin zones, and opposite localization
directions responsible for the bipolar skin effect, in agreement with
recent studies of spinful non-Hermitian gauge lattices
\cite{Sanahal2025,ZhangSU22025}. In the non-Abelian regime, the spin
sectors hybridize, and the spectrum is governed collectively by four
generalized Bloch factors. The resulting braided spectra and
spin-hybridized skin modes reproduce those reported for non-Abelian
spinful lattices
\cite{Pang2024,ChenNonAbelian2026,WuNonAbelian2026}. Recovering these
independently established Abelian and non-Abelian spectral features
provides a nontrivial benchmark of the channel-resolved transfer
matrix before the same construction is applied to the transport
boundary conditions that form the central focus of this work.

Connecting the finite chain to spin-degenerate featureless leads
extends the same construction directly to transport. We derive
finite-length conditions for quasibound states and lasing, scattering
zeros and coherent perfect absorption, left- and
right-reflectionlessness, strict transparency, and invisibility. In
the Abelian case, the transfer matrix reduces to exact spin-resolved
expressions involving Chebyshev polynomials. At fixed onsite
parameters and reciprocal propagation scale, the pole, zero, and
reflectionless conditions are independent of the directed-amplification
factor. The latter instead controls the left-right transmission
asymmetry, enters the transparency and invisibility conditions, and is
directly related to the spin-resolved skin localization.

The non-Abelian transport problem remains intrinsically matrix valued,
and its boundary conditions select coherent spin-hybridized
eigenchannels. Our numerical results demonstrate spin-hybridized
lasing and coherent-perfect-absorption channels, Zeeman-field control
of the non-Abelian lasing threshold, and direction- and field-dependent
reflectionless spin conversion. We further distinguish
reflectionlessness from strict transparency and invisibility. In
particular, invisibility requires reflection suppression and
phase-matched transmission for the same spinor; separate zeros of the
reflection and transparency determinants, or unit transmitted
intensity alone, are not sufficient. A non-Abelian invisible channel
may therefore be a spin-hybridized state, although strict invisibility
requires its transmitted spinor to reproduce the incident spinor up
to the reference propagation phase. These results expose transport
signatures of the non-Abelian gauge structure that cannot be inferred
from spectral braiding or skin localization alone.

The central advance is therefore a common boundary-sensitive framework
in which the same bulk transfer matrix determines the non-Bloch
spectrum and skin localization of the closed chain and, after the lead
contacts are incorporated, the spin-resolved scattering properties of
the finite system.\\ The remainder of this paper is organized as follows. Section~II summarizes the reduced-channel transfer matrix
(Sec.~II~A) and the lead construction with the multichannel
boundary-condition classification (Sec.~II~B); the complete derivation
is given in Appendix~A. Section~III introduces the spinful model and its
transfer matrix (Sec.~III~A) and then obtains the periodic- and
open-boundary spectra, generalized Brillouin zones, and skin effects in
the Abelian and non-Abelian limits (Sec.~III~B). Section~IV treats
finite-length transport: Sec.~IV~A derives the exact Abelian transfer
blocks, and Secs.~IV~B--IV~E analyze quasibound states and lasing,
scattering zeros and coherent perfect absorption, reflectionless
transport, and transparency and invisibility, comparing Abelian and
non-Abelian gauge fields in each case. Section~V discusses the sensing
implications of the resulting transport eigenchannels, and Sec.~VI
summarizes the conclusions and possible extensions.
\section{Mathematical Framework}
\label{sectionII}

This section fixes the notation and summarizes the transfer-matrix
construction used throughout the paper. The complete derivation,
including the reduction to the active hopping subspaces, the
featureless-lead modes, and the full set of transport boundary
conditions, is given in Appendix~\ref{app:mathematical_framework}.
\subsection{Reduced-channel transfer matrix}

We consider a one-dimensional tight-binding lattice with $q$ internal
degrees of freedom (spin, orbital, or sublattice) per unit cell and
hopping of range $\fr$. Grouping $\fr$ adjacent unit cells into a
supercell of dimension $\fn=q\fr$ reduces the Schr\"odinger equation to
the nearest-neighbor recursion
\be
E\Psi_{n}=\bK_n\Psi_{n}+\bJ_{R,n}\Psi_{n-1}+\bJ_{L,n}\Psi_{n+1},
\qquad \Psi_{n}\in\CC^{\fn},
\label{recursion rel.}
\ee
where $\bK_n$ contains all terms within supercell $n$, and $\bJ_{L}$
($\bJ_{R}$) contains the left- (right-) directed hopping between
neighboring supercells. The supercell construction generally makes
$\bJ_{L}$ rank deficient, so a conventional transfer matrix cannot be
obtained by inverting it. We therefore retain only the active hopping
subspaces and use the paired factorization
\be
\bJ_{L}=VX_{L}W^{\dagger},\qquad
\bJ_{R}=WX_{R}V^{\dagger},\qquad
V^{\dagger}W=0 .
\label{SVD}
\ee
Here the columns of $V$ and $W$ are orthonormal bases of the two active
subspaces, $\rho=\mathrm{rank}\,\bJ_{L}=\mathrm{rank}\,\bJ_{R}$ is the
number of active channels, and $X_{L}$ and $X_{R}$ are invertible
$\rho\times\rho$ hopping blocks. A singular value decomposition supplies
such bases, but $X_{L,R}$ need not be diagonal. For independent
nonreciprocal hopping matrices, the shared bases and the orthogonality
$V^{\dagger}W=0$ are assumptions on the hopping structure; they are
satisfied by the spinful model of Sec.~III.

Expanding $\Psi_n=V\alpha_n+W\beta_n+Y\zeta_n$, where $Y$ spans the
inactive intracell sector, and eliminating $\zeta_n$ through a Schur
complement~\cite{ZhangSchur2005} leaves the active-channel blocks
\be
\overline{G}_{ab}=G_{ab}-G_{ay}G_{yy}^{-1}G_{yb},
\qquad a,b\in\{v,w\},
\label{eq:Schur}
\ee
where $G_{ab}=A^{\dagger}(EI_{\fn}-\bK_n)B$ denotes the projection onto
the bases $A,B\in\{V,W,Y\}$ labeled by $a,b\in\{v,w,y\}$. With the
state vector $\Phi_n=(\beta_n,\alpha_{n-1})^{T}$, the recursion becomes
$\Phi_{n+1}=M_n(E)\Phi_n$, with the $2\rho\times2\rho$ transfer matrix
\begin{equation}
	\arraycolsep=2pt
	M_n=
	\begin{pmatrix}
		X_L^{-1}\bigl[\overline{G}_{vw}-\overline{G}_{vv}\overline{G}_{wv}^{-1}\overline{G}_{ww}\bigr]
		&
		X_L^{-1}\overline{G}_{vv}\overline{G}_{wv}^{-1}X_R
		\\[1.4ex]
		-\overline{G}_{wv}^{-1}\overline{G}_{ww}
		&
		\overline{G}_{wv}^{-1}X_R
	\end{pmatrix}.
	\label{eq:generalSVDtransfer}
\end{equation}
Its dimension is set by the number of active channels rather than by the
supercell dimension $\fn$. When the active subspaces span the complete
supercell space ($\fn=2\rho$), the $Y$ sector is absent and
$\overline{G}_{ab}=G_{ab}$. For nearest-neighbor hopping with an
invertible left-hopping matrix, a two-site supercell is sufficient, and
Eq.~\eqref{eq:generalSVDtransfer} equals the product of the conventional
one-site transfer matrices of its two sites. Its eigenvalues are
therefore the squares of the one-site transfer eigenvalues, and
Secs.~III and~IV work with the
one-site matrices directly. In either case, the transfer matrix of a
finite sample is the ordered product
\be
M=M_{N}M_{N-1}\cdots M_{1},
\label{total transfer matrix}
\ee
over its $N$ propagation steps (supercells, or physical sites when
one-site matrices are used).

\subsection{Leads and transport boundary conditions}

We connect the sample to two semi-infinite featureless leads in the
wide-band limit of Refs.~\cite{ghaemiPRA2021, ghaemiPRB2023}, specified
in channel space by
\begin{align}
	K_{vv}&=K_{ww}=E\,I_{\rho}, & K_{vw}&=K_{wv}=w\,I_{\rho},
	\nonumber\\
	X_L&=X_R=w\,I_{\rho}, & w&<0.
	\label{eq:wideband}
\end{align}
Each active channel then supports one right- and one left-propagating
lead wave, with amplitudes $\phi^{(+)}$ and
$\phi^{(-)}\in\CC^{\rho}$. In the wide-band phase convention used here,
these waves acquire the factors $i$ and $-i$, respectively, per lattice
site of the lead (Appendix~\ref{app:mathematical_framework}). The
propagating-wave transfer matrix
\be
\mathcal{M}=\mathcal{Q}^{-1}M\mathcal{Q},
\qquad
\mathcal{Q}=
\begin{pmatrix}
	iI_{\rho} & I_{\rho}\\
	I_{\rho} & iI_{\rho}
\end{pmatrix},
\label{eq:Qtransform}
\ee
relates the lead amplitudes on the two sides of the sample, and the
scattering matrix follows from the same amplitudes:
\begin{align}
	\begin{pmatrix}
		\phi^{(+,R)}\\
		\phi^{(-,R)}
	\end{pmatrix}
	&=
	\mathcal{M}
	\begin{pmatrix}
		\phi^{(+,L)}\\
		\phi^{(-,L)}
	\end{pmatrix},
	\label{eq:Mprop}
	\\
	\begin{pmatrix}
		\phi^{(-,L)}\\
		\phi^{(+,R)}
	\end{pmatrix}
	&=
	\begin{pmatrix}
		r_L & t_{R\rightarrow L}\\
		t_{L\rightarrow R} & r_R
	\end{pmatrix}
	\begin{pmatrix}
		\phi^{(+,L)}\\
		\phi^{(-,R)}
	\end{pmatrix}.
	\label{eq:Smatrix}
\end{align}
For invertible $\mathcal{M}_{22}$, the $\rho\times\rho$ blocks of
$\mathcal{M}$ give
\begin{align}
	&r_L =
	-\mathcal{M}_{22}^{-1}\mathcal{M}_{21}, &&t_{L\rightarrow R} =
	\mathcal{M}_{11}
	-\mathcal{M}_{12}\mathcal{M}_{22}^{-1}\mathcal{M}_{21},
	\nonumber\\
	&r_R =
	\mathcal{M}_{12}\mathcal{M}_{22}^{-1}, &&t_{R\rightarrow L} =
	\mathcal{M}_{22}^{-1}.
	\label{eq:Sblocks}
\end{align}

A transport channel is an independent propagating degree of freedom in
the active subspace, so the channel amplitudes are vectors in
$\CC^{\rho}$; for the spinful model, they are two-component spinors. We
denote incoming amplitudes by $\chi_L\equiv\phi^{(+,L)}$ and
$\chi_R\equiv\phi^{(-,R)}$, and outgoing amplitudes by
$\eta_L\equiv\phi^{(-,L)}$ and $\eta_R\equiv\phi^{(+,R)}$
(Table~\ref{tab:channel_amplitude_notation}). A boundary condition is called
\emph{strong} when it holds as a matrix identity in the full
active-channel space and therefore applies to every incident channel.
More generally, the same effect may occur only for a particular nonzero
channel spinor, which then defines a transport eigenchannel. For
transparency and invisibility, we compare with the reference
transmission amplitude $\tau_0$ acquired through a segment of the
featureless lead with the same number of sites as the sample; for the
uniform $N$-site sample considered below, $\tau_0=i^{N}$.
Table~\ref{tab:transport_BC} lists the closed-chain spectral conditions
and the strong transport conditions in terms of $M$ and $\mathcal{M}$;
they are derived in Appendix~\ref{app:transport_boundary_conditions}.
\section{Spinful model}
\subsection{Model and transfer matrix} 
\begin{figure}[t]
	\centering
	\includegraphics[width=0.99\columnwidth]
	{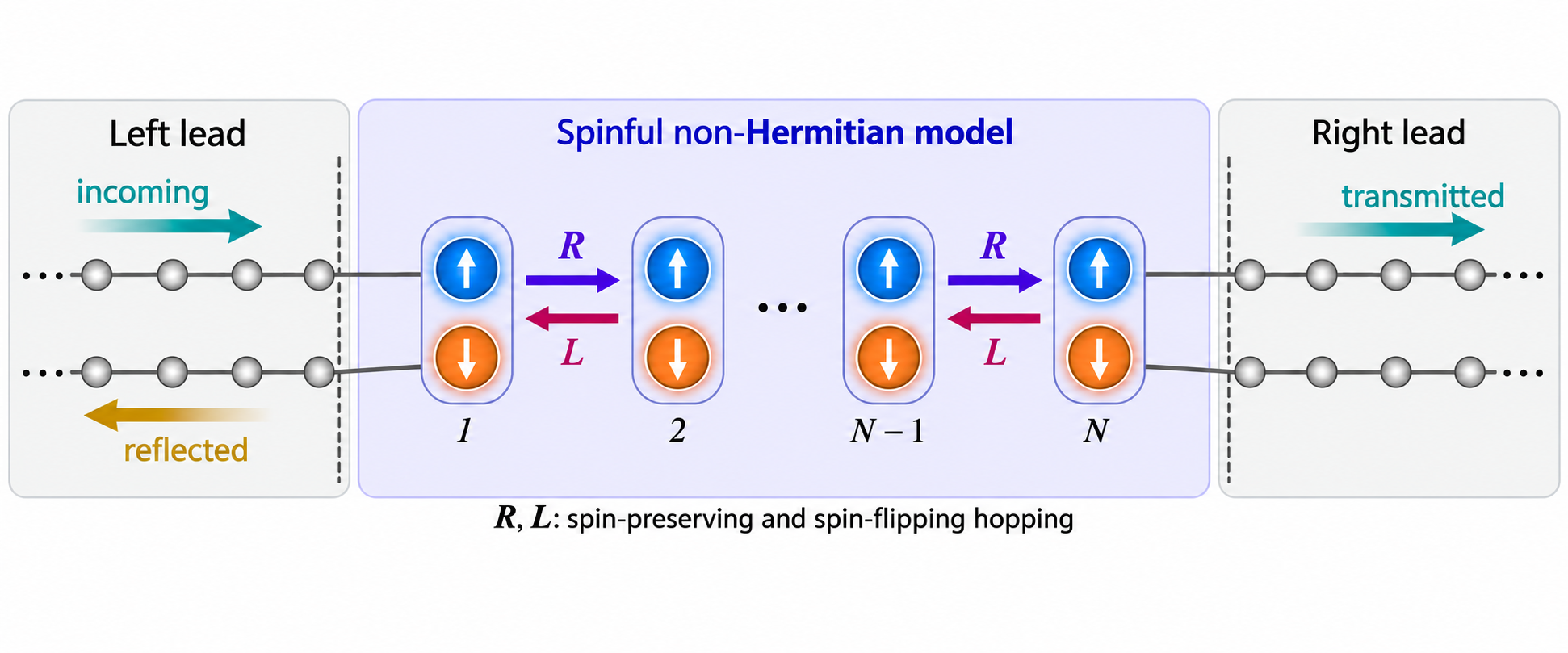}
	\caption{
		Schematic of the transport geometry for the spinful non-Hermitian
		tight-binding chain. Each physical site \(n\) hosts the two-component
		spinor \(\mathbf{c}_n=(c_{n,+},c_{n,-})^{T}\). The arrows \(R\) and
		\(L\) represent the right- and left-directed matrix hoppings
		\(t_{2}T_{\nu}(\phi)\) and \(t_{1}T_{\mu}(\theta)\), respectively.
		Since these matrices act on the complete spin doublet, each link
		generally contains both spin-preserving and spin-flipping hopping
		amplitudes. The finite sample is connected to spin-degenerate leads,
		where a state incident from the left generates reflected and
		transmitted components.
	}
	\label{fig:spinful_transport_geometry}
\end{figure}
We now consider the spinful extension of the one-dimensional tight-binding lattice, schematically shown in figure \ref{fig:spinful_transport_geometry}, where the particle carries an internal spin-$\tfrac{1}{2}$ degree of freedom. Each physical site therefore hosts a two-component spinor, so that the local Hilbert-space dimension is $q=2$. Spin-dependent gauge fields are incorporated through matrix-valued hopping amplitudes.
For the uniform nearest-neighbor model, we take the left- and right-directed hopping matrices and the onsite term as
\begin{align}
\mathbf{t}^{L}_{1,n} &= t_1 T_{\mu}(\theta),
&
\mathbf{t}^{R}_{1,n} &= t_2 T_{\nu}(\phi),
&
\mathbf{t}_{0,n} &= \mathbf{V}_n,
\label{eq:spinful_hoppings}
\end{align}
where
\begin{equation}
T_{\mu}(\theta)=e^{i\theta\sigma_{\mu}},
\qquad
T_{\nu}(\phi)=e^{i\phi\sigma_{\nu}},
\qquad
\mathbf{V}_n=V_nI_2+\mathbf{B}_n\cdot\boldsymbol{\sigma}.
\label{eq:spinful_gauge_matrices}
\end{equation}
Here $\mu,\nu\in\{x,y,z\}$, $\boldsymbol{\sigma}=(\sigma_x,\sigma_y,\sigma_z)$ denotes the vector of Pauli matrices, and $\mathbf{B}_n$ represents an external Zeeman field. The parameters $t_1$ and $t_2$ determine the hopping amplitudes in the two directions, while $\theta$ and $\phi$ are the corresponding spin-dependent gauge parameters. They may be real or complex, thereby allowing both unitary spin rotations and non-Hermitian spin-dependent amplification or attenuation.

With these definitions, the tight-binding Hamiltonian reads
\begin{equation}
H=\sum_{n=1}^{N}
\left[
c_n^{\dagger} t_1T_{\mu}(\theta)c_{n+1}
+c_{n+1}^{\dagger} t_2T_{\nu}(\phi)c_n
+c_n^{\dagger}\mathbf{V}_n c_n
\right],
\label{eq:spinful_hamiltonian}
\end{equation}
where $c_n=(c_{n,+},c_{n,-})^T$ is the spinor of annihilation operators in the $\sigma_z$ basis and
$c_n^{\dagger}=(c_{n,+}^{\dagger},c_{n,-}^{\dagger})$. In the general non-Hermitian and nonreciprocal setting, the two directed hopping matrices are independent (i.e., $\mathbf{t}^{R}_{1,n}\neq (\mathbf{t}^{L}_{1,n})^{\dagger}$).

The Abelian or non-Abelian character of the gauge field is determined by the commutation of the two spin-dependent link matrices. Using
\begin{equation}
T_{\mu}(\theta)
=
\cos\theta\,I_2+i\sin\theta\,\sigma_{\mu},
\end{equation}
one obtains
\begin{equation}
\left[T_{\mu}(\theta),T_{\nu}(\phi)\right]
=
-\sin\theta\sin\phi\,[\sigma_{\mu},\sigma_{\nu}]
=
-2i\sin\theta\sin\phi\,
\epsilon_{\mu\nu\lambda}\sigma_{\lambda}.
\label{eq:gauge_commutator}
\end{equation}
Thus, the gauge field is Abelian when the two link matrices commute, for example for $\mu=\nu$, and is generically non-Abelian for $\mu\neq\nu$ when $\sin\theta\sin\phi\neq0$. For real $\theta$ and $\phi$, the link matrices $T_\mu$ and
$T_\nu$ are unitary and belong to $SU(2)$. For complex gauge
parameters, they are generally nonunitary but remain unimodular ($	\det T_{\mu,\nu}
=1$), because the Pauli matrices are traceless. Thus, complex gauge parameters produce reciprocal amplification and
attenuation between the spin components without changing the
determinant of the link matrix. This property will later be used to
determine the determinant of the spinful transfer matrix and the
product of its generalized Bloch factors.\\
To apply the transfer-matrix formalism, we group two adjacent physical sites $(n-1,n)$ into one supercell. The supercell wave function has four components because each of the two sites carries a two-component spinor. The corresponding hopping and onsite matrices are
\begin{align}
\mathbf{J}_{L}
&=
\begin{pmatrix}
0_2 & 0_2\\
t_1T_{\mu}(\theta) & 0_2
\end{pmatrix},
&
\mathbf{J}_{R}
&=
\begin{pmatrix}
0_2 & t_2T_{\nu}(\phi)\\
0_2 & 0_2
\end{pmatrix},
\label{eq:spinful_JLR}
\\[1mm]
\mathbf{K}_n
&=
\begin{pmatrix}
\mathbf{V}_{n-1} & t_1T_{\mu}(\theta)\\
t_2T_{\nu}(\phi) & \mathbf{V}_{n}
\end{pmatrix}.
\label{eq:spinful_K}
\end{align}
The placement of the off-diagonal blocks in $\mathbf{K}_n$ follows directly from the directed hopping terms in Eq. \eqref{eq:spinful_hamiltonian}.
The block structure of $J_L$ and $J_R$ already identifies their
two-dimensional active subspaces. The transfer-matrix construction
requires only orthonormal bases for these subspaces and the associated
reduced hopping blocks; it does not require the latter to be diagonal.
An explicit SVD would merely introduce parameter-dependent rotations
within the spin space and would obscure the physical spin basis.
We therefore use the exact constant-basis factorization
\begin{equation}
	J_L=V_0X_LW_0^\dagger,
	\qquad
	J_R=W_0X_RV_0^\dagger,
	\label{eq:constant_basis_factorization}
\end{equation}
with
\begin{align}
V_0=
\begin{pmatrix}
0_2\\
I_2
\end{pmatrix},
\qquad
W_0=
\begin{pmatrix}
I_2\\
0_2
\end{pmatrix},
\qquad
\\\nonumber
\\
X_L=t_1T_{\mu}(\theta),
\qquad
X_R=t_2T_{\nu}(\phi).
\label{eq:spinful_XLR}
\end{align}
These channel matrices satisfy
\begin{equation}
V_0^{\dagger}V_0
=
W_0^{\dagger}W_0
=
I_2,
\qquad
V_0^{\dagger}W_0
=
0_2.
\label{eq:spinful_orthogonality}
\end{equation}
Because the supercell dimension is four and the two active subspaces each have dimension two, they already span the complete supercell Hilbert space. Hence, the complementary $Y$ sector introduced in the general construction is absent in this model.

Using the definitions
\begin{equation}
K_{ab}=A^{\dagger}\mathbf{K}_nB,
\qquad
G_{ab}=a^{\dagger}(EI_4-\mathbf{K}_n)b,
\end{equation}
with $a,b\in\{V_0,W_0\}$, the projected blocks are
\begin{align}
K_{vv}&=\mathbf{V}_{n},
&
K_{ww}&=\mathbf{V}_{n-1},
\\
K_{wv}&=t_1T_{\mu}(\theta),
&
K_{vw}&=t_2T_{\nu}(\phi),
\label{eq:spinful_projected_K}
\\
G_{vv}&=EI_2-\mathbf{V}_{n},
&
G_{ww}&=EI_2-\mathbf{V}_{n-1},
\\
G_{wv}&=-t_1T_{\mu}(\theta),
&
G_{vw}&=-t_2T_{\nu}(\phi).
\label{eq:spinful_projected_G}
\end{align}
Since there is no $Y$ sector, the blocks $\overline{G}_{ab}$ of Eq.~\eqref{eq:Schur} introduced in the general formalism reduce simply to the corresponding $G_{ab}$ blocks.

Substituting Eqs. \eqref{eq:spinful_XLR}--\eqref{eq:spinful_projected_G} into the general transfer matrix gives

\begin{align}
\mathbf{M}_{11}
&=
\frac{1}{t_1^2}
T_{\mu}^{-1}(\theta)
\left(EI_2-\mathbf{V}_{n}\right)
T_{\mu}^{-1}(\theta)
\left(EI_2-\mathbf{V}_{n-1}\right)
\\[1mm] \nonumber
&-\frac{t_2}{t_1}
T_{\mu}^{-1}(\theta)T_{\nu}(\phi),
\\[1mm]
\mathbf{M}_{12}
&=
-\frac{t_2}{t_1^2}
T_{\mu}^{-1}(\theta)
\left(EI_2-\mathbf{V}_{n}\right)
T_{\mu}^{-1}(\theta)T_{\nu}(\phi),
\\[1mm]
\mathbf{M}_{21}
&=
\frac{1}{t_1}
T_{\mu}^{-1}(\theta)
\left(EI_2-\mathbf{V}_{n-1}\right),
\\[1mm]
\mathbf{M}_{22}
&=
-\frac{t_2}{t_1}
T_{\mu}^{-1}(\theta)T_{\nu}(\phi).
\end{align}
In the above equations, the order of the factors is essential. In the presence of a non-Abelian gauge field,
\begin{equation}
[T_{\mu}(\theta),T_{\nu}(\phi)]\neq0,
\end{equation}
and the matrices cannot generally be rearranged as scalar factors. Moreover, when the Zeeman term is not aligned with the gauge axis, $\mathbf{V}_n$ may also fail to commute with the hopping matrices. The inverse $T_{\mu}^{-1}(\theta)$, rather than the Hermitian conjugate $T_{\mu}^{\dagger}(\theta)$, appears because the transfer relation requires the inverse of the forward hopping matrix.

The ordered product of two gauge matrices can be simplified using the Pauli-matrix algebra. In particular,
\begin{equation}
T_{\mu}(\theta)T_{\nu}(\phi)
=
e^{i\gamma\hat{\mathbf{n}}\cdot\boldsymbol{\sigma}},
\label{eq:gauge_composition}
\end{equation}
where
\begin{equation}
\cos\gamma
=
\cos\theta\cos\phi
-
(\hat{\mathbf e}_{\mu}\cdot\hat{\mathbf e}_{\nu})
\sin\theta\sin\phi,
\label{eq:effective_angle}
\end{equation}
and
\begin{align}
\hat{\mathbf n}\sin\gamma
={}&
\sin\theta\cos\phi\,\hat{\mathbf e}_{\mu}
+
\cos\theta\sin\phi\,\hat{\mathbf e}_{\nu}
\nonumber\\
&-
\sin\theta\sin\phi
(\hat{\mathbf e}_{\mu}\times\hat{\mathbf e}_{\nu}).
\label{eq:effective_axis}
\end{align}
For real gauge parameters, Eq. \eqref{eq:gauge_composition} is the standard composition law for $SU(2)$ rotations. For complex $\theta$ or $\phi$, the same algebraic relations remain valid with generally complex $\gamma$ and $\hat{\mathbf n}$. In the transfer-matrix blocks, the ordered combination that appears directly is $T_{\mu}^{-1}(\theta)T_{\nu}(\phi)$, which follows from the same composition formulas after the replacement $\theta\rightarrow-\theta$.
\subsection{Periodic and open-boundary spectra from the transfer matrix}
\label{sec:pbc_obc_gbz}

Before discussing transport through the connected system to the leads, we first analyze the
isolated-chain spectrum under periodic and open boundary conditions.  This is
useful because the same transfer matrix that later determines the scattering
coefficients also contains the generalized Bloch factors responsible for the
non-Hermitian skin effect (NHSE) \cite{YaoWang2018,Yokomizo2019,KunstDwivedi2019}. Indeed, in the following sections, we analysis connection between the transfer matrix method in this work and well-known approach to derive spectral information of spinful non-hermitian system.\\
For a translationally invariant spinful chain, the bulk recursion relation is
\begin{equation}
E\psi_n=\mathbf{V}\psi_n+t_1T_{\mu}(\theta)\psi_{n+1}
+t_2T_{\nu}(\phi)\psi_{n-1},
\label{eq:uniform_bulk_recursion_new}
\end{equation}
where $\psi_n\in\mathbb{C}^2$ and $\mathbf{V}$ is independent of $n$ \footnote{General site-dependent onsite
terms, including spatially modulated $\mathcal{PT}$-symmetric
gain--loss profiles, are described by the ordered product of the local
transfer matrices. Periodic profiles can be treated by enlarging the
unit cell.}.
Introducing the generalized Bloch ansatz
\begin{equation}
\psi_n=\beta^n u,
\qquad u\in\mathbb{C}^2,
\label{eq:generalized_bloch_new}
\end{equation}
gives the non-Bloch Hamiltonian
\begin{equation}
H(\beta)=\mathbf{V}
+t_1T_{\mu}(\theta)\beta
+t_2T_{\nu}(\phi)\beta^{-1}.
\label{eq:Hbeta_new}
\end{equation}
The allowed pairs $(E,\beta)$ satisfy
\begin{equation}
P(E,\beta):=\det\!\left[H(\beta)-EI_2\right]=0.
\label{eq:char_beta_new}
\end{equation}
For the nearest-neighbor spin-$1/2$ problem, multiplication by $\beta^2$
turns Eq. \eqref{eq:char_beta_new} into a fourth-order polynomial in $\beta$.
Thus, for a generic energy there are four generalized Bloch factors
$\beta_j(E)$, counted with algebraic multiplicity.

The connection with the transfer matrix is immediate.  For the uniform model,
the one-site transfer matrix introduced above is
\begin{equation}
\mathbf{M}_{n}(E)=
\begin{pmatrix}
\dfrac{1}{t_1}T_{\mu}^{-1}(\theta)(EI_2-\mathbf{V})
&-\dfrac{t_2}{t_1}T_{\mu}^{-1}(\theta)T_{\nu}(\phi)
\\[1.2ex]
I_2&0_2
\end{pmatrix}.
\label{eq:T_one_site_spectral}
\end{equation}
Substitution of Eq. \eqref{eq:generalized_bloch_new} into the transfer
relation gives
\begin{equation}
\beta
\begin{pmatrix}\beta u\\u\end{pmatrix}=
\mathbf{M}_n(E)
\begin{pmatrix}\beta u\\u\end{pmatrix}
.
\label{eq:beta_eigenvalue_transfer_new}
\end{equation}
Therefore, the generalized Bloch factors are precisely the eigenvalues of the
one-site transfer matrix.\\ Under periodic boundary conditions (PBC), $\beta=e^{ik}$ lies on the unit
circle.  Hence the PBC bands follow from
\begin{equation}
\det\!\left[H(e^{ik})-EI_2\right]=0,
\label{eq:PBC_general_new}
\end{equation}
or equivalently from the transfer-matrix condition
\begin{equation}
\det\!\left[\mathbf M(E)-e^{ik}I_4\right]=0.
\label{eq:PBC_transfer_new}
\end{equation}
For open boundary conditions (OBC), the generalized Bloch factors are not in
general restricted to $|\beta|=1$.  Ordering the four roots according to
\begin{equation}
|\beta_1|\leq|\beta_2|\leq|\beta_3|\leq|\beta_4|,
\end{equation}
the thermodynamic OBC bulk spectrum is selected, for the generic two-component
nearest-neighbor problem, by the generalized-Brillouin-zone (GBZ) condition
\cite{Yokomizo2019}
\begin{equation}
|\beta_2(E)|=|\beta_3(E)|.
\label{eq:GBZ_general_new}
\end{equation}
On the other hand, the corresponding finite-chain spectrum is obtained exactly from the transfer
matrix.  For a chain of $N$ physical sites, the open boundary condition leads to $\det M_{1,1}=0,$ where $M$ is the total transfer matrix which is given by \ref{total transfer matrix}. This gives the exact secular
condition for a finite open chain and remains valid for a nonuniform
system through the ordered product of its local transfer matrices.
In contrast, as mentioned, Eq. \eqref{eq:GBZ_general_new} assumes a translationally
invariant bulk and determines the OBC bulk-band continuum in the
thermodynamic limit. In the following sections, we study the exact spectra of open and periodic boundary conditions for the abelian and non-abelian spinful model following two approaches.
\subsubsection{Abelian limit: exact PBC and OBC spectra}
\label{sec:abelian_closed_spectrum}

We first consider the Abelian case $\mu=\nu$ and assume that the onsite
matrix commutes with the gauge axis (i.e., $[\mathbf{V},\sigma_{\mu}]=0)$. The problem then separates into the two spin eigenstates
\begin{equation}
	\sigma_{\mu}|s\rangle=s|s\rangle,
	\qquad s=\pm1.
\end{equation}
Defining
\begin{equation}
	V_s=\langle s|\mathbf{V}|s\rangle,
	\qquad
	a_s=t_1e^{is\theta},
	\qquad
	b_s=t_2e^{is\phi},
	\label{eq:abelian_as_bs}
\end{equation}
the lattice equation in each spin sector becomes
\begin{equation}
	E\psi_{n,s}
	=
	V_s\psi_{n,s}
	+a_s\psi_{n+1,s}
	+b_s\psi_{n-1,s}.
	\label{eq:abelian_scalar_recursion}
\end{equation}

\textit{Direct spectral derivation---}
We first derive the PBC and OBC spectra directly, without using a transfer
matrix. Substituting the generalized Bloch form
$\psi_{n,s}=\beta^n u_s$ into
Eq. \eqref{eq:abelian_scalar_recursion} gives
\begin{equation}
	E=V_s+a_s\beta+b_s\beta^{-1}.
	\label{eq:abelian_Ebeta}
\end{equation}

Under periodic boundary conditions, $\psi_{n+N,s}=\psi_{n,s}$, so that
$\beta^N=1$. Writing $\beta=e^{ik}$ gives the two spin-resolved PBC bands
\begin{equation}
	E_s^{\mathrm{PBC}}(k)
	=
	V_s+t_1e^{i(k+s\theta)}
	+t_2e^{-ik+is\phi}.
	\label{eq:abelian_PBC_new}
\end{equation}
For complex gauge parameters, these bands generally form closed loops in
the complex-energy plane.

For OBC, multiplication of Eq. \eqref{eq:abelian_Ebeta} by $\beta$ gives
the quadratic equation
\begin{equation}
	a_s\beta^2-(E-V_s)\beta+b_s=0,
	\label{eq:abelian_quadratic}
\end{equation}
whose two roots are
\begin{equation}
	\beta_{s,\pm}(E)
	=
	\frac{(E-V_s)\pm
		\sqrt{(E-V_s)^2-4a_sb_s}}{2a_s}.
	\label{eq:abelian_beta_roots}
\end{equation}
Their product is
\begin{equation}
	\beta_{s,+}\beta_{s,-}
	=
	\frac{b_s}{a_s}.
	\label{eq:abelian_root_product}
\end{equation}
For each decoupled spin sector, the thermodynamic OBC bulk spectrum is
selected by the equal-modulus condition
$|\beta_{s,+}|=|\beta_{s,-}|$ \cite{Yokomizo2019}. Using
Eq. \eqref{eq:abelian_root_product}, the corresponding spin-resolved GBZ
is a circle of radius
\begin{equation}
		r_s\equiv|\beta_s|
		=
		\sqrt{\left|\frac{b_s}{a_s}\right|}
		=
		\sqrt{\left|\frac{t_2}{t_1}\right|}
		\exp\!\left[
		\frac{s}{2}
		\left(
		\operatorname{Im}\theta-\operatorname{Im}\phi
		\right)
		\right].
	\label{eq:abelian_GBZ_radius_new}
\end{equation}
The two roots on this circle can be parametrized as
\begin{equation}
	\beta_s^{(\pm)}(k)
	=
	\sqrt{\frac{b_s}{a_s}}\,e^{\pm ik},
	\label{eq:abelian_beta_param}
\end{equation}
where in general, $b_s/a_s$ is complex, and its square root has two
possible values. We choose one consistently; the other choice only
shifts $k$ by $\pi$ and leaves the results unchanged. Substitution
into Eq. \eqref{eq:abelian_Ebeta} gives the thermodynamic OBC bulk
dispersion
\begin{equation}
		E_s^{\mathrm{OBC}}(k)
		=
		V_s+2\sqrt{a_sb_s}\cos k
		=
		V_s+2\sqrt{t_1t_2e^{is(\theta+\phi)}}\cos k.
	\label{eq:abelian_OBC_continuum_new}
\end{equation}

The exact finite-chain spectrum follows directly by imposing
\begin{equation}
	\psi_{0,s}=\psi_{N+1,s}=0.
	\label{eq:abelian_direct_OBC}
\end{equation}
For distinct roots, the general solution is
\begin{equation}
	\psi_{n,s}
	=
	A_s\left[\beta_s^{(+)}(k)\right]^n
	+
	B_s\left[\beta_s^{(-)}(k)\right]^n.
\end{equation}
The boundary condition at $n=0$ gives $B_s=-A_s$, while the condition at
$n=N+1$ requires
\begin{equation}
	\sin[(N+1)k]=0.
\end{equation}
Consequently,
\begin{equation}
	k_m=\frac{m\pi}{N+1},
	\qquad
	m=1,\ldots,N,
	\label{eq:abelian_quantized_k}
\end{equation}
and the exact finite-chain OBC eigenvalues are
\begin{equation}
		E_{m,s}^{\mathrm{OBC}}
		=
		V_s+
		2\sqrt{t_1t_2e^{is(\theta+\phi)}}
		\cos\!\left(\frac{m\pi}{N+1}\right),
		\quad
		m=1,\cdots,N.
	\label{eq:abelian_exact_OBC_new}
\end{equation}
The corresponding spatial profiles have the form
\begin{equation}
	\psi_{n,s}
	\propto
	\left(\frac{b_s}{a_s}\right)^{n/2}
	\sin\!\left(\frac{m\pi n}{N+1}\right).
	\label{eq:abelian_exact_eigenstate}
\end{equation}
Equations \eqref{eq:abelian_PBC_new} and
\eqref{eq:abelian_exact_OBC_new} agree with the spin-resolved spectra
reported for the Abelian gauge model in Ref. \cite{Sanahal2025}.

\textit{Transfer-matrix recovery---}
We now show that the transfer-matrix method reproduces the same analytical
results. Equation \eqref{eq:abelian_scalar_recursion} can be written as
\begin{equation}
	\begin{pmatrix}
		\psi_{n+1,s}\\
		\psi_{n,s}
	\end{pmatrix}
	=
	\mathbf{M}_s(E)
	\begin{pmatrix}
		\psi_{n,s}\\
		\psi_{n-1,s}
	\end{pmatrix},
\end{equation}
where the one-site transfer matrix in spin sector $s$ is
\begin{equation}
	\mathbf{M}_s(E)
	=
	\begin{pmatrix}
		\dfrac{E-V_s}{a_s} & -\dfrac{b_s}{a_s}\\[1.2ex]
		1 & 0
	\end{pmatrix}.
	\label{eq:abelian_scalar_transfer_new}
\end{equation}
Its characteristic equation is
\begin{equation}
	\det\!\left[\mathbf{M}_s(E)-\beta I_2\right]
	=
	\beta^2-\frac{E-V_s}{a_s}\beta+\frac{b_s}{a_s}
	=0,
	\label{eq:abelian_transfer_characteristic}
\end{equation}
which is identical to Eq. \eqref{eq:abelian_quadratic}. Therefore, the
two transfer-matrix eigenvalues are precisely the generalized Bloch roots
$\beta_{s,\pm}(E)$ in Eq. \eqref{eq:abelian_beta_roots}. The PBC
condition
\begin{equation}
	\det\!\left[
	\mathbf{M}_s(E)-e^{ik}I_2
	\right]=0
	\label{eq:abelian_transfer_PBC}
\end{equation}
then immediately reproduces the PBC dispersion
\eqref{eq:abelian_PBC_new}.

For a uniform chain of $N$ sites, the total spin-resolved transfer matrix
is $\mathbf{M}_s^N(E)$. The OBC condition is
\begin{equation}
	\left[\mathbf{M}_s^N(E)\right]_{11}=0.
	\label{eq:abelian_transfer_OBC}
\end{equation}
To evaluate this condition analytically, we define
\begin{equation}
	d_s
	=
	\det\mathbf{M}_s
	=
	\frac{b_s}{a_s},
	\qquad
	\widetilde{\mathbf{M}}_s
	=
	d_s^{-1/2}\mathbf{M}_s,
	\label{eq:abelian_transfer_normalization}
\end{equation}
so that $\det\widetilde{\mathbf{M}}_s=1$, and introduce
\begin{equation}
	z_s
	=
	\frac{1}{2}\operatorname{Tr}\widetilde{\mathbf{M}}_s
	=
	\frac{E-V_s}{2\sqrt{a_sb_s}}.
	\label{eq:abelian_zs}
\end{equation}
The $N$th power can then be expressed through Chebyshev polynomials of
the second kind,
\begin{equation}
	\mathbf{M}_s^N
	=
	d_s^{N/2}
	\left[
	U_{N-1}(z_s)\widetilde{\mathbf{M}}_s
	-
	U_{N-2}(z_s)I_2
	\right].
	\label{eq:Chebyshev_transfer_new}
\end{equation}
Substitution into Eq. \eqref{eq:abelian_transfer_OBC} gives $U_N(z_s)=0$. The zeros of $U_N$ are
\begin{equation}
	z_s=\cos\!\left(\frac{m\pi}{N+1}\right),
	\qquad
	m=1,\ldots,N.
\end{equation}
Using Eq. \eqref{eq:abelian_zs}, we recover exactly the finite-chain
spectrum in Eq. \eqref{eq:abelian_exact_OBC_new}. Thus, the direct
boundary-condition calculation and the transfer-matrix construction give
the same PBC and OBC spectra.

\textit{Transfer-matrix origin of the bipolar NHSE---}
The localization direction follows directly from the eigenvalues of
$\mathbf{M}_s(E)$. If
\begin{equation}
	\mathbf{M}_s(E)\Phi_{s,\alpha}
	=
	\beta_{s,\alpha}(E)\Phi_{s,\alpha},
\end{equation}
then propagation through $n$ lattice sites gives
\begin{equation}
	\mathbf{M}_s^n(E)\Phi_{s,\alpha}
	=
	\beta_{s,\alpha}^{\,n}(E)\Phi_{s,\alpha}.
	\label{eq:abelian_transfer_growth}
\end{equation}
Thus, the modulus of a transfer eigenvalue directly determines the
spatial growth or decay of the corresponding mode. On the GBZ, the two
eigenvalues in spin sector $s$ have the common modulus $r_s$ given by
Eq. \eqref{eq:abelian_GBZ_radius_new}. The corresponding growth and
localization exponents are
\begin{equation}
	\gamma_s
	=
	\ln r_s
	=
	\frac{1}{2}\ln\left|\frac{t_2}{t_1}\right|
	+
	\frac{s}{2}
	\left(
	\operatorname{Im}\theta-\operatorname{Im}\phi
	\right),
	\qquad
	\kappa_s=-\gamma_s.
	\label{eq:kappa_abelian_new}
\end{equation}
When $r_s>1$, repeated application of $\mathbf{M}_s$ amplifies the mode
toward increasing $n$, leading to accumulation at the right boundary.
When $r_s<1$, the mode decays toward increasing $n$ and accumulates at
the left boundary.

For equal hopping magnitudes, $|t_1|=|t_2|$, the two GBZ radii satisfy
\begin{equation}
	r_{\pm}
	=
	|\beta_{\pm}|
	=
	\exp\!\left[
	\pm\frac{1}{2}
	\left(
	\operatorname{Im}\theta-\operatorname{Im}\phi
	\right)
	\right],
	\qquad
	r_+r_-=1.
	\label{eq:opposite_spin_radii}
\end{equation}
Therefore, if
$\operatorname{Im}\theta-\operatorname{Im}\phi>0$, the $s=+1$
transfer eigenmodes have $r_+>1$ and accumulate at the right boundary,
whereas the $s=-1$ modes have $r_-<1$ and accumulate at the left
boundary. Reversing the sign of the imaginary gauge mismatch exchanges
the two localization directions. This reciprocal relation between the
two spin-resolved transfer-eigenvalue moduli is the transfer-matrix
origin of the bipolar NHSE.\\
In the spin basis that diagonalizes $\sigma_\mu$, the full Abelian
transfer matrix is block diagonal,
$\mathbf{M}_{\mathrm{A}}(E)
=\mathbf{M}_{+}(E)\oplus\mathbf{M}_{-}(E)$,
where
$\mathbf{M}_{\pm}(E)\equiv\mathbf{M}_{s=\pm1}(E)$
are the one-site transfer matrices of the two spin sectors. Their
determinants are
\begin{equation}
	d_{\pm}
	\equiv
	\det\mathbf{M}_{\pm}(E)
	=
	\frac{t_2}{t_1}e^{\pm i(\phi-\theta)}.
\end{equation}
Therefore, the determinant of the complete spinful transfer matrix is
\begin{equation}
	\det\mathbf{M}_{\mathrm{A}}(E)
	=
	d_+d_-
	=
	\left(\frac{t_2}{t_1}\right)^2.
	\label{eq:abelian_full_transfer_determinant}
\end{equation}
For $t_1=t_2$, the full transfer matrix is unimodular, although
$r_+\neq1$ and $r_-\neq1$ whenever
$\operatorname{Im}\theta\neq\operatorname{Im}\phi$. This result demonstrates that, unlike in a single-channel nearest-neighbor chain, unimodularity of the complete multichannel
transfer matrix is not sufficient to exclude the NHSE. It constrains
only the product of all transfer eigenvalues, or equivalently the sum of their spatial growth exponents, while allowing nonzero and opposite
growth rates in different spin sectors. The skin effect must therefore be diagnosed from the individual transfer-eigenvalue moduli or the spin-resolved GBZs.

\begin{figure*}[t]
    \centering
    \includegraphics[width=0.99\textwidth]
    {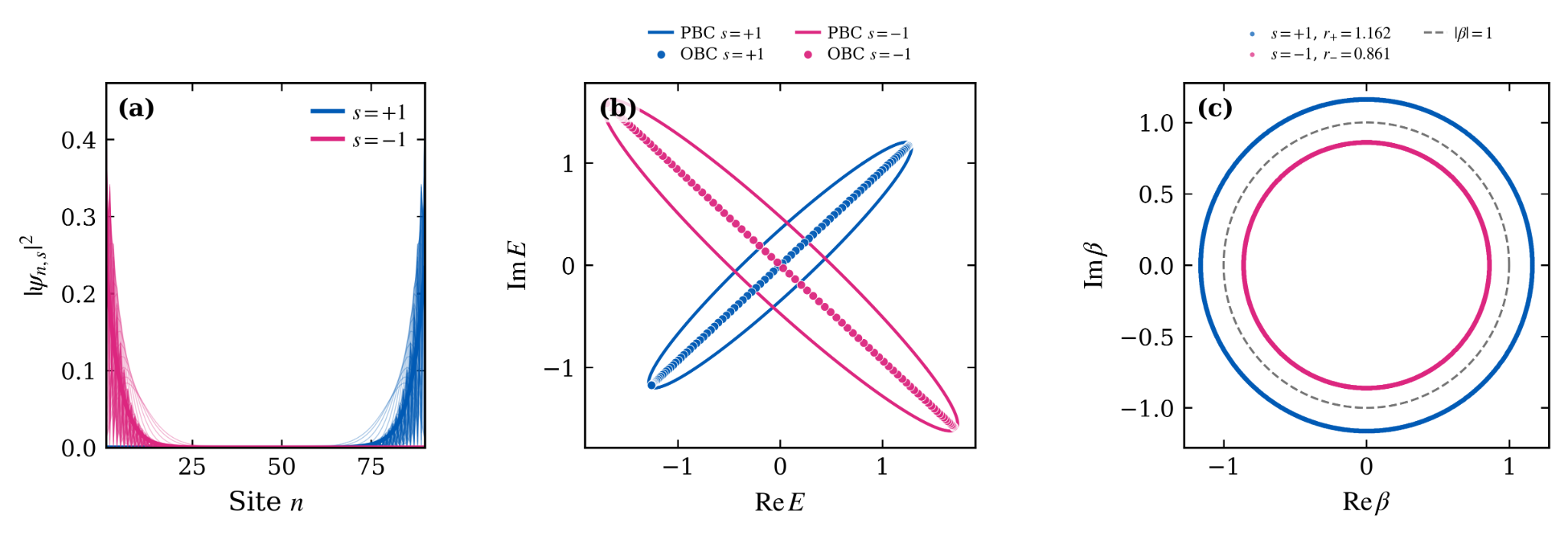}
    \caption{
    Transfer-matrix reproduction of the PBC and OBC spectra and the
    bipolar NHSE previously reported for the Abelian spinful model in
    Ref. \protect\cite{Sanahal2025}. All plots are generated using the
    spin-resolved transfer matrices $\mathbf{M}_s(E)$ for
    $N=90$, $t_1=t_2=1$, $V=0$, $\mu=\nu=z$,
    $\theta=0.3i$, and $\phi=1.5$.
    (a) Normalized spatial profiles of the $2N$ OBC eigenstates.
    Thin curves represent individual states and thick curves represent
    the spin-sector averages, showing accumulation of the $s=+1$ and
    $s=-1$ sectors at the right and left boundaries, respectively.
    (b) PBC spectral loops obtained from
    $\det[\mathbf{M}_s(E)-e^{ik}I_2]=0$ and exact finite-chain OBC
    eigenvalues obtained from
    $[\mathbf{M}_s^N(E)]_{11}=0$. Colors distinguish the two spin
    sectors.
    (c) Spin-resolved GBZs obtained from the transfer eigenvalues
    satisfying $|\beta_{s,+}|=|\beta_{s,-}|$. The dashed curve denotes
    the conventional Bloch unit circle. The radii
    $r_+\simeq1.162$ and $r_-\simeq0.861$ account for the opposite
    localization directions of the two spin sectors.
    }
    \label{fig:abelian_tm_spectrum_skin}
\end{figure*}
Figure \ref{fig:abelian_tm_spectrum_skin} reproduces, using the
spin-resolved transfer matrices, the PBC and OBC spectra and the
bipolar skin localization previously reported for the Abelian gauge
model in Ref. \cite{Sanahal2025}. Plot (a) shows that the two spin
sectors accumulate at opposite boundaries, while plot (b) compares
the corresponding PBC spectral loops with the exact finite-chain OBC
eigenvalues. The colors identify the two decoupled spin sectors
without introducing an additional topological classification.

The transfer eigenvalues provide the direct localization criterion.
For the parameters in Fig. \ref{fig:abelian_tm_spectrum_skin},
Eq. \eqref{eq:opposite_spin_radii} gives
$r_+\simeq1.162$ and $r_-\simeq0.861$, in agreement with the
spin-resolved GBZs in plot (c). These radii are obtained directly
from the moduli of the eigenvalues of $\mathbf{M}_+(E)$ and
$\mathbf{M}_-(E)$. Consequently, Eq. \eqref{eq:abelian_transfer_growth}
shows that the $s=+1$ modes grow toward increasing $n$ and accumulate
at the right boundary, whereas the $s=-1$ modes decay in the same
direction and accumulate at the left boundary. The figure therefore
directly verifies the transfer-matrix mechanism underlying the
spin-resolved bipolar NHSE.
\subsubsection{Non-Abelian limit: PBC and OBC spectra from the
	transfer matrix}
\label{sec:nonabelian_closed_spectrum}

For $\mu\neq\nu$ and generic
$\sin\theta\sin\phi\neq0$, the two hopping matrices do not
commute:
\begin{equation}
	\bigl[T_{\mu}(\theta),T_{\nu}(\phi)\bigr]
	=
	-2i\sin\theta\sin\phi\,
	\epsilon_{\mu\nu\lambda}\sigma_{\lambda}
	\neq0.
	\label{eq:nonabelian_commutator}
\end{equation}
Consequently, the two spin channels cannot be diagonalized
simultaneously by a position-independent spin rotation. For
analytical simplicity, we first take a spin-independent onsite
matrix (i.e., $\mathbf{V}=VI_2$). If a uniform Zeeman field is included through
$\mathbf{V}=VI_2+\mathbf{B}\cdot\boldsymbol{\sigma}$, the only
modification below is
$\mathbf{h}(\beta)\rightarrow\mathbf{B}+\mathbf{h}(\beta)$.
We set $\mathbf{B}=0$ because the additional cross terms obscure
the expressions without changing the transfer-matrix construction.

\textit{Direct spectral derivation---}
Using
\begin{equation}
	T_{\mu}(\theta)
	=
	\cos\theta\,I_2+i\sin\theta\,\sigma_{\mu},
	\label{eq:gauge_pauli_expansion_nonabelian}
\end{equation}
the non-Bloch Hamiltonian in Eq. \eqref{eq:Hbeta_new} can be
written as
\begin{equation}
	H(\beta)
	=
	h_0(\beta)I_2+
	\mathbf{h}(\beta)\cdot\boldsymbol{\sigma},
	\label{eq:nonabelian_H_decomposition}
\end{equation}
where
\begin{align}
	h_0(\beta)
	&=
	V+t_1\beta\cos\theta
	+t_2\beta^{-1}\cos\phi,
	\label{eq:h0_nonabelian_new}\\
	\mathbf{h}(\beta)
	&=
	it_1\beta\sin\theta\,\hat{\mathbf e}_{\mu}
	+it_2\beta^{-1}\sin\phi\,\hat{\mathbf e}_{\nu}.
	\label{eq:hvec_nonabelian_new}
\end{align}
The Pauli-matrix identity
\begin{equation}
	\bigl(\mathbf{h}\cdot\boldsymbol{\sigma}\bigr)^2
	=
	\bigl(\mathbf{h}\cdot\mathbf{h}\bigr)I_2
	\label{eq:pauli_vector_square}
\end{equation}
follows because the antisymmetric contribution from
$\epsilon_{abc}\sigma_c$ vanishes when contracted with the
symmetric product $h_a h_b$. Therefore,
\begin{equation}
	\det\!\left[H(\beta)-EI_2\right]
	=
	\bigl[h_0(\beta)-E\bigr]^2
	-\mathbf{h}(\beta)\cdot\mathbf{h}(\beta),
	\label{eq:nonabelian_characteristic_identity}
\end{equation}
and the two non-Bloch energy branches are
\begin{equation}
	E_{\pm}(\beta)
	=
	h_0(\beta)
	\pm
	\sqrt{\mathbf{h}(\beta)\cdot\mathbf{h}(\beta)}.
	\label{eq:nonabelian_Ebeta_new}
\end{equation}
Here
\begin{equation}
	\mathbf{h}\cdot\mathbf{h}
	\equiv
	\sum_{a=x,y,z}h_a^2,
\end{equation}
is an algebraic product without complex conjugation. This is
required because Eq. \eqref{eq:nonabelian_Ebeta_new} follows from
the characteristic polynomial rather than from a Hermitian norm.

Defining
\begin{equation}
	c_{\mu\nu}
	=
	\hat{\mathbf e}_{\mu}\cdot\hat{\mathbf e}_{\nu},
\end{equation}
we obtain
\begin{align}
	\mathbf{h}(\beta)\cdot\mathbf{h}(\beta)
	={}&
	-t_1^2\beta^2\sin^2\theta
	-t_2^2\beta^{-2}\sin^2\phi
	\nonumber\\
	&-2t_1t_2c_{\mu\nu}\sin\theta\sin\phi.
	\label{eq:hdoth_nonabelian_new}
\end{align}
For two distinct Cartesian Pauli axes,
$c_{\mu\nu}=0$.

Under PBC, $\psi_{n+N}=\psi_n$ and therefore
$\beta=e^{ik_m}$, where
\begin{equation}
	k_m=\frac{2\pi m}{N},
	\qquad m=0,\ldots,N-1.
\end{equation}
In the thermodynamic limit, $k$ becomes continuous and the PBC
bands are
\begin{equation}
	E_{\pm}^{\mathrm{PBC}}(k)
	=
	h_0(e^{ik})
	\pm
	\sqrt{
		\mathbf{h}(e^{ik})\cdot\mathbf{h}(e^{ik})
	}.
	\label{eq:nonabelian_PBC_new}
\end{equation}
Unlike the Abelian bands, these two branches cannot generally be
assigned to fixed spin projections. Their spinors vary with $k$
because the noncommuting hopping matrices rotate the spin about
different axes \cite{WuNonAbelian2026}. For OBC, let $\varepsilon=E-V$. Because $P(E,\beta)$ in Eq. \eqref{eq:char_beta_new} contains
negative powers of $\beta$, we define the corresponding quartic
polynomial as
\begin{equation}
	\mathcal{P}(E,\beta)
	\equiv
	\beta^2P(E,\beta).
\end{equation}
Using Eqs. \eqref{eq:h0_nonabelian_new} and
\eqref{eq:hdoth_nonabelian_new}, one obtains
\begin{align}
	0=&\mathcal{P}(E,\beta)
	=
	t_1^2\beta^4
	-2t_1\varepsilon\cos\theta\,\beta^3
	\nonumber\\
	&+
	\Bigl[
	\varepsilon^2
	+2t_1t_2
	\bigl(
	\cos\theta\cos\phi
	+c_{\mu\nu}\sin\theta\sin\phi
	\bigr)
	\Bigr]\beta^2
	\nonumber\\
	&-2t_2\varepsilon\cos\phi\,\beta
	+t_2^2.
	\label{eq:nonabelian_quartic_new}
\end{align}
For $t_1t_2\neq0$, this equation has four nonzero roots
$\beta_j(E)$, counted with algebraic multiplicity. In the Abelian
limit $\mu=\nu$ and $c_{\mu\nu}=1$, the quartic factorizes into
the two spin-resolved quadratic equations obtained previously.

Let $u_j$ denote the spinor associated with $\beta_j$:
\begin{equation}
	\left[H(\beta_j)-EI_2\right]u_j=0.
	\label{eq:nonabelian_null_spinors}
\end{equation}
For distinct roots, the bulk solution is
\begin{equation}
	\psi_n
	=
	\sum_{j=1}^{4}
	c_j\beta_j^n u_j.
	\label{eq:nonabelian_general_solution}
\end{equation}
Imposing $\psi_0=\psi_{N+1}=0$ gives
\begin{equation}
	\mathcal{B}_N(E)\mathbf{c}=0,
\end{equation}
where $\mathbf{c}=(c_1,c_2,c_3,c_4)^T$ and
\begin{equation}
	\mathcal{B}_N(E)=
	\begin{pmatrix}
		u_1 & u_2 & u_3 & u_4\\
		\beta_1^{N+1}u_1 &
		\beta_2^{N+1}u_2 &
		\beta_3^{N+1}u_3 &
		\beta_4^{N+1}u_4
	\end{pmatrix}.
	\label{eq:nonabelian_boundary_matrix}
\end{equation}
Each displayed entry is a two-component column, so that
$\mathcal{B}_N(E)$ is a $4\times4$ matrix. The direct finite-chain
OBC secular equation is therefore
\begin{equation}
	\det\mathcal{B}_N(E)=0.
	\label{eq:nonabelian_finite_boundary_det}
\end{equation}

\textit{Transfer-matrix recovery.---}
To connect these results explicitly with the transfer-matrix
formalism, define the left- and right-directed hopping matrices
\begin{equation}
	L=t_1T_{\mu}(\theta),
	\qquad
	R=t_2T_{\nu}(\phi).
	\label{eq:nonabelian_LR_definition}
\end{equation}
Then, the one-site transfer matrix is given by:
\begin{equation}
	\mathbf{M}(E)=
	\begin{pmatrix}
		L^{-1}(EI_2-VI_2) & -L^{-1}R\\
		I_2 & 0_2
	\end{pmatrix}.
	\label{eq:nonabelian_one_site_transfer}
\end{equation}
Substitution of $\psi_n=\beta^n u$ gives
\begin{equation}
	\mathbf{M}(E)
	\begin{pmatrix}
		\beta u\\
		u
	\end{pmatrix}
	=
	\beta
	\begin{pmatrix}
		\beta u\\
		u
	\end{pmatrix}.
	\label{eq:nonabelian_transfer_eigenmode}
\end{equation}
Thus, the four generalized Bloch factors are exactly the
eigenvalues of the $4\times4$ one-site transfer matrix.

More explicitly, block-determinant reduction gives
\begin{align}
	\det\!\left[\mathbf{M}(E)-\beta I_4\right]
	&=
	\frac{\beta^2}{\det L}
	\det\!\left[H(\beta)-EI_2\right]
	\nonumber\\
	&=
	\frac{1}{t_1^2}\,
	\mathcal{P}(E,\beta),
	\label{eq:nonabelian_transfer_characteristic}
\end{align}
where $\det L=t_1^2$. Therefore,
Eq. \eqref{eq:nonabelian_quartic_new} is precisely the
characteristic equation of the transfer matrix, rather than an
independent Bloch-Hamiltonian construction.

For a periodic ring of $N$ sites, propagation around the ring
must reproduce the initial state. The exact PBC condition is
\begin{equation}
	\det\!\left[\mathbf{M}^N(E)-I_4\right]=0.
	\label{eq:nonabelian_finite_PBC_transfer}
\end{equation}
Equivalently, one of the transfer eigenvalues must satisfy
$\beta^N=1$, giving
\begin{equation}
	\det\!\left[
	\mathbf{M}(E)-e^{ik_m}I_4
	\right]=0.
	\label{eq:nonabelian_PBC_transfer}
\end{equation}
Using Eq. \eqref{eq:nonabelian_transfer_characteristic},
this condition reproduces the PBC bands in
Eq. \eqref{eq:nonabelian_PBC_new}.

For an open chain, write
\begin{equation}
	\mathbf{M}^N(E)
	=
	\begin{pmatrix}
		\mathbf{M}_{11}^{(N)}(E) &
		\mathbf{M}_{12}^{(N)}(E)\\
		\mathbf{M}_{21}^{(N)}(E) &
		\mathbf{M}_{22}^{(N)}(E)
	\end{pmatrix},
	\label{eq:nonabelian_transfer_power_blocks}
\end{equation}
where every block is $2\times2$. Starting with $\psi_0=0$ and an arbitrary nonzero boundary spinor $\psi_1=u_{\rm OBC}$, we can obtain the following
\begin{equation}
\left[\mathbf{M}^{N}(E)\right]_{11}u_{\rm OBC}=0,\qquad u_{\rm OBC}\neq0. 
\end{equation}
A nonzero spinor $u_{\rm OBC}$ exists if and only if
\begin{equation}
		\det\!
		\left[\mathbf{M}^N(E)\right]_{11}
		=0.
	\label{eq:nonabelian_OBC_transfer_exact}
\end{equation}
Equation \eqref{eq:nonabelian_OBC_transfer_exact} is the exact
finite-chain OBC secular equation. For a uniform diagonalizable
transfer matrix, expanding $\mathbf{M}^N$ in its four
eigenmodes reduces this condition to
Eq. \eqref{eq:nonabelian_finite_boundary_det}. The transfer-matrix
form is more general because it remains valid at defective points
and, after replacing $\mathbf{M}^N$ by an ordered product, for
nonuniform chains.

To obtain the thermodynamic OBC bulk spectrum, order the four
transfer eigenvalues according to
\begin{equation}
	|\beta_1|
	\leq
	|\beta_2|
	\leq
	|\beta_3|
	\leq
	|\beta_4|.
\end{equation}
Each transfer eigenmode grows or decays with the spatial rate
$\ln|\beta_j|$. In the large-$N$ boundary determinant, a
continuous OBC band can form only when the two middle transfer
modes have the same exponential weight:
\begin{equation}
	|\beta_2(E)|=|\beta_3(E)|
	\quad\Longleftrightarrow\quad
	\ln|\beta_2(E)|=\ln|\beta_3(E)|.
	\label{eq:nonabelian_middle_growth}
\end{equation}
Otherwise, one of these contributions exponentially dominates,
and cancellation at the distant boundary occurs only at isolated
energies. The non-Abelian GBZ and its OBC energies are therefore
obtained from
\begin{equation}
	\mathcal{P}(E,\beta_2)
	=
	\mathcal{P}(E,\beta_3)
	=
	0,
	\qquad
	|\beta_2|=|\beta_3|.
	\label{eq:nonabelian_GBZ_pair}
\end{equation}
The equal-modulus condition does not mean
$\beta_2=\beta_3$; it means only that these two transfer
eigenmodes have equal spatial growth rates.

In the Abelian limit, $\mathbf{M}(E)$ separates into two scalar
$2\times2$ transfer matrices. The determinant of each block fixes
the product of one pair of roots, and the equal-modulus condition
then produces a circular spin-resolved GBZ. In the non-Abelian
case, no constant spin basis block diagonalizes $\mathbf{M}(E)$.
Only the collective product of all four transfer eigenvalues is
fixed, so the common modulus of $\beta_2$ and $\beta_3$ may vary
with energy. This is why the non-Abelian GBZ is generally
noncircular and why its transfer eigenvectors possess
energy-dependent spin composition.

An instructive analytically solvable non-Abelian limit is obtained
for two orthogonal Pauli axes and
\begin{equation}
	\theta=\phi=\frac{\pi}{2},
	\qquad
	t_1,t_2>0.
	\label{eq:maximal_nonabelian_limit}
\end{equation}
The hopping matrices are then
$L=it_1\sigma_{\mu}$ and $R=it_2\sigma_{\nu}$, with
$[L,R]\neq0$. The corresponding transfer matrix is
\begin{equation}
	\mathbf{M}_{\perp}(E)
	=
	\begin{pmatrix}
		-\dfrac{i(E-V)}{t_1}\sigma_{\mu}
		&
		-\dfrac{t_2}{t_1}\sigma_{\mu}\sigma_{\nu}\\[1.2ex]
		I_2&0_2
	\end{pmatrix}.
	\label{eq:nonabelian_special_transfer}
\end{equation}
Its characteristic equation,
Eq. \eqref{eq:nonabelian_transfer_characteristic}, reduces to
\begin{equation}
	t_1^2\beta^4
	+(E-V)^2\beta^2
	+t_2^2=0.
	\label{eq:nonabelian_special_quartic}
\end{equation}
Introducing $y=\beta^2$ gives
\begin{equation}
t_1^2y^2+(E-V)^2y+t_2^2=0,
\end{equation}
whose roots satisfy
\begin{equation}
y_1y_2=\frac{t_2^2}{t_1^2}.
\end{equation}
The middle-root OBC condition requires
\begin{equation}
|y_1|=|y_2|=\frac{t_2}{t_1},
\qquad
|\beta|=\sqrt{\frac{t_2}{t_1}}.
\label{eq:nonabelian_special_radius}
\end{equation}
We may therefore parameterize the two roots as
\begin{equation}
y_{1,2}
=
\frac{t_2}{t_1}e^{\pm2ik},
\qquad 0\leq k<\pi.
\end{equation}
Using their sum in the quadratic equation gives the
thermodynamic OBC bands
\begin{equation}
E_{\pm}^{\mathrm{OBC}}(k)
=
V\pm
\sqrt{-2t_1t_2\cos(2k)}.
\label{eq:nonabelian_special_OBC}
\end{equation}
For a finite chain, the exact eigenvalues are still determined by
Eq. \eqref{eq:nonabelian_OBC_transfer_exact} with
$\mathbf{M}=\mathbf{M}_{\perp}$.

Under PBC, setting $\beta=e^{ik}$ in the same transfer
characteristic equation gives
\begin{equation}
E_{\pm}^{\mathrm{PBC}}(k)
=
V\pm i
\sqrt{
	t_1^2e^{2ik}
	+t_2^2e^{-2ik}
}.
\label{eq:nonabelian_special_PBC}
\end{equation}
Thus, both the PBC and OBC spectra in this genuinely non-Abelian
limit follow from different boundary conditions imposed on the
same transfer matrix.

\begin{figure*}[t]
    \centering
    \includegraphics[width=0.99\textwidth]
    {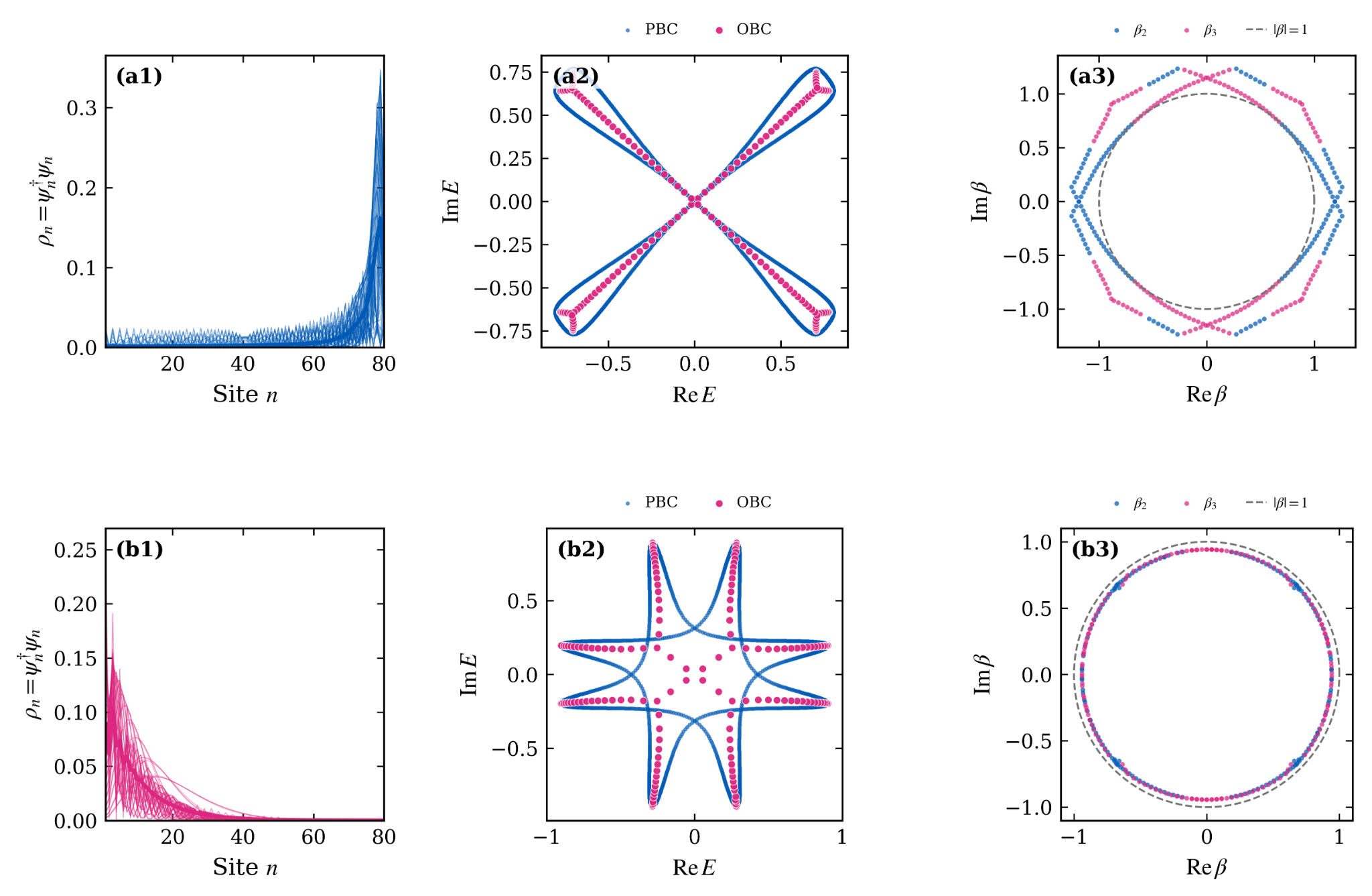}
    \caption{
    Transfer-matrix PBC and OBC spectra, skin profiles, and generalized
    Brillouin zones of the non-Abelian spinful chain with $N=80$, $V=0$,
    $R=J_R e^{i\theta_R\sigma_x}$, and
    $L=J_L e^{i\theta_L\sigma_y}$. The upper row uses
    $J_L=J_R=0.6$, $\theta_L=0.5$, and $\theta_R=-1.5$, while the lower
    row uses $J_L=0.7$, $J_R=0.6$, $\theta_L=1.2$, and
    $\theta_R=-1.5$.
    (a1,b1) Normalized spin-summed densities of all finite-chain OBC
    eigenstates.
    (a2,b2) PBC spectra obtained from
    $\det[\mathbf{M}(E)-e^{ik}I_4]=0$ and exact finite-chain OBC
    eigenvalues obtained from
    $\det\{[\mathbf{M}^{N}(E)]_{11}\}=0$.
    (a3,b3) Non-Abelian GBZs formed by the two middle transfer eigenvalues
    satisfying $|\beta_2|=|\beta_3|$; the dashed curve is the unit circle.
    The two parameter sets exhibit predominantly right- and left-boundary
    accumulation, respectively.
    }
    \label{fig:nonabelian_tm_benchmark}
\end{figure*}

To benchmark the transfer-matrix construction against the non-Hermitian
non-Abelian spinful lattice studied in Ref. \cite{WuNonAbelian2026}, we take
\begin{equation}
R=J_R e^{i\theta_R\sigma_x},
\qquad
L=J_L e^{i\theta_L\sigma_y}.
\label{eq:nonabelian_benchmark_links}
\end{equation}
In the notation introduced above, this corresponds to
$(t_1,\mu,\theta)=(J_L,y,\theta_L)$ and
$(t_2,\nu,\phi)=(J_R,x,\theta_R)$. All results in
Fig. \ref{fig:nonabelian_tm_benchmark} are generated from the one-site
transfer matrix in Eq. \eqref{eq:nonabelian_one_site_transfer}.

The PBC spectrum follows from
Eq. \eqref{eq:nonabelian_PBC_transfer} by restricting a transfer eigenvalue
to $\beta=e^{ik}$. For the finite-chain OBC calculation, the upper-left
block of $\mathbf{M}^{N}(E)$ is generated directly from the transfer
recurrence. Defining $F_0(E)=0_2$ and $F_1(E)=I_2$, we use
\begin{equation}
F_{n+1}(E)
=
L^{-1}(EI_2-VI_2)F_n(E)-L^{-1}RF_{n-1}(E),
\label{eq:nonabelian_benchmark_recurrence}
\end{equation}
for which
\begin{equation}
F_{N+1}(E)=\left[\mathbf{M}^{N}(E)\right]_{11}.
\label{eq:nonabelian_benchmark_boundary_block}
\end{equation}
The matrix polynomial $F_{N+1}(E)$ is linearized only to obtain numerical
root seeds; every root is subsequently refined and verified using the exact
OBC condition in Eq. \eqref{eq:nonabelian_OBC_transfer_exact}. No
finite-chain Hamiltonian diagonalization is used.

For each OBC eigenvalue, the boundary spinor $u_{\mathrm{OBC}}$ is obtained
as the right null vector of $[\mathbf{M}^{N}(E)]_{11}$. Iterating the same
one-site transfer matrix from
$(\psi_1,\psi_0)^T=(u_{\mathrm{OBC}},0)^T$ reconstructs the normalized
spin-summed density
$\rho_n=\psi_n^{\dagger}\psi_n$. The non-Abelian GBZ is obtained from the
four transfer eigenvalues ordered by modulus. The finite-chain OBC roots are
used as initial samples, which are then refined onto the middle-root locus in
Eq. \eqref{eq:nonabelian_GBZ_pair}.

Figure \ref{fig:nonabelian_tm_benchmark} compares the two parameter sets
using the same three transfer-matrix diagnostics as in the Abelian case.
For the symmetric amplitudes, the OBC states accumulate predominantly at the
right boundary, as shown in plot (a1). Changing to the second parameter set
reorganizes both the complex-energy spectrum and the transfer eigenmodes,
producing predominantly left-localized OBC states in plot (b1). Because the
two gauge links do not commute, these profiles are spin-hybridized total
densities rather than a superposition of two independently conserved spin
sectors.

Plots (a2) and (b2) show that the PBC loops and the exact finite-chain OBC
eigenvalues follow from different boundary conditions imposed on the same
one-site transfer matrix. Plots (a3) and (b3) display the corresponding
middle-root loci. Their generally noncircular form and parameter dependence
reflect the collective competition among all four transfer eigenmodes. The
agreement between the OBC spectral displacement, the boundary accumulation,
and the condition $|\beta_2|=|\beta_3|$ provides a direct transfer-matrix
description of the non-Abelian NHSE.
\section{Analytical transport boundary conditions}
\label{sec:analytic_transport}

We now specialize the multichannel boundary-condition classification
summarized in Table \ref{tab:transport_BC} to a finite spinful chain.
Our aim is to derive explicit finite-length conditions for the poles,
zeros, reflectionless states, transparency, and invisibility in the
Abelian and non-Abelian regimes
\cite{ghaemiPRA2021, ghaemiPRB2023}. The bulk transfer matrix is the same one
used to determine the PBC and OBC spectra in
Sec. \ref{sec:pbc_obc_gbz}; coupling the sample to the leads modifies
only its first and last propagation steps. This connects the transport
conditions to the transfer eigenvalues and generalized Bloch factors
of the closed system
\cite{Mostafazadeh2020,KunstDwivedi2019}.

Consider a uniform sample of $N\geq2$ sites with
\begin{equation}
	L=t_1T_{\mu}(\theta),
	\qquad
	R=t_2T_{\nu}(\phi),
	\label{eq:transport_LR}
\end{equation}
uniform onsite matrix $\mathbf{V}$, and spin-independent lead
couplings $wI_2$, with $w<0$. Using one physical site as the
propagation step, the transfer matrices at the left interface, in the
bulk, and at the right interface are
\begin{align}
	\mathbf{T}_1(E)
	&=
	\begin{pmatrix}
		L^{-1}(EI_2-\mathbf{V}) & -wL^{-1}\\
		I_2&0_2
	\end{pmatrix},
	\label{eq:Tleft_transport}
	\\
	\mathbf{T}(E)
	&=
	\begin{pmatrix}
		L^{-1}(EI_2-\mathbf{V}) & -L^{-1}R\\
		I_2&0_2
	\end{pmatrix},
	\label{eq:Tbulk_transport}
	\\
	\mathbf{T}_N(E)
	&=
	\begin{pmatrix}
		w^{-1}(EI_2-\mathbf{V}) & -w^{-1}R\\
		I_2&0_2
	\end{pmatrix}.
	\label{eq:Tright_transport}
\end{align}
The real-space transfer matrix across the complete scattering region
is therefore
\begin{equation}
	\mathbf{M}_N(E)
	=
	\mathbf{T}_N(E)\,
	\mathbf{T}^{N-2}(E)\,
	\mathbf{T}_1(E).
	\label{eq:finite_transport_realM}
\end{equation}
In particular, the bulk factor $\mathbf{T}^{N-2}(E)$ contains the
same transfer eigenvalues that determine the generalized Bloch
factors in the closed-chain problem.

Specializing the wave-matching transformation introduced previously
to the two spin channels gives $	\mathcal{M}_N(E)= \mathcal{Q}^{-1}\mathbf{M}_N(E)\mathcal{Q}$, where $\mathcal{Q}$ is given in Eq.~\eqref{eq:Qtransform}.\\ The following subsections evaluate the block conditions listed in
Table \ref{tab:transport_BC} for the Abelian and non-Abelian gauge
fields.

\subsection{Abelian gauge field: exact finite-length transport equations}
\label{sec:abelian_analytic_transport}

For $\mu=\nu$ and $[\mathbf{V},\sigma_{\mu}]=0$, the two spin sectors decouple,
as in the Abelian gauge model of Ref. \cite{Sanahal2025}.
Let $s=\pm1$ denote the eigenvalue of $\sigma_{\mu}$ and define
\begin{align}
a_s&=t_1e^{is\theta},
&b_s&=t_2e^{is\phi},
&V_s&=\langle s|\mathbf{V}|s\rangle,
\label{eq:ab_transport_abs}
\\
q_s&=\sqrt{a_sb_s},
&d_s&=\frac{b_s}{a_s},
&z_s&=\frac{E-V_s}{2q_s},
&\omega_s&=\frac{q_s}{w}.
\label{eq:ab_transport_qd}
\end{align}
A consistent branch of the square root is understood.  The two combinations
$q_s$ and $d_s$ have a useful physical interpretation,
\begin{equation}
q_s=\sqrt{t_1t_2}\,e^{is(\theta+\phi)/2},
\qquad
d_s=\frac{t_2}{t_1}e^{is(\phi-\theta)}.
\label{eq:ab_qd_gauge}
\end{equation}
Thus $q_s$ depends on the product of the directed hoppings and on the sum of the gauge parameters, whereas $d_s$ measures the directed amplification and contains their difference. Ref. \cite{ ghaemiPRA2021} showed that these parameters play an important role in understanding the transport effects.
The scalar bulk transfer matrix of spin sector $s$ is
\begin{equation}
\mathbf{T}_s(E)=
\begin{pmatrix}
\dfrac{E-V_s}{a_s}&-\dfrac{b_s}{a_s}\\[1ex]
1&0
\end{pmatrix}.
\end{equation}
Using the Chebyshev representation of $\mathbf{T}_s^{N-2}$ and including the
two lead contacts exactly, Eq. \eqref{eq:finite_transport_realM} reduces to
\begin{equation}
\mathbf{M}_s(E)
=d_s^{(N-1)/2}
\begin{pmatrix}
\omega_s U_N(z_s)&-U_{N-1}(z_s)\\
U_{N-1}(z_s)&-\omega_s^{-1}U_{N-2}(z_s)
\end{pmatrix}.
\label{eq:abelian_total_real_transport}
\end{equation}
Here $U_n(z)$ is the Chebyshev polynomial of the second kind.  Equation
\eqref{eq:abelian_total_real_transport} is the direct spin-resolved analogue
of the factorization used in Refs. \cite{ghaemiPRA2021, ghaemiPRB2023}: the factor
$d_s^{(N-1)/2}$ contains the directed amplification, while the normalized
matrix depends only on $q_s$, $w$, and $E-V_s$.

For compactness, define
\begin{align}
\mathcal{F}_s^{\pm}(E)
&=\omega_sU_N(z_s)-\omega_s^{-1}U_{N-2}(z_s)
\pm2iU_{N-1}(z_s),
\label{eq:Fpm_abelian_transport}
\\
\mathcal{R}_s(E)
&=\omega_sU_N(z_s)+\omega_s^{-1}U_{N-2}(z_s).
\label{eq:R_abelian_transport}
\end{align}
Transforming Eq. \eqref{eq:abelian_total_real_transport} to the propagating
basis gives the four scalar transfer blocks
\begin{align}
\mathcal{M}_{11}^{(s)}
&=\frac{d_s^{(N-1)/2}}{2}\,\mathcal{F}_s^{+},
&
\mathcal{M}_{22}^{(s)}
&=\frac{d_s^{(N-1)/2}}{2}\,\mathcal{F}_s^{-},
\label{eq:abelian_A_D}
\\
\mathcal{M}_{12}^{(s)}
&=-\frac{i\,d_s^{(N-1)/2}}{2}\,\mathcal{R}_s,
&
\mathcal{M}_{21}^{(s)}
&=\frac{i\,d_s^{(N-1)/2}}{2}\,\mathcal{R}_s.
\label{eq:abelian_B_C}
\end{align}
These expressions immediately yield all transport boundary conditions.
\subsection{Quasibound states and lasing}
\label{sec:quasibound_lasing}

A quasibound state has no incoming components, $\chi_L=\chi_R=0$, and therefore satisfies 
\begin{equation} \mathcal{M}_{22}(E)\eta_L=0,\qquad \eta_L\neq0,\qquad \eta_R=\mathcal{M}_{12}(E)\eta_L. 
\end{equation}
For the time convention $e^{-iEt}$, a pole
$E=E_R+iE_I$ with $E_I<0$ is a decaying quasibound resonance,
whereas $E_I=0$ gives a stationary lasing state. A pole with
$E_I>0$ instead represents a linearly unstable outgoing mode that
already grows in time.
In the Abelian case, the two spin sectors decouple and
\begin{equation}
	\mathcal{M}_{22}^{(s)}
	=
	\frac{d_s^{(N-1)/2}}{2}\,
	\mathcal{F}_s^{-}(E).
\end{equation}
The quasibound and lasing condition in sector $s$ is therefore $\mathcal{F}_s^{-}(E)=0$. Because $d_s\neq0$, the prefactor $d_s^{(N-1)/2}$ cannot create or
remove a zero of $\mathcal{M}_{22}^{(s)}$. Consequently, at fixed
$q_s$, $V_s$, $w$, and $N$, the quasibound energies and lasing
threshold are independent of $d_s$. The factor $d_s$ nevertheless
controls the spatial amplification and direction-dependent
transmission. This invariance does not apply when the microscopic
parameters are varied in a way that changes both $q_s$ and $d_s$.

To follow the poles toward the real axis without introducing an
additional spin direction, we supplement the sample by a uniform
passive loss and a spin-independent pump,
\begin{equation}
	V(g)=-i\Gamma I_2+igI_2.
	\label{eq:scalar_pump}
\end{equation}
Since
$EI_2-V(g)=[E+i(\Gamma-g)]I_2$, every pole is shifted vertically
according to
\begin{equation}
	E_{\rm pole}(g)=E_{\rm pole}(0)+ig.
	\label{eq:pole_vertical_shift}
\end{equation}
Thus, a damped pole initially located at
$E_{\rm pole}(0)=E_R-i\gamma$ reaches the real axis at the positive
pump value $g=\gamma$. An upper-half-plane pole instead reaches the
real axis at negative $g$, corresponding to the addition of loss
rather than gain.

For real $E$ and $g$, we locate these crossings numerically using the
singular values of $\mathcal{M}_{22}$. Specifically, let
\begin{equation}
	\mathcal{M}_{22}
	=
	U
	\begin{pmatrix}
		\sigma_{\max} & 0\\
		0 & \sigma_{\min}
	\end{pmatrix}
	W^{\dagger},
	\qquad
	\sigma_{\max}\geq\sigma_{\min}\geq0,
	\label{eq:M22_svd}
\end{equation}
be its singular-value decomposition. Equivalently,
\begin{equation}
	\sigma_{\max,\min}
	=
	\sqrt{
		\lambda_{\max,\min}
		\!\left(
		\mathcal{M}_{22}^{\dagger}\mathcal{M}_{22}
		\right)},
	\label{eq:M22_singular_values}
\end{equation}
where $\lambda_{\max}$ and $\lambda_{\min}$ denote the largest and
smallest eigenvalues, respectively. We define the normalized
real-axis residual
\begin{equation}
	\mathcal{R}_{\rm pole}(E,g)
	= \frac{\sigma_{\min}[\mathcal{M}_{22}(E,g)]}
	{\sigma_{\max}[\mathcal{M}_{22}(E,g)]}
	.
	\label{eq:lasing_residual}
\end{equation}
For the generic rank-one singularities considered here,
$\sigma_{\max}>0$, and
\begin{equation}
	\mathcal{R}_{\rm pole}=0
	\quad\Longleftrightarrow\quad
	\sigma_{\min}=0
	\quad\Longleftrightarrow\quad
	\det\mathcal{M}_{22}=0.
\end{equation}
The right singular vector associated with $\sigma_{\min}$ approaches the left-outgoing null vector $\eta_L$ at the crossing.
To characterize its spin polarization, let $\eta_{L,c}$ be the
normalized right-null vector of the selected positive-energy
crossing in case $c\in\{{\rm A},{\rm NA}\}$:
\begin{align}
\mathcal{M}_{22}(E_{\rm L}^{c},g_c)\eta_{L,c}&=0,\qquad \eta_{L,c}^{\dagger}\eta_{L,c}=1
\end{align}
Its Bloch vector is defined by
\begin{equation}
	\mathbf{S}_{c}
	\equiv
	\frac{
		\eta_{L,c}^{\dagger}
		\boldsymbol{\sigma}
		\eta_{L,c}
	}{
		\eta_{L,c}^{\dagger}\eta_{L,c}
	}
	=
	\left(
	S_{c x},S_{c y},S_{c z}
	\right),
	\qquad
	c\in\{{\rm A},{\rm NA}\},
	\label{eq:lasing_bloch_vector}
\end{equation}
where
$\boldsymbol{\sigma}=(\sigma_x,\sigma_y,\sigma_z)$. Thus
$S_{c j}=\eta_{L,c}^{\dagger}\sigma_j\eta_{L,c}$ for the
normalized spinors used here. Both Abelian and non-Abelian channels
are pure spinors and therefore satisfy
$|\mathbf{S}_{c}|=1$; spin hybridization means that the Bloch
vector is not aligned with a fixed gauge axis, rather than that the
state is an incoherent mixture.

\begin{figure*}[t]
	\centering
	\includegraphics[width=0.99\textwidth]
	{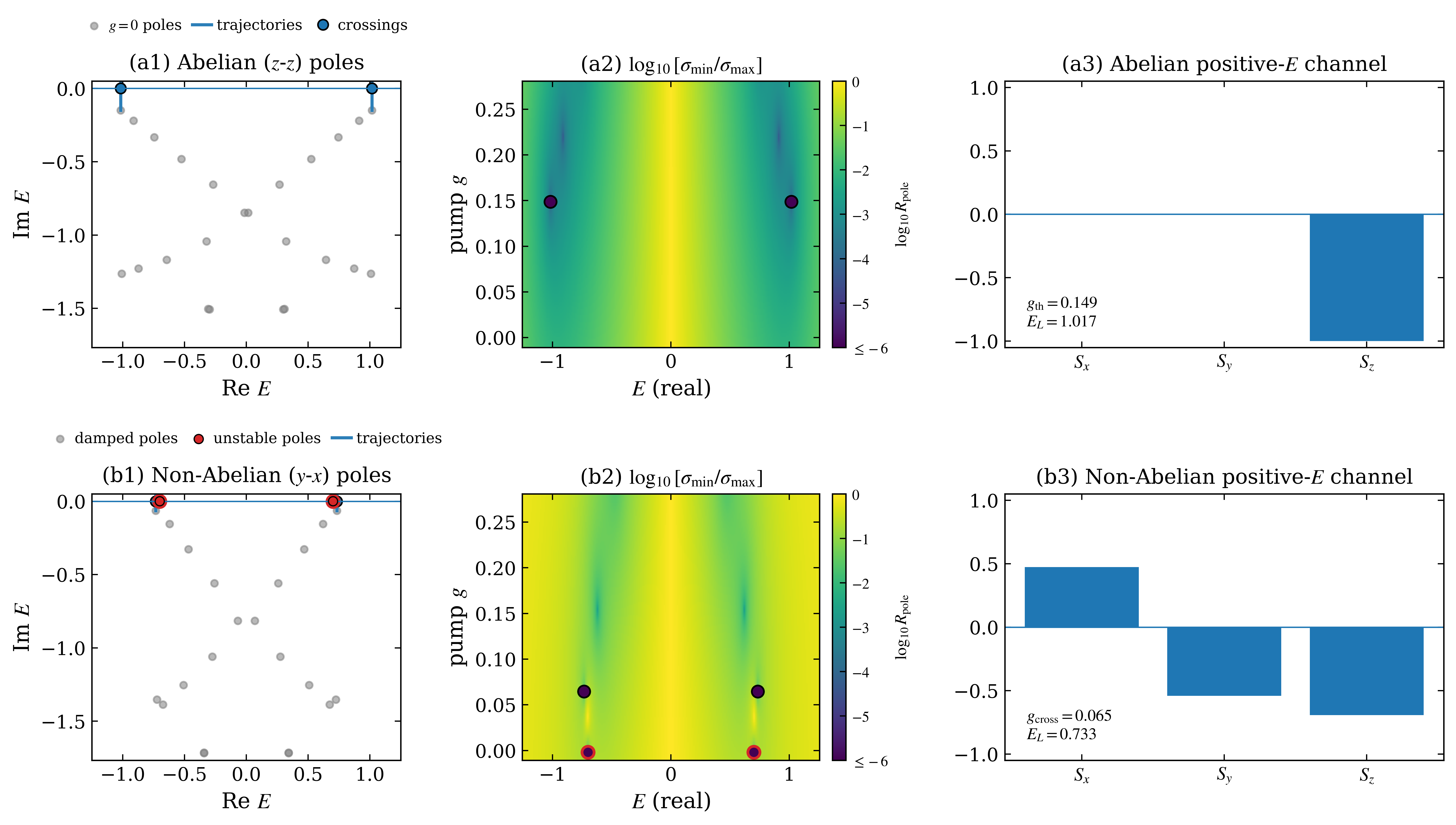}
	\caption{
		Quasibound poles and real-axis outgoing-state crossings obtained
		from $\det\mathcal{M}_{22}=0$ for
		$N=12$, $J_R=J_L=0.6$, $\theta_R=-1.5$,
		$\theta_L=0.5$, $\Gamma=0.7$, and $w=-1$, with
		$V(g)=-i\Gamma I_2+igI_2$.
		The Abelian row uses
		$L=J_Le^{i\theta_L\sigma_z}$ and
		$R=J_Re^{i\theta_R\sigma_z}$, whereas the non-Abelian row uses
		$L=J_Le^{i\theta_L\sigma_y}$ and
		$R=J_Re^{i\theta_R\sigma_x}$.
		(a1,b1) Complete sets of $2N=24$ poles at $g=0$, together with
		the vertical trajectories of the selected pole pairs as $g$ is
		varied.
		(a2,b2) Real-axis residual
		$\log_{10}\mathcal{R}_{\rm pole}$, with
		$\mathcal{R}_{\rm pole}$ defined in
		Eq. \eqref{eq:lasing_residual}. Blue dots mark the positive-g
		crossings of initially damped poles. The red dots in (b2) mark
		the negative-g crossings of the upper-half-plane poles shown in
		red in (b1).
		Plots (a3,b3) show the Bloch-vector components of the normalized left-outgoing null spinors $\eta_{L,\mathrm{A}}$ and $\eta_{L,\mathrm{NA}}$ for the selected positive-energy crossings.
	}
	\label{fig:quasibound_lasing}
\end{figure*}

Figure \ref{fig:quasibound_lasing} compares the Abelian and
non-Abelian cases. In the Abelian control, the two gauge links share
the same spin axis,
\begin{equation}
	L=J_Le^{i\theta_L\sigma_z},
	\qquad
	R=J_Re^{i\theta_R\sigma_z}.
\end{equation}
The scattering problem consequently separates into the two
$\sigma_z$ spin sectors. The damped pole pair closest to the real
axis is
$E_{\rm QB}^{\rm A}\simeq\pm1.0173-0.1488i$ and reaches the real
axis at $g_{\rm th}^{\rm A}\simeq0.1488$, with
$E_L^{\rm A}\simeq\pm1.0173$. The zeros of
$\mathcal{R}_{\rm pole}$ in plot (a2) independently identify the
same crossings.

Plot (a3) displays the positive-energy channel, for which
\begin{equation}
	\mathbf{S}_{\rm A}=(0,0,-1).
	\label{eq:abelian_lasing_spin}
\end{equation}
The negative-energy partner has the opposite polarization,
$-\mathbf{S}_{\rm A}=(0,0,1)$. Hence the two Abelian crossings can
be assigned unambiguously to the two eigenstates of the common
$\sigma_z$ gauge axis.

For the non-Abelian case,
\begin{equation}
	L=J_Le^{i\theta_L\sigma_y},
	\qquad
	R=J_Re^{i\theta_R\sigma_x},
\end{equation}
so that $[L,R]\neq0$. The transfer block
$\mathcal{M}_{22}$ cannot generally be diagonalized in an
energy-independent spin basis. The nearest damped pole pair is
$E_{\rm QB}^{(\pm)}
\simeq\pm0.7334-0.06450i$ and reaches the real axis at
$g_{\rm cross}^{\rm NA}\simeq0.06450$, with
$E_L^{(\pm)}\simeq\pm0.7334$. These crossings produce the blue
minima in plot (b2).

The complete non-Abelian pole spectrum also contains the pair
$E_{\rm up}^{(\pm)}
\simeq\pm0.7011+0.001817i$. These poles are retained in plot (b1)
but are not classified as quasibound resonances because their
imaginary parts are positive. For the convention $e^{-iEt}$, they
are already growing outgoing modes at $g=0$. Equation
\eqref{eq:pole_vertical_shift} moves them onto the real axis at
$g\simeq-0.001817$, producing the red dots in plot (b2).
Consequently, $g_{\rm cross}^{\rm NA}\simeq0.06450$ denotes the
positive-g crossing of the selected damped pole pair, rather than
the onset of stability loss of the complete non-Abelian system.

For the positive-energy non-Abelian channel, the normalized null
vector gives
\begin{equation}
	\mathbf{S}_{\rm NA}
	\simeq
	(0.474,-0.542,-0.694),
	\label{eq:nonabelian_lasing_spin}
\end{equation}
whereas the negative-energy partner has
$-\mathbf{S}_{\rm NA}\simeq(-0.474,0.542,0.694)$. Neither channel
is an eigenstate of $\sigma_x$, $\sigma_y$, or $\sigma_z$.
Instead, its polarization is selected by the full matrix-valued
propagation through the sample.

The robust Abelian--non-Abelian distinction is therefore the spin
structure of the outgoing eigenchannel, rather than the numerical
ordering of the real-axis crossings. Commuting links produce
independent pure-spin lasing sectors, whereas noncommuting links
select coherent spin-hybridized channels through the full matrix
condition $\det\mathcal{M}_{22}=0$.\\
\paragraph*{Effect of a Zeeman field.---}
We finally introduce a real uniform Zeeman field along the $z$ axis
through
\begin{equation}
	\mathbf V(g,B_z)
	=
	-i\Gamma I_2+igI_2+B_z\sigma_z .
	\label{eq:zeeman_pumped_onsite}
\end{equation}
In the Abelian $z$--$z$ configuration,
$[L,\sigma_z]=[R,\sigma_z]=0$, so the transfer problem remains
separated into the two spin sectors. The poles then satisfy
\begin{equation}
	E_{p,s}(B_z,g)
	=
	E_{p,s}(0,0)+sB_z+ig ,
	\label{eq:abelian_zeeman_poles}
\end{equation}
where $s=\pm1$. Thus, a real Zeeman field shifts the two
spin-resolved lasing frequencies in opposite directions without
changing their imaginary parts or threshold gains.

For the non-Abelian links
$L=J_Le^{i\theta_L\sigma_y}$ and
$R=J_Re^{i\theta_R\sigma_x}$, neither hopping matrix commutes with
$\sigma_z$. The Zeeman field therefore modifies the full
matrix-valued transfer product and changes both the real and
imaginary parts of the quasibound poles. We characterize this change
using the signed real-axis-crossing gain
\begin{equation}
	g_{\star}(B_z)
	=
	-\max_{1\leq j\leq 2N}
	\operatorname{Im}E_j(B_z,0),
	\label{eq:zeeman_signed_crossing}
\end{equation}
where $E_j(B_z,0)$ are all zero-pump outgoing poles determined by
$\det\mathcal M_{22}(E,B_z)=0$. If $g_{\star}>0$, the uppermost pole
lies below the real axis and the physical positive-pump threshold is
$g_{\rm th}=g_{\star}$. If $g_{\star}<0$, the uppermost pole already
lies above the real axis at zero pump. The negative value should
therefore not be interpreted as a negative lasing threshold; instead,
$-g_{\star}$ is the additional loss required to return that pole to
the real axis.

\begin{figure}[t]
	\centering
	\includegraphics[width=0.98\columnwidth]
	{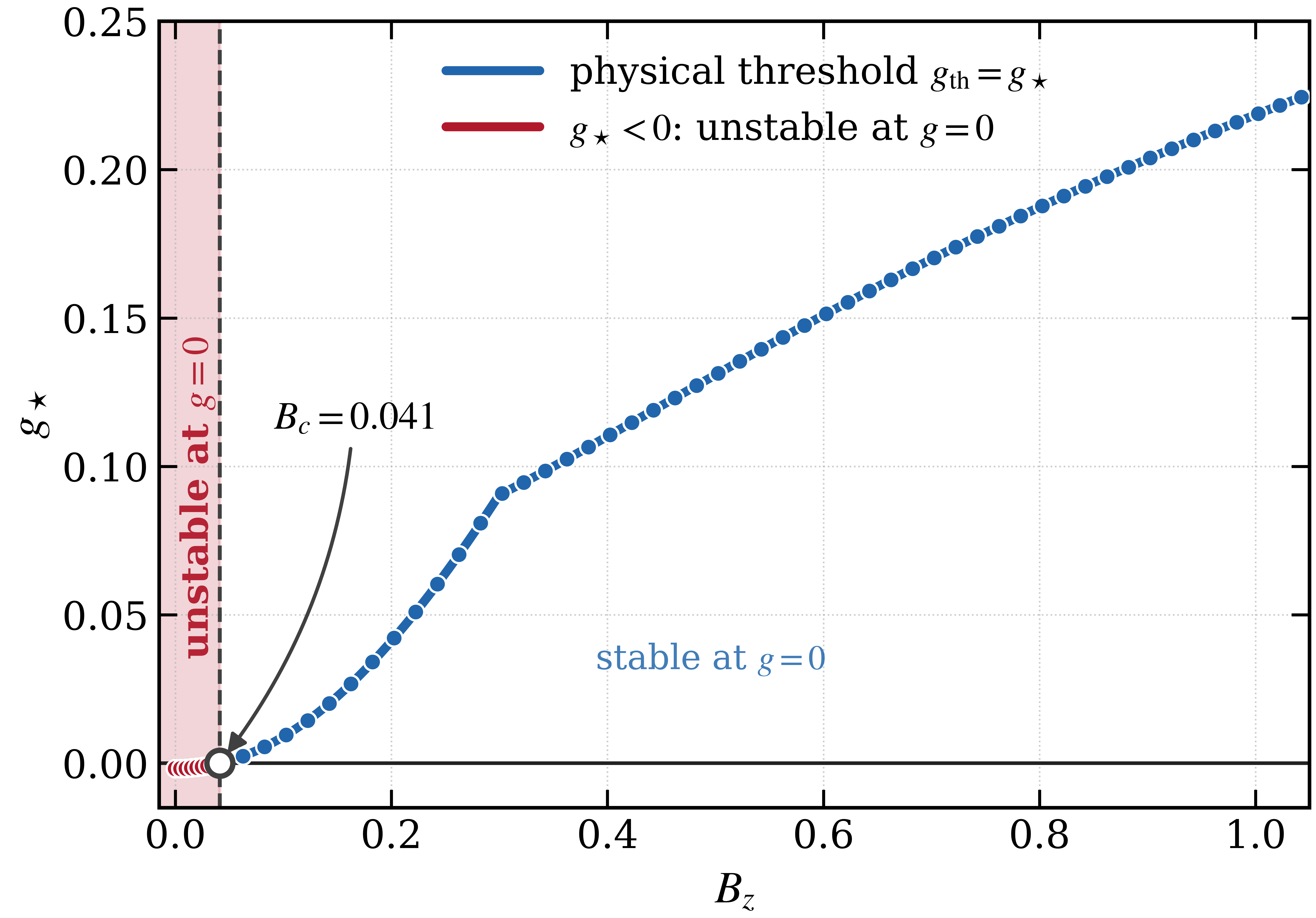}
	\caption{
		Signed real-axis-crossing gain $g_{\star}(B_z)$ for the
		non-Abelian chain, calculated from
		$\det\mathcal M_{22}=0$ with $N=12$, $J_R=J_L=0.6$,
		$\theta_R=-1.5$, $\theta_L=0.5$, $\Gamma=0.7$, and $w=-1$.
		Only $B_z\geq0$ is shown. In the red region,
		$0\leq B_z<B_c$, the signed quantity is negative because an
		outgoing pole already lies above the real axis at zero pump;
		this is not a physical negative-pump threshold. At
		$B_c\simeq0.04115$ the uppermost pole lies on the real axis.
		For $B_z>B_c$, the blue curve gives the physical threshold
		$g_{\rm th}=g_{\star}>0$. The slope change near
		$B_z\simeq0.30$ marks a switch in the pole pair determining the
		first lasing threshold.
	}
	\label{fig:nonabelian_zeeman_threshold}
\end{figure}

As shown in Fig. \ref{fig:nonabelian_zeeman_threshold},
$g_{\star}(0)\simeq-0.00182$ for the present parameters. It reaches
zero at $B_c\simeq0.04115$ and becomes a positive threshold for
$B_z>B_c$, increasing to approximately $0.2185$ at $B_z=1$. The
change in slope near $B_z\simeq0.30$ results from a switch between the
quasibound-pole pairs controlling the first crossing. Unlike in the
Abelian case, the Zeeman field can therefore change the stability and
lasing threshold of the non-Abelian system.

\subsection{Scattering zeros and coherent perfect absorption}
\label{sec:scattering_zeros_cpa}

A scattering zero satisfies purely incoming boundary conditions. In
the propagating-wave basis, it is therefore selected by
\begin{equation}
\mathcal{M}_{11}(E)\chi_L=0,\qquad \chi_L\neq0,\qquad \chi_R=\mathcal{M}_{21}(E)\chi_L. 
\end{equation} 
Here $\chi_L$ is the independent left-incident spinor, while $\chi_R$ is the required right-incident component. Thus, $\chi_L$ determines the complete coherent two-sided input. A
scattering zero on the real-energy axis defines a CPA eigenchannel.\\
In an Abelian spin sector, Eq. \eqref{eq:abelian_A_D} gives
$\mathcal{M}_{11}^{(s)}=d_s^{(N-1)/2}\mathcal{F}_s^{+}/2$, so the
scattering-zero and CPA condition reduces to $\mathcal{F}_s^{+}(E)=0$.
Because $d_s^{(N-1)/2}\neq0$ is only a multiplicative prefactor, it
does not affect the zero condition at fixed $q_s$, $V_s$, $w$, and
$N$. As in the pole problem, this invariance does not apply to a
microscopic parameter variation that changes $q_s$ and $d_s$
simultaneously.

To move the scattering zeros toward the real axis without selecting a
spin direction, we introduce the spin-independent loss $\mathbf{V}(\gamma)=-i\gamma I_2$. Since $EI_2-\mathbf{V}(\gamma)=(E+i\gamma)I_2$, every zero undergoes
the exact vertical shift
\begin{equation}
	E_{{\rm zero},j}(\gamma)
	=
	E_{{\rm zero},j}(0)-i\gamma.
	\label{eq:cpa_vertical_shift}
\end{equation}
Consequently, an upper-half-plane zero
$E_{{\rm zero},j}(0)=E_R+i\gamma_0$, with $\gamma_0>0$, reaches CPA at
$\gamma_{\rm CPA}=\gamma_0$ and $E_{\rm CPA}=E_R$.

Let $\sigma_{\min}$ and $\sigma_{\max}$ denote the smallest and
largest singular values of $\mathcal{M}_{11}(E,\gamma)$ as we did for quasibound state in the previous section. Therefore, in analogy
with the pole residual, we define
\begin{equation}
	R_{\rm zero}(E,\gamma)
	=
	\frac{\sigma_{\min}[\mathcal{M}_{11}(E,\gamma)]}
	{\sigma_{\max}[\mathcal{M}_{11}(E,\gamma)]}.
	\label{eq:cpa_residual}
\end{equation}
For the generic rank-one zeros considered here,
$R_{\rm zero}=0$ is equivalent to
$\det\mathcal{M}_{11}=0$. For the spin plots, we characterize the
normalized left-incident null vector of the positive-energy CPA state
by
\begin{equation}
	\mathbf{S}_{\rm CPA}^{c}
	=
	\frac{\chi_{L,c}^{\dagger}
		\boldsymbol{\sigma}\chi_{L,c}}
	{\chi_{L,c}^{\dagger}\chi_{L,c}},
	\qquad
	c\in\{\mathrm{A},\mathrm{NA}\},
	\label{eq:cpa_bloch_vector}
\end{equation}
where $\boldsymbol{\sigma}=(\sigma_x,\sigma_y,\sigma_z)$.

\begin{figure*}[t]
	\centering
	\includegraphics[width=0.96\textwidth]
	{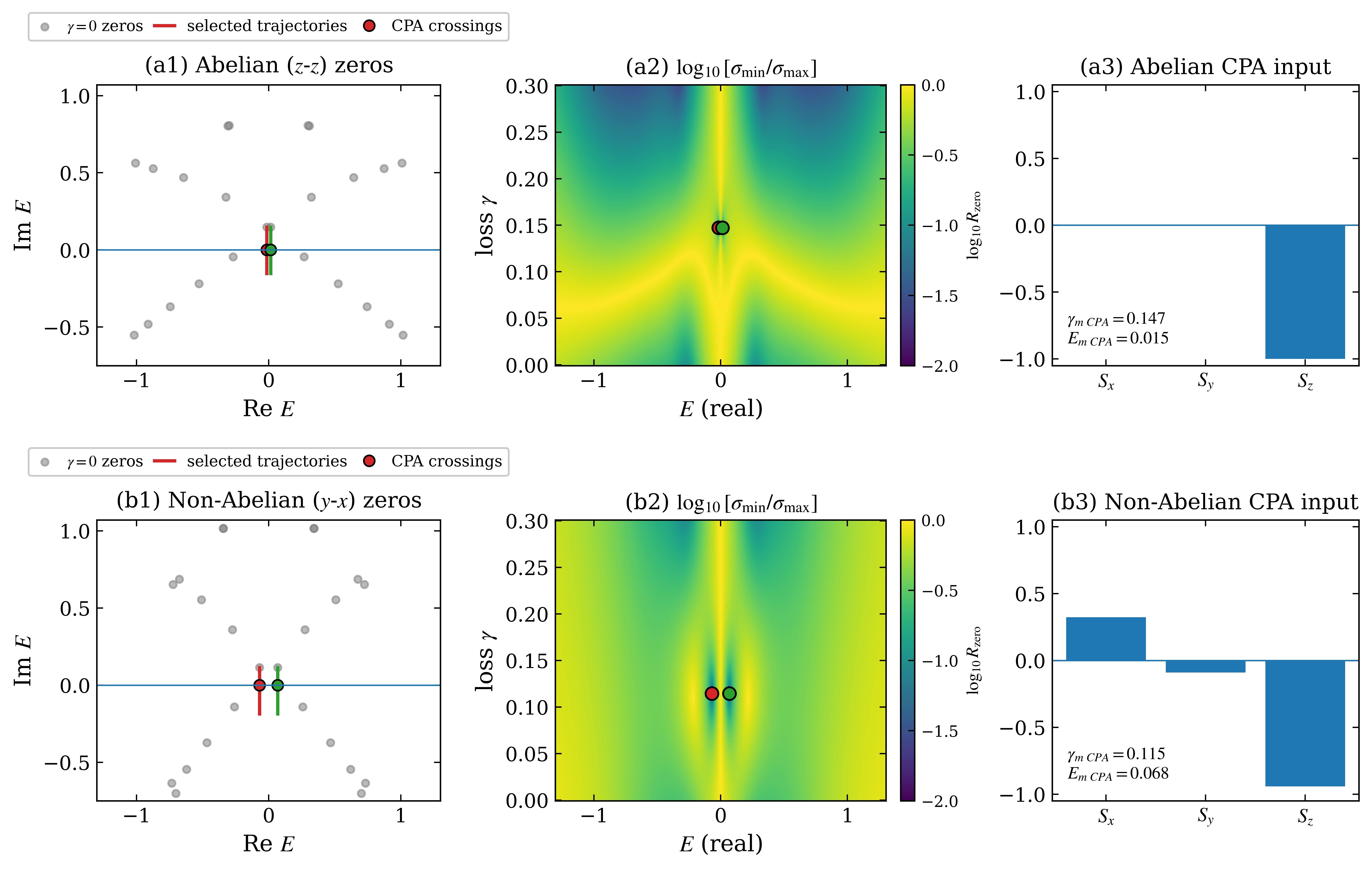}
	\caption{
		Scattering zeros and CPA obtained from
		$\det\mathcal{M}_{11}(E,\gamma)=0$ for $N=12$,
		$J_L=J_R=0.6$, $\theta_R=-1.5$, $\theta_L=0.5$, and
		$w=-1$, with the scalar loss
		$\mathbf{V}(\gamma)=-i\gamma I_2$. The upper row is the Abelian
		configuration
		$L=J_Le^{i\theta_L\sigma_z}$,
		$R=J_Re^{i\theta_R\sigma_z}$, and the lower row is the
		non-Abelian configuration
		$L=J_Le^{i\theta_L\sigma_y}$,
		$R=J_Re^{i\theta_R\sigma_x}$.
		(a1,b1) Complete sets of $2N=24$ zeros at $\gamma=0$ and the
		vertical trajectories of the upper-half-plane pairs closest to the
		real axis. The colored dots mark their CPA crossings.
		(a2,b2) Real-energy maps of $\log_{10}R_{\rm zero}$, with the same
		color scale in both rows; the dots mark the exact zeros of
		$R_{\rm zero}$.
		(a3,b3) Bloch-vector components of the left-incident spinors for
		the selected positive-energy CPA states. The Abelian input is a
		$\sigma_z$ eigenstate, whereas the non-Abelian input is a coherent
		spin-hybridized state.
	}
	\label{fig:cpa_transport}
\end{figure*}

Figure \ref{fig:cpa_transport} compares the two gauge structures. In
the Abelian control, the two spin sectors remain decoupled and the
closest upper-half-plane zero pair is
$E_{{\rm zero}}^{(\pm)}(0)\simeq
\pm0.01464+0.14737i$. Both zeros therefore reach the real axis at
$\gamma_{\rm CPA}^{\rm A}\simeq0.14737$, with
$E_{\rm CPA}^{(\pm)}\simeq\pm0.01464$. The positive-energy input has
$\mathbf{S}_{\rm CPA}^{\rm A}=(0,0,-1)$, while its negative-energy
partner has the opposite polarization. Thus, the two CPA channels are
unambiguously associated with the two eigenstates of the common
$\sigma_z$ gauge axis.

For the non-Abelian configuration, the two links do not commute and
$\mathcal{M}_{11}$ cannot generally be diagonalized in an
energy-independent spin basis. The nearest zero pair is
$E_{{\rm zero}}^{(\pm)}(0)\simeq
\pm0.06842+0.11461i$ and reaches CPA at
$\gamma_{\rm CPA}^{\rm NA}\simeq0.11461$, with
$E_{\rm CPA}^{(\pm)}\simeq\pm0.06842$. For the positive-energy input,
$\mathbf{S}_{\rm CPA}^{\rm NA}\simeq
(0.324,-0.090,-0.942)$; the negative-energy partner has
$-\mathbf{S}_{\rm CPA}^{\rm NA}$. Neither channel is an eigenstate of
$\sigma_x$, $\sigma_y$, or $\sigma_z$.

Although the non-Abelian CPA loss is smaller for this parameter set,
the numerical ordering of the two thresholds is not universal. The
robust distinction is the input-spin structure: commuting links select
independent fixed-axis spin sectors, whereas noncommuting links select
coherent spin-hybridized CPA channels through the full matrix-valued
propagation problem.
\subsection{Reflectionless transport}
\label{sec:reflectionless_transport}

Away from scattering poles, where $\mathcal{M}_{22}$ is invertible,
the scattering relations imply
\begin{align}
\mathcal{M}_{21}(E)\chi_L=0,\quad &\chi_L\neq0 \quad\Longleftrightarrow\quad \det\mathcal{M}_{21}(E)=0,\label{left-reflectionless}\\
\mathcal{M}_{12}(E)\eta_L=0,\quad &\eta_L\neq0 \quad\Longleftrightarrow\quad \det\mathcal{M}_{12}(E)=0\label{right-reflectionless}. 
\end{align}
The first relation selects the physical left-incident spinor $\chi_L$. For right incidence, $\eta_L$ is the null vector of $\mathcal{M}_{12}$, while the physical incident spinor is obtained from $\chi_R=\mathcal{M}_{22}\eta_L$. Thus, the two determinant
conditions identify reflectionless eigenchannels rather than the
strong matrix conditions $\mathcal{M}_{21}=0_2$ and
$\mathcal{M}_{12}=0_2$, which would apply to every incident spinor.

To locate these conditions numerically, we define the normalized
residuals
\begin{equation}
	\begin{aligned}
		\rho_L(E,B_z)
		&=
		\frac{\sigma_{\min}[\mathcal{M}_{21}(E,B_z)]}
		{\sigma_{\max}[\mathcal{M}_{N}(E,B_z)]},
		\\
		\rho_R(E,B_z)
		&=
		\frac{\sigma_{\min}[\mathcal{M}_{12}(E,B_z)]}
		{\sigma_{\max}[\mathcal{M}_{N}(E,B_z)]}.
	\end{aligned}
	\label{eq:reflectionless_residuals}
\end{equation}
Here $\sigma_{\min}(X)$ and $\sigma_{\max}(X)$ denote the smallest
and largest singular values of $X$. The normalization by the complete
propagating-wave transfer matrix $\mathcal{M}_N$ remains well defined
also when an entire reflection block approaches zero. In other words, the full-matrix normalization is particularly useful here because,
at the Abelian symmetry point $B_z=0$ with
$\theta_L+\theta_R=0$, the two spin-resolved reflectionless
conditions coincide, and the complete reflection block can vanish. Consequently, $\rho_L=0$ and $\rho_R=0$ reproduce the two
conditions in Eqs. \eqref{left-reflectionless} and \eqref{right-reflectionless}; away from their zeros,
these quantities are boundary-condition residuals rather than
reflection probabilities.

For a uniform Abelian chain, Eqs. \eqref{eq:abelian_B_C} give
$\mathcal{M}_{12}^{(s)}=-\mathcal{M}_{21}^{(s)}$. Left- and
right-reflectionlessness therefore coincide exactly in each spin
sector and are determined by
\begin{equation}
	\mathcal{R}_s(E)
	=
	\omega_sU_N(z_s)+\omega_s^{-1}U_{N-2}(z_s)
	=0.
	\label{eq:abelian_reflectionless_condition}
\end{equation}
At fixed $q_s$, $V_s$, $w$, and $N$, this condition is independent of
the directed-amplification factor $d_s$. As in the lasing and CPA
conditions, this statement does not apply to microscopic parameter
variations that change $q_s$ and $d_s$ simultaneously. A nonuniform
scalar chain, or a non-Abelian chain, need not retain the left-right
coincidence.

We compare the Abelian and non-Abelian cases using the same Zeeman
control $\mathbf{V}=B_z\sigma_z$. It commutes with the Abelian links
$L=J_Le^{i\theta_L\sigma_z}$ and
$R=J_Re^{i\theta_R\sigma_z}$, but not with the non-Abelian links
$L=J_Le^{i\theta_L\sigma_y}$ and
$R=J_Re^{i\theta_R\sigma_x}$. For a normalized left-incident
reflectionless spinor $\chi_{L}$, we use
\begin{align} 
&\eta_R=t_{L\rightarrow R}\chi_L,&R_L=\left|r_L\chi_L\right|^2,\\
\nonumber\\
&T_L=\left|\eta_R\right|^2,&\hat{\eta}_R=\frac{\eta_R}{|\eta_R|}. 
\end{align}
For any nonzero spinor $u$, define 
\begin{equation}
\mathbf{S}[u]=\frac{u^{\dagger}\boldsymbol{\sigma}u}{u^{\dagger}u}. 
\end{equation}
For left incidence, $\mathbf{S}_{\rm in}=\mathbf{S}[\chi_L]$ and $\mathbf{S}_{\rm out}=\mathbf{S}[\hat{\eta}_R]$.
\begin{figure*}[t]
	\centering
	\includegraphics[width=0.958\textwidth]
	{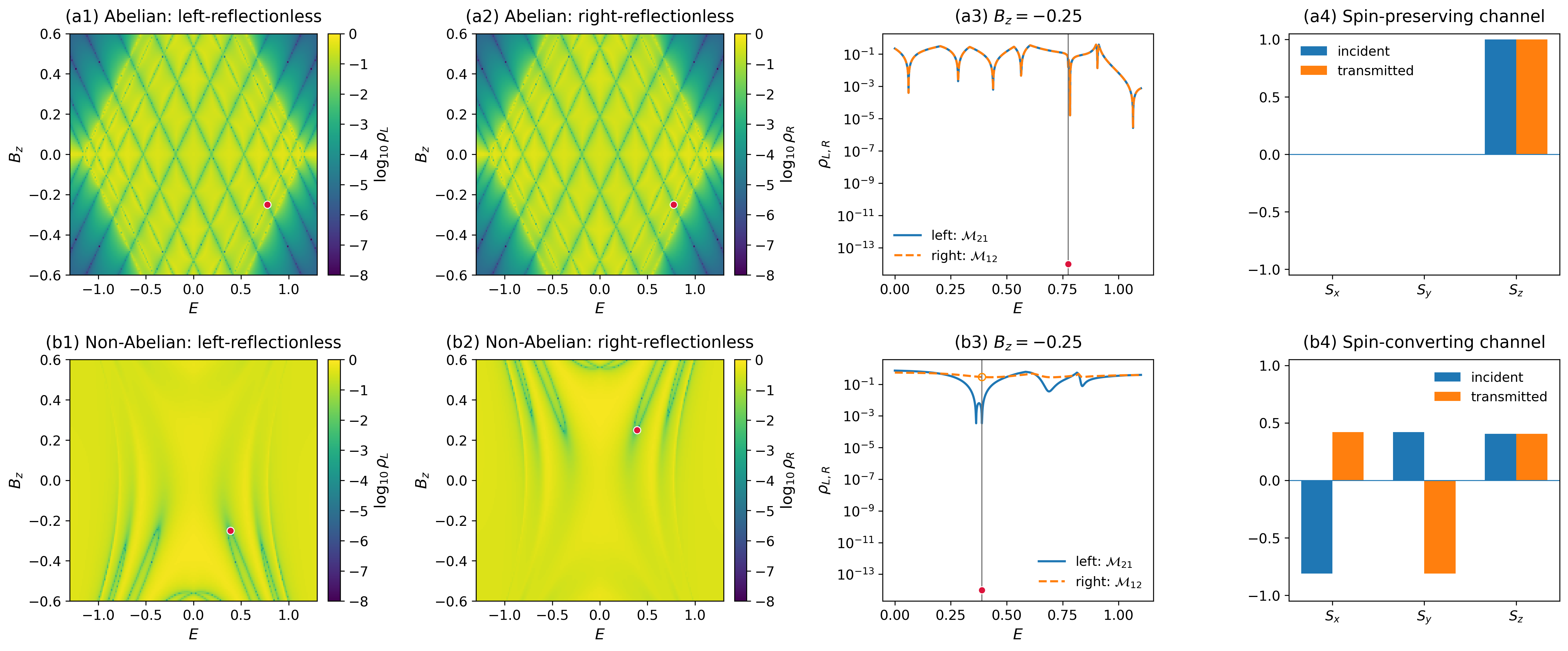}
	\caption{
		Reflectionless eigenchannels and their transmission through a
		spinful chain with $N=12$, $J_L=J_R=0.6$,
		$\theta_L=1.5$, $\theta_R=-1.5$, $w=-1$, and
		$\mathbf{V}=B_z\sigma_z$. The upper row is the Abelian
		configuration $L=J_Le^{i\theta_L\sigma_z}$ and
		$R=J_Re^{i\theta_R\sigma_z}$; the lower row is the non-Abelian
		configuration $L=J_Le^{i\theta_L\sigma_y}$ and
		$R=J_Re^{i\theta_R\sigma_x}$.
		(a1,b1) Maps of the left residual $\log_{10}\rho_L$.
		(a2,b2) Maps of the right residual $\log_{10}\rho_R$, plotted
		with the same color scale. The red markers in (a1,a2) denote the
		common Abelian root at $B_z=-0.25$; those in (b1,b2) denote the
		non-Abelian left root at $B_z=-0.25$ and its right-reflectionless
		counterpart at $B_z=+0.25$, respectively.
		(a3,b3) Cuts at $B_z=-0.25$. The filled dots mark the exact
		left-reflectionless roots; the open circle in (b3) shows the
		finite right residual at the selected non-Abelian left root.
		(a4,b4) Bloch-vector components of the incident and transmitted
		spinors for the selected left-reflectionless channels. The
		Abelian channel preserves its spin, whereas the non-Abelian
		channel undergoes spin conversion.
	}
	\label{fig:reflectionless_transport}
\end{figure*}

Figure \ref{fig:reflectionless_transport} summarizes the resulting
reflectionless structure. In the Abelian case the two spin sectors
remain decoupled, and the left and right maps in plots (a1) and (a2)
coincide, as required by
Eq. \eqref{eq:abelian_reflectionless_condition}. At $B_z=-0.25$, we
select the positive-energy $s=+1$ root
$E_{\rm refl}^{\rm A}\simeq0.7750$. It has
$R_L\simeq 0$, $T_L\simeq1$, and
$\mathbf{S}_{\rm in}^{\rm A}=\mathbf{S}_{\rm out}^{\rm A}=(0,0,1)$.
The Abelian reflectionless channel is therefore spin preserving.

For the non-Abelian links, the transfer problem cannot be decomposed
into fixed spin sectors. At $B_z=-0.25$, a representative
left-reflectionless root occurs at
$E_{\rm refl,L}^{\rm NA}\simeq0.3895$, where
$R_L\simeq3.8\times10^{-32}$ and $T_L\simeq1$. At the same point the
right residual remains finite, $\rho_R\simeq0.288$, so this state is
reflectionless only for incidence from the left. For the present
parameter choice, the right-reflectionless counterpart occurs at the
same energy after reversing the Zeeman field to $B_z=+0.25$; numerically,
$\rho_L(E,B_z)=\rho_R(E,-B_z)$.

The non-Abelian incident and transmitted spinors have
$\mathbf{S}_{\rm in}^{\rm NA}\simeq(-0.812,0.420,0.405)$ and
$\mathbf{S}_{\rm out}^{\rm NA}\simeq(0.420,-0.812,0.405)$,
respectively. Their spin fidelity is
nearly $0.241$,
corresponding to a Bloch-sphere rotation of approximately
$121.2^\circ$. Where by spin fidelity we mean
\begin{equation}
F_{\rm spin}
=
\left\vert
\left\langle
\chi_L
\middle|
\hat{\eta}_R
\right\rangle
\right\vert^2.
\end{equation}
Thus, the selected non-Abelian channel combines
vanishing reflection and unit transmitted intensity with pronounced
spin conversion.

Unit transmitted intensity should not be confused with the stronger
transparency condition
$\eta_R=t_{L\to R}\chi_{L}=\tau_0\chi_{L}$. Here
$\tau_0=i^N=1$. For the Abelian root,
$t_{L\to R}\chi_{L}\simeq(0.2188-0.9758i)\chi_{L}$, so the
spin is preserved but the reference phase is not. The non-Abelian
root additionally fails the proportionality condition because the
spinor is rotated. The robust distinction is therefore that commuting
links select independent fixed-axis reflectionless sectors, whereas
noncommuting links permit direction- and field-dependent
reflectionless spin conversion.
\subsection{Transparency and invisibility}

Let $\tau_0$ denote the transmission amplitude through a reference lead
segment of the same length as the sample. With our lead-phase convention,
$\tau_0=i^N$. Strict
transparency requires
\begin{equation}
\begin{aligned}
\eta_R=t_{L\rightarrow R}\chi_L
&=\tau_0\chi_L
&&\text{(left transparency)},\\
\eta_L=t_{R\rightarrow L}\chi_R
&=\tau_0\chi_R
&&\text{(right transparency)}.
\end{aligned}
\label{eq:transparent_channel_condition}
\end{equation}
These conditions are stronger than unit transmitted intensity,
because they also require the correct transmission phase and
preservation of the incident spinor. Requiring
$t_{L\rightarrow R}=\tau_0 I_2$ or
$t_{R\rightarrow L}=\tau_0 I_2$ would instead make every spin
channel transparent; here we focus on transparent eigenchannels.

In the Abelian limit, the two spin sectors are independent, and
Eq. \eqref{eq:transparent_channel_condition} reduces to
$t_{L\rightarrow R,s}=\tau_0$ or
$t_{R\rightarrow L,s}=\tau_0$. From
Eqs. \eqref{eq:abelian_A_D} and \eqref{eq:abelian_B_C}, together
with the scattering relations, the spin-resolved transmission
amplitudes are
\begin{equation}
t_{L\rightarrow R,s}
=
\frac{2d_s^{(N-1)/2}}{\mathcal{F}_s^{-}},
\qquad
t_{R\rightarrow L,s}
=
\frac{2d_s^{-(N-1)/2}}{\mathcal{F}_s^{-}}.
\label{eq:abelian_transmission_amplitudes}
\end{equation}
Using Eq. \eqref{eq:abelian_transmission_amplitudes} gives
\begin{align}
\mathcal{F}_s^{-}
&=
2d_s^{(N-1)/2}\tau_0^{-1}
&&\text{(left transparent)},
\label{eq:abelian_left_transparent}
\\
\mathcal{F}_s^{-}
&=
2d_s^{-(N-1)/2}\tau_0^{-1}
&&\text{(right transparent)}.
\label{eq:abelian_right_transparent}
\end{align}
Unlike the pole, zero, and reflectionless conditions, these
equations depend explicitly on the directed-amplification factor
$d_s$. In particular,
Eq. \eqref{eq:abelian_transmission_amplitudes} implies
\begin{equation}
\left|
\frac{t_{L\rightarrow R,s}}
     {t_{R\rightarrow L,s}}
\right|
=
|d_s|^{N-1}
=
r_s^{\,2(N-1)},
\label{eq:abelian_directional_amplification}
\end{equation}
where $r_s=\sqrt{|d_s|}$ is the spin-resolved GBZ radius. Thus, the
same quantity that determines the skin-localization length also
controls the directional imbalance of the transmission amplitudes.
Simultaneous left and right transparency of the same spin sector
requires the stronger complex condition $d_s^{N-1}=1$.

An invisible channel must be reflectionless and transparent for the
same incident spinor:
\begin{equation}
	r_{\alpha}\chi_{\alpha}=0,
	\qquad
	\bigl(t_{\alpha}-\tau_0I_2\bigr)\chi_{\alpha}=0,
	\label{eq:scattering_invisibility_condition}
\end{equation}
where $\alpha\in\{L,R\}$. Consequently, separate zeros of the reflection and transparency
determinants do not establish invisibility unless their null vectors
coincide. In terms of the propagating-wave transfer matrix, the two
conditions can be combined into
\begin{align}
	\mathcal{C}_{L}(E)
	&=
	\begin{pmatrix}
		\mathcal{M}_{21}\\
		\mathcal{M}_{11}-\tau_0I_2
	\end{pmatrix},
	\label{eq:left_invisibility_matrix}
	\\
	\mathcal{C}_{R}(E)
	&=
	\begin{pmatrix}
		\mathcal{M}_{12}\\
		\mathcal{M}_{22}-\tau_0^{-1}I_2
	\end{pmatrix}.
	\label{eq:right_invisibility_matrix}
\end{align}
Left invisibility requires
$\mathcal{C}_{L}\chi_L=0$. For right incidence, the null vector of
$\mathcal{C}_{R}$ is proportional to the transmitted spinor
$\eta_L=t_{R\rightarrow L}\chi_R$; at transparency it has the same spin polarization
as the incident state and differs only by the reference amplitude ($\eta_L=\tau_0\chi_R$).

For an Abelian spin channel, these matrix conditions reduce to
\begin{align}
	\mathcal{R}_s&=0,
	&
	\mathcal{F}_s^{-}
	&=
	2d_s^{(N-1)/2}\tau_0^{-1}
	&&\text{(left invisible)},
	\label{eq:abelian_left_invisible}
	\\
	\mathcal{R}_s&=0,
	&
	\mathcal{F}_s^{-}
	&=
	2d_s^{-(N-1)/2}\tau_0^{-1}
	&&\text{(right invisible)}.
	\label{eq:abelian_right_invisible}
\end{align}
In the non-Abelian case there is no fixed spin basis in which these
conditions reduce to independent scalar equations. The invisible
spinor is instead selected by the common null space of the full
matrix-valued reflection and transparency conditions.

For the numerical calculations, we monitor the strict-transparency
residuals
\begin{equation}
	\rho_{\mathrm{tr}}^{\alpha}
	=
	\frac{
		\sigma_{\min}\!\left(t_{\alpha}-\tau_0I_2\right)
	}{
		\sigma_{\max}(t_{\alpha})+|\tau_0|
	},
	\label{eq:transparency_residual}
\end{equation}
and the invisibility residuals
\begin{equation}
	\rho_{\mathrm{inv}}^{\alpha}
	=
	\frac{
		\sigma_{\min}(\mathcal{C}_{\alpha})
	}{
		\sigma_{\max}(\mathcal{C}_{\alpha})+1
	}.
	\label{eq:invisibility_residual}
\end{equation}
Here $\sigma_{\min}$ and $\sigma_{\max}$ denote the smallest and
largest singular values of the indicated matrix, and $\mathcal{C}_{L}$ and $\mathcal{C}_{R}$ act on $\chi_L$ and $\eta_L$ respectively. The normalizations
make the residuals dimensionless without changing their zeros.
Transparency requires $\rho_{\mathrm{tr}}^{\alpha}=0$, whereas
invisibility requires the stronger common-spinor condition
$\rho_{\mathrm{inv}}^{\alpha}=0$.
Moreover, in the above relations, the denominators provide nonzero reference scales that keep the
residuals finite and bounded, even when the complete transparency or
invisibility constraint matrix vanishes. Here the constant $1$
corresponds to the unit reference amplitude $|\tau_0|=1$; these
normalizations do not change the locations of the exact zeros.
\begin{figure*}[t]
	\centering
	\includegraphics[width=0.98\textwidth]
	{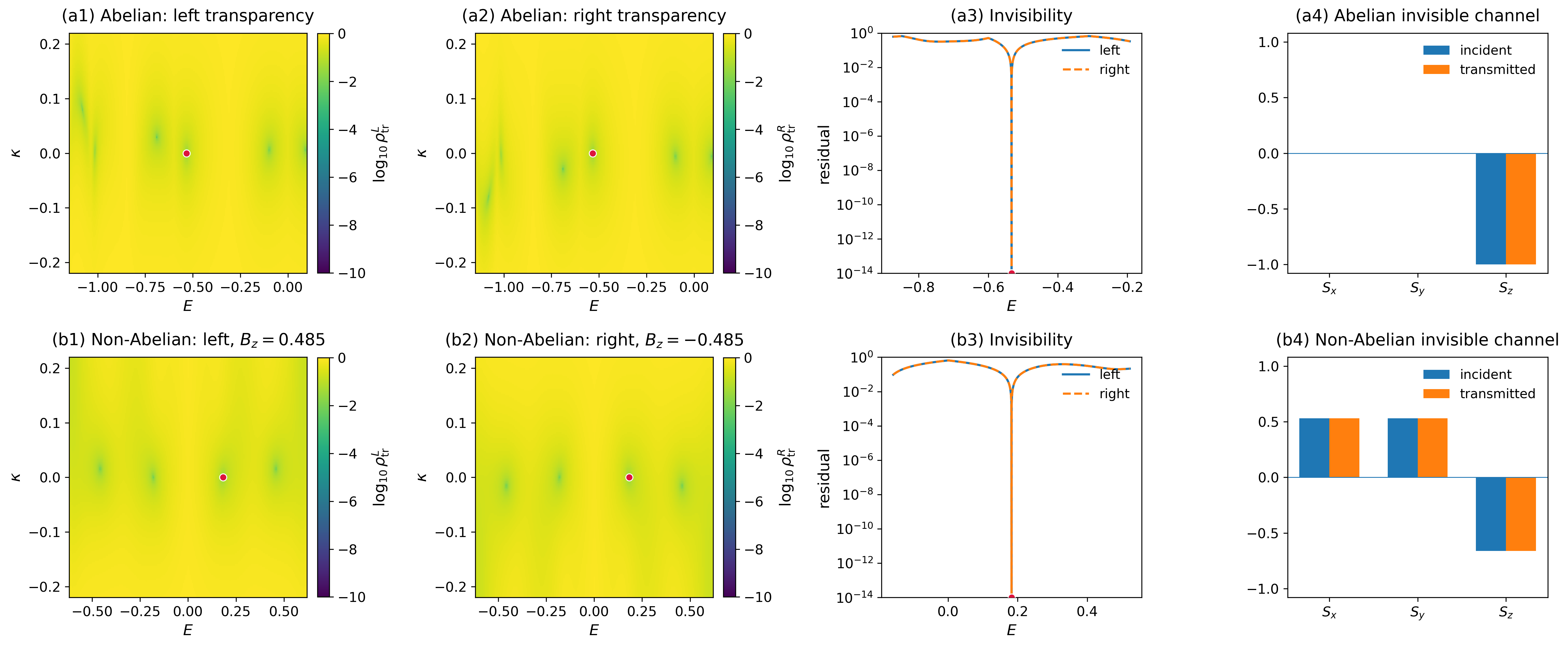}
\caption{
Strict transparency and invisibility obtained from the transfer-matrix
formulation for $N=12$, $J=0.6$, $w=-1$, and
$J_L=Je^{-\kappa}$, $J_R=Je^{\kappa}$, so that
$J_LJ_R=J^2$ remains fixed while the directional asymmetry is varied.
The upper row shows the Abelian links
$L=J_Le^{i\theta_L\sigma_z}$ and
$R=J_Re^{i\theta_R\sigma_z}$, with
$\theta_L=-0.365261$, $\theta_R=0.365261$, and $B_z=0$.
The lower row shows the non-Abelian links
$L=J_Le^{i\theta_L\sigma_y}$ and
$R=J_Re^{i\theta_R\sigma_x}$, with
$\theta_L=-0.739699$ and $\theta_R=0.739699$.
(a1,a2) Left- and right-transparency residuals for the Abelian chain.
(b1,b2) Non-Abelian left and right transparency for the
symmetry-related fields $B_z=+0.484598$ and $B_z=-0.484598$,
respectively. Red circles mark points that also satisfy the
corresponding invisibility condition.
(a3) Abelian left- and right-invisibility residuals at
$\kappa=B_z=0$.
(b3) Non-Abelian invisibility residuals for the field-reversed left
and right channels.
(a4,b4) Bloch-vector components of the incident and transmitted
left-invisible channels. Their equality confirms that strict
invisibility restores the incident spinor, although the selected
non-Abelian channel is spin hybridized.
}
	\label{fig:transparency_invisibility}
\end{figure*}
Figure \ref{fig:transparency_invisibility} compares representative
invisible channels in the two gauge regimes. The parameterization
$J_L=Je^{-\kappa}$ and $J_R=Je^{+\kappa}$ changes the directed
amplification while keeping the reciprocal scale $J_LJ_R$ fixed.
The red markers identify common zeros of the reflection and
transparency conditions rather than points selected only by
unit transmitted intensity.

For the Abelian chain, an invisible point occurs at
$E_{\mathrm{inv}}^{\mathrm A}\simeq-0.533090$ and $\kappa=0$.
The left-invisible channel is the pure spin state
$\mathbf S_{\mathrm A}^{L}=(0,0,-1)$, whereas the corresponding
right-invisible channel is the opposite eigenstate
$\mathbf S_{\mathrm A}^{R}=(0,0,1)$. Numerically,
$R\simeq0$, $T=1$ to machine precision, and the transmitted
state has the reference amplitude $\tau_0=i^{12}=1$. Thus,
invisibility in the Abelian model remains resolved into pure spin
sectors of the common gauge axis.

For the non-Abelian chain, the left-invisible point occurs at
$E_{\mathrm{inv}}^{\mathrm{NA}}\simeq0.182234$, $\kappa=0$, and
$B_z\simeq+0.484598$. Its Bloch vector is
$\mathbf S_{\mathrm{NA}}^{L}\simeq
(0.530,0.530,-0.662)$, demonstrating that the selected incident
channel is a coherent spin-hybridized state. Reversing the Zeeman
field produces the right-invisible partner with
$\mathbf S_{\mathrm{NA}}^{R}=-\mathbf S_{\mathrm{NA}}^{L}$.
At both points the numerical calculation gives
$R\lesssim10^{-30}$, $T=1$, and
$|\langle\chi_{\mathrm{in}}|\chi_{\mathrm{out}}\rangle|^2=1$.

The last equality distinguishes invisibility from the reflectionless
spin conversion discussed above. A non-Abelian reflectionless channel
may rotate the transmitted spin, whereas a strictly invisible channel
must reproduce the incident spinor up to the reference phase.
Non-Abelian propagation can nevertheless select a hybridized input
state through internal spin mixing. The Abelian and non-Abelian
examples are therefore chosen to demonstrate the different channel
structures rather than to compare universal transparency thresholds,
which remain parameter and length dependent.

\section{Implications for sensing}
\label{sec:sensing}

The transport conditions derived above also define useful operating points
for parameter sensing. A weak perturbation that modifies the onsite matrix or
either hopping matrix displaces the scattering poles and zeros and changes the
spinor selected by the corresponding transfer-matrix block. In a photonic realization, the two spin components may represent
polarization or two coupled modes, while the perturbation may arise
from a refractive-index change \cite{Chow2004RefractiveIndex},
birefringence \cite{Huang2008Birefringence}, magneto-optically
induced mode splitting \cite{Ruan2025MagnetoOptical}, or a change in
the intermode coupling strength \cite{Zhu2010ModeSplitting}. The
measurable responses can therefore include shifts of the lasing or
CPA frequency, changes in the threshold pump or loss, directional
reflection contrast, or rotation of the output Stokes vector. These responses rely on scattering poles, zeros, and transport
eigenchannels and do not require an exceptional point; an
exceptional-point interpretation would require independent
verification of eigenvalue and eigenvector coalescence.\\Energies below are expressed in units of $|w|$.

\textit{a. Directional-reflection readout}:
For the uniform chain with identical leads and
$\mathbf V=V I_2+\mathbf B\cdot\boldsymbol{\sigma}$, transposition followed by
the spatial reflection $n\mapsto N+1-n$ relates the two reflection matrices.
Let $U$ satisfy
$U\sigma_\mu^{T}U^\dagger=\sigma_\mu$ and
$U\sigma_\nu^{T}U^\dagger=\sigma_\nu$, and define
$\hat{\mathbf e}_{\perp}=\hat{\mathbf e}_{\mu}\times
\hat{\mathbf e}_{\nu}$. Then
\begin{align}
r_L(E;\mathbf B)
&=
U\,r_R^{T}(E;\widetilde{\mathbf B})\,U^\dagger,
\label{eq:sensing_mirror_transpose}
\\
\widetilde{\mathbf B}
&=
\mathbf B
-2(\mathbf B\cdot\hat{\mathbf e}_{\perp})
\hat{\mathbf e}_{\perp}.
\nonumber
\end{align}
For the $(\mu,\nu)=(y,x)$ configuration used in the numerical transport
figures, one may take $U=\sigma_x$ and
$\hat{\mathbf e}_{\perp}=-\hat{\mathbf e}_z$. Consequently, the reflectances
for an unpolarized incident state,
$R_\alpha=\tfrac12\operatorname{Tr}(r_\alpha^\dagger r_\alpha)$, define the
contrast
\begin{equation}
\mathcal{A}_{\rm refl}(E;\mathbf B)
=
\frac{R_L(E;\mathbf B)-R_R(E;\mathbf B)}
{R_L(E;\mathbf B)+R_R(E;\mathbf B)},
\label{eq:sensing_directional_contrast}
\end{equation}
which is odd under reversal of the field component perpendicular to the two
gauge axes. It therefore vanishes at $B_\perp=0$, even in the presence of
field components in the gauge-axis plane. Within the uniform Abelian model, a
field commuting with the common gauge axis also gives $r_L=r_R$ in each spin
sector and hence $\mathcal{A}_{\rm refl}=0$. Thus, within the symmetric-contact geometry
considered here, a nonzero $\mathcal{A}_{\rm refl}$ supplies a sign-sensitive null
measurement of $B_\perp$ without polarization-resolved detection. For the
non-Abelian parameters of the reflectionless-transport calculation and
$N=12$, the numerical calculation gives
$|\partial\mathcal \mathcal{A}_{\rm refl}/\partial B_z|_{B_z=0}\simeq2.584$ near
$E\simeq0.286$, where $R_L+R_R\simeq0.697$. Calculations for different $N$
do not show a monotonic size enhancement; the useful feature of this readout
is the symmetry-enforced zero at the reference field.\\
The principal sensing outcome is therefore a symmetry-protected
directional-reflection null: a perpendicular Zeeman perturbation
generates a signed left-right reflection asymmetry about
$\mathcal{A}_{\rm refl}=0$. This provides a differential readout
of both the magnitude and sign of the perturbation without requiring
polarization-resolved detection, although it does not by itself imply
enhanced measurement precision.\\
\textit{b. Polarimetric readout near invisibility}:
Let
\begin{equation}
\chi_0\equiv\chi_L(\delta=0),
\qquad
\chi_0^\dagger\chi_0=1,
\end{equation}
denote the normalized left-invisible incident spinor of the
unperturbed sample at the calibrated energy $E_0$. Here $\delta$ denotes a small deviation of the sensed parameter from its calibrated value, such as $\delta B_z$. Under a
perturbation $\delta$, the incident energy and spinor are held fixed,
while the transmitted spinor is
\begin{equation}
\eta_R(\delta)
=
t_{L\rightarrow R}(E_0;\delta)\chi_0.
\end{equation}
We define its normalized form and output Bloch vector by
\begin{equation}
\hat{\eta}_R(\delta)
=
\frac{\eta_R(\delta)}
{\sqrt{\eta_R^\dagger(\delta)\eta_R(\delta)}},
\qquad
\mathbf{S}_{\rm out}(\delta)
=
\hat{\eta}_R^\dagger(\delta)
\boldsymbol{\sigma}
\hat{\eta}_R(\delta).
\end{equation}
At the unperturbed invisible point,
$t_{L\rightarrow R}(0)\chi_0=\tau_0\chi_0$, and therefore
\begin{equation}
\mathbf{S}_0
\equiv
\mathbf{S}_{\rm out}(0)
=
\chi_0^\dagger\boldsymbol{\sigma}\chi_0.
\end{equation}

Because $r_L(0)\chi_0=0$, a weak perturbation produces
\begin{align}
&r_L(\delta)\chi_0
=
\delta
\left.
\frac{\partial r_L}{\partial\delta}
\right|_{\delta=0}
\chi_0
+
O(\delta^2),
\\
&\left\|r_L(\delta)\chi_0\right\|^2
=
O(\delta^2),
\nonumber\\
&\mathbf{S}_{\rm out}(\delta)
=
\mathbf{S}_0
+
\delta
\left.
\frac{\partial\mathbf{S}_{\rm out}}{\partial\delta}
\right|_{\delta=0}
+
O(\delta^2).
\label{eq:polarimetric_response}
\end{align}
Thus, the reflected intensity begins quadratically in the
perturbation, whereas the transmitted polarization can change at
first order.

At the non-Abelian invisible point
$E_0\simeq0.182234$, $B_z\simeq0.484598$,
$\theta_L=-0.739699$, and $\theta_R=0.739699$, the
finite-difference calculation gives
\begin{equation}
\left|
\frac{\partial\mathbf{S}_{\rm out}}{\partial B_z}
\right|
\simeq1.731,
\qquad
\left|
\frac{\partial\mathbf{S}_{\rm out}}{\partial V}
\right|
\simeq6.463,
\label{eq:polarimetric_sensitivities}
\end{equation}
with energies measured in units of $|w|$. The same calculation shows
that the two oriented response vectors form an angle of approximately
$139.1^\circ$ on the Bloch sphere. They are therefore not collinear
and can, in principle, be distinguished through a full Stokes-vector
measurement.

By contrast, an Abelian invisible channel is a fixed-axis spin
eigenstate. A scalar onsite shift or a Zeeman perturbation along the
common gauge axis changes its scattering amplitudes but does not
rotate its normalized polarization. The non-Abelian invisible
channel can therefore convert a weak perturbation into a linear,
polarization-resolved signal while its reflected intensity begins
quadratically. Normalizing the Stokes vector rejects ideal
common-mode power fluctuations, although it does not remove detector
or quantum noise.

The principal sensing outcome is thus a first-order polarimetric
readout around a non-Abelian invisible point. Moreover, the direction
of the Stokes-vector response can distinguish perturbations that
couple differently to the onsite matrix, providing information beyond
a scalar intensity measurement.\\
\textit{c. Threshold and frequency readouts}:
In the Abelian $z$-$z$ configuration, a longitudinal Zeeman term shifts the
two fixed-spin lasing frequencies in opposite directions while leaving their
imaginary parts, and hence their threshold gains, unchanged. The resulting
frequency splitting provides a natural magnetic or birefringent readout,
analogous to mode-splitting measurements in optical microlaser sensors
\cite{He2011}. In the non-Abelian configuration, the same perturbation also
changes the imaginary parts of the poles. The signed crossing gain obtained
above changes sign at $B_c\simeq0.04115$. At a calibrated stable operating
point $B_0>B_c$, where $g_{\rm th}=g_\star>0$, a small perturbation obeys
\begin{equation}
\delta g_{\rm th}
\simeq
\left.
\frac{\partial g_\star}{\partial B_z}
\right|_{B_0}
\delta B_z.
\label{eq:sensing_threshold_response}
\end{equation}
Thus the perturbation can be inferred from either the lasing frequency or the
pump value at which a pole reaches the real axis. Here $B_z$ is the Zeeman
energy appearing in the onsite matrix; conversion to a physical magnetic
field requires the appropriate gyromagnetic or magneto-optical coefficient.
The linear transfer matrix determines the threshold location but not the
above-threshold output intensity, which would require gain saturation. The CPA
condition provides the time-reversed counterpart, with the displacement of a
dark absorption channel serving as an additional readout \cite{Chong2010}.

The directional-amplification relation derived in the Abelian sector,
$|t_{L,s}/t_{R,s}|=|d_s|^{N-1}$, provides another possible length-dependent
transduction signal when the perturbation changes $d_s$. However, a large
response or a steep threshold curve does not by itself establish improved
measurement precision. A quantum-sensing claim requires a specified
input--output measurement, the noise introduced by loss and gain, and a
precision measure such as the signal-to-noise ratio or Fisher information
\cite{LauClerk2018,McDonald2020}. In particular, nonreciprocity can be a useful
sensing resource, but the NHSE alone does not guarantee a quantum advantage.
The present transfer-matrix construction supplies the channel-resolved
response functions and operating points needed for such an extension. Natural
platforms include polarization-coupled waveguides and resonators, electrical
circuits with non-Abelian links \cite{ChenNonAbelian2026}, and spinor
atoms in synthetic gauge fields \cite{Goldman2014,Aidelsburger2018}.

\section{Conclusions}
\label{sec:conclusions}

We have developed a transfer-matrix framework that treats the
non-Bloch spectrum, skin localization, and spin-resolved transport of
one-dimensional non-Hermitian gauge lattices within a common
boundary-sensitive description. Starting from a general finite-range
multichannel lattice, we reduced the propagation problem to the
active channel subspace selected by the inter-supercell hopping
matrices. The inactive intracell sector was eliminated through a
Schur complement, so that the transfer-matrix dimension is determined
by the number of active channels rather than by the complete
supercell Hilbert space. For the spin-$1/2$ gauge chain, this
construction produces a $4\times4$ transfer matrix that accommodates
independent forward and backward hoppings and preserves the ordering
of noncommuting spin rotations.

For the closed chain, the transfer eigenvalues were identified with
the generalized Bloch factors. Periodic- and open-boundary spectra,
finite-chain quantization, generalized Brillouin zones, and the NHSE
therefore follow from different boundary conditions imposed on the
same local propagation matrix. In the Abelian limit, the problem
separates into two fixed spin sectors, allowing the finite-chain
spectra, GBZ radii, and localization exponents to be obtained
analytically. Opposite spin projections can consequently acquire
opposite spatial growth rates and accumulate at different
boundaries, providing a transfer-matrix interpretation of the bipolar
NHSE. In the non-Abelian regime, no energy-independent spin basis
decouples the propagation. The spectrum is instead determined
collectively by four generalized Bloch factors, whose middle-modulus
condition selects the thermodynamic OBC continuum, while their
eigenvectors carry the energy-dependent spin texture of the skin
modes. This also shows why a unimodular total transfer matrix does not
preclude nontrivial skin localization: its determinant fixes only the
collective product of the transfer eigenvalues, not their individual
growth rates.

Connecting the finite sample to spin-degenerate featureless leads
extends the same transfer construction directly to scattering. In
the Abelian limit, the exact finite-length transfer matrix can be
expressed through Chebyshev polynomials, yielding analytical
conditions for quasibound states and lasing, scattering zeros and
CPA, reflectionlessness, transparency, and invisibility. These
expressions separate the reciprocal propagation scale from the
directed-amplification factor. At fixed reciprocal scale and onsite
parameters, the latter does not shift the pole, zero, or
reflectionless conditions, but it controls the directional
transmission amplitudes, the transparency conditions, and the
spin-resolved GBZ radius. This establishes an explicit connection
between scattering asymmetry and skin localization.

The non-Abelian transport problem remains intrinsically matrix
valued. Its transport conditions select particular null spinors of
the transfer blocks rather than independent fixed-axis spin sectors.
The corresponding lasing and CPA eigenchannels are coherent
spin-hybridized states, and a Zeeman field can modify both the
stability and the lasing threshold because it does not commute with
the two gauge links. The reflectionless calculations further reveal
a directional effect: left- and right-reflectionless channels need
not coincide, and a wave can be transmitted with negligible
reflection while undergoing a pronounced spin rotation. Such a state
must not be confused with a transparent or invisible channel, since
unit transmitted intensity does not guarantee preservation of the
complex transmission amplitude or of the incident spinor.

Our analysis of strict invisibility makes this distinction explicit.
Reflectionlessness and phase-matched transparency must hold for the
same channel vector; separate determinant zeros are not sufficient if
they select different spinors. In the Abelian model, invisible
channels remain pure eigenstates of the common gauge axis. In the
non-Abelian model, the selected invisible channel can instead be a
spin-hybridized coherent superposition. Nevertheless, strict
invisibility restores this spinor after transmission, up to the
reference propagation phase. Non-Abelian reflectionlessness can
therefore produce net spin conversion, whereas non-Abelian
invisibility selects a hybridized input polarization but leaves no
net spin rotation in the outgoing state.

The central result is a direct connection between the transfer modes
governing non-Bloch spectral physics and the spinors selected by
finite-length scattering boundary conditions. Pole and zero
trajectories, directional reflectionless channels, Zeeman-controlled
lasing thresholds, and transport-eigenchannel polarizations provide
observable signatures that complement spectral braiding and boundary
localization. The framework is not restricted to the uniform
two-axis model considered here. It can be extended to spatially
varying gauge fields, longer-range hopping, disorder, additional
orbital channels, and quasi-one-dimensional reductions of
higher-dimensional lattices. Incorporating interactions, nonlinear
gain saturation, or dynamical noise would provide further directions
for studying how non-Abelian gauge structure, non-Bloch propagation,
and multichannel transport cooperate in open non-Hermitian systems.

\begin{acknowledgments}
	The authors acknowledge financial support from the Federal Ministry of
	Education and Research of Germany through the program ``Souver\"an.
	Digital. Vernetzt.'', joint project 6G-life (project identification
	number 16KISK002), and through projects 16KISQ039, 16KISQ093, 16KIS0948,
	and 16KIS1598K, as well as from the DFG Emmy-Noether program under Grant
	No.~NO 1129/2-1. This work was also supported by the Munich Quantum
	Valley, which is supported by the Bavarian state government with funds
	from the Hightech Agenda Bayern Plus.
	
	The authors used GPT-6 Astra (OpenAI) and Claude (Anthropic) to improve
	the wording and presentation of the manuscript, including consistency
	checks of the text, and to assist with the visual presentation of the
	figures. All scientific ideas, derivations, numerical calculations, and
	conclusions are the authors' own. The authors reviewed all AI-assisted
	material and take full responsibility for the content of the paper.
\end{acknowledgments}
\section*{Data availability}
The data that support the findings of this article are not publicly available. The data are available from the authors upon reasonable request.
\appendix
\section{Detailed Mathematical Framework}
\label{app:mathematical_framework}

\subsection{Generalized tight-binding model and active-channel reduction}

We consider a generalized tight-binding Hamiltonian with finite-range hopping
and $q$ internal degrees of freedom, such as spin, orbital, or sublattice
components, per unit cell:
\begin{align}
\mathcal{H}
={}&
\sum_{n=1}^{N}\sum_{\alpha,\beta=1}^{q}
\left[
\sum_{l=1}^{\fr}
\left(
c^{\dagger}_{n,\alpha}
[\bt^{L}_{l,n}]_{\alpha\beta}
c_{n+l,\beta}
+
c^{\dagger}_{n+l,\alpha}
[\bt^{R}_{l,n}]_{\alpha\beta}
c_{n,\beta}
\right)
\right.
\nonumber\\
&\left.
\hspace{3.3cm}
+c^{\dagger}_{n,\alpha}
[\bt_{0,n}]_{\alpha\beta}
c_{n,\beta}
\right].
\label{app:eq:general_hamiltonian}
\end{align}
Here $\bt^{L(R)}_{l,n}$ denotes the hopping matrix to the left (right),
$\bt_{0,n}$ is the intracell term, and $\fr$ is the hopping range.

To obtain a nearest-neighbor recursion, we group $\fr$ adjacent unit cells
into a supercell of dimension $\fn=q\fr$. The superlattice Hamiltonian can
then be written as
\begin{equation}
H=
\sum_n
\left(
\bc_n^{\dagger}\bJ_L\bc_{n+1}
+\bc_n^{\dagger}\bK_n\bc_n
+\bc_{n+1}^{\dagger}\bJ_R\bc_n
\right),
\label{app:eq:supercell_hamiltonian}
\end{equation}
which gives
\begin{equation}
E\Psi_n
=
\bK_n\Psi_n
+\bJ_{R,n}\Psi_{n-1}
+\bJ_{L,n}\Psi_{n+1},
\qquad
\Psi_n\in\mathbb{C}^{\fn}.
\label{app:eq:supercell_recursion}
\end{equation}
The matrices $\bJ_L$ and $\bJ_R$ contain the directed hopping between
neighboring supercells, while $\bK_n$ contains all terms within supercell
$n$.

For hopping range $\fr=2$, their block structure is
\begin{align}
\bJ_{L,n}
&=
\begin{pmatrix}
\bt^{L}_{2,n-1} & 0\\
\bt^{L}_{1,n} & \bt^{L}_{2,n}
\end{pmatrix},
&
\bJ_{R,n}
&=
\begin{pmatrix}
\bt^{R}_{2,n-1} & \bt^{R}_{1,n}\\
0 & \bt^{R}_{2,n}
\end{pmatrix},
\nonumber\\[1mm]
\bK_n
&=
\begin{pmatrix}
\bt_{0,n-1} & \bt^{L}_{1,n-1}\\
\bt^{R}_{1,n-1} & \bt_{0,n}
\end{pmatrix}.
\label{app:eq:range_two_blocks}
\end{align}

We define
\begin{equation}
G_n(E)=EI_{\fn}-\bK_n,
\label{app:eq:G_definition}
\end{equation}
so that Eq. \eqref{app:eq:supercell_recursion} becomes
\begin{equation}
G_n(E)\Psi_n
=
\bJ_{R,n}\Psi_{n-1}
+\bJ_{L,n}\Psi_{n+1}.
\label{app:eq:G_recursion}
\end{equation}
If $\bJ_{L,n}$ is invertible, a conventional transfer matrix follows by
solving directly for $\Psi_{n+1}$. The supercell construction, however,
generally produces rank-deficient inter-supercell hopping matrices. We
therefore retain only the active hopping subspaces.
Let $\rho=\mathrm{rank}\,\bJ_{L}=\mathrm{rank}\,\bJ_{R}$ denote the number
of active channels. The paired hopping factorization is
\begin{equation}
	\bJ_{L}=VX_{L}W^{\dagger},\qquad
	\bJ_{R}=WX_{R}V^{\dagger},\qquad
	V^{\dagger}W=0,
	\label{app:eq:SVD}
\end{equation}
where the columns of $V$ and $W$ are orthonormal bases of the two active
subspaces and $X_{L}$ and $X_{R}$ are the corresponding invertible
$\rho\times\rho$ hopping blocks. A singular value decomposition of
$\bJ_{L}$ supplies such bases and makes $X_{L}$ diagonal, but $X_{R}$ is
then not diagonal in general, and the construction below does not require
either block to be diagonal. The shared bases and the orthogonality
$V^{\dagger}W=0$ are assumptions on the hopping structure. A matrix $Y$ completes the
orthogonal basis of the full supercell space, and the state is expanded as
\begin{equation}
\Psi_n=V\alpha_n+W\beta_n+Y\zeta_n.
\label{app:eq:supercell_expansion}
\end{equation}
The vectors $\alpha_n,\beta_n\in\mathbb{C}^{\rho}$ are active-channel
amplitudes, whereas $\zeta_n$ belongs to the inactive intracell sector.
Using Eqs. \eqref{app:eq:SVD} and \eqref{app:eq:supercell_expansion}, the
hopping terms reduce to
\begin{align}
\bJ_R\Psi_{n-1}&=WX_R\alpha_{n-1},
&
\bJ_L\Psi_{n+1}&=VX_L\beta_{n+1}.
\label{app:eq:projected_hoppings}
\end{align}

With $G_{ab}=A^{\dagger}G_nB$, where $A,B\in\{V,W,Y\}$ correspond to
$a,b\in\{v,w,y\}$, projection of Eq. \eqref{app:eq:G_recursion} gives
\begin{align}
G_{vv}\alpha_n+G_{vw}\beta_n+G_{vy}\zeta_n
&=X_L\beta_{n+1},
\label{app:eq:V_projection}
\\
G_{wv}\alpha_n+G_{ww}\beta_n+G_{wy}\zeta_n
&=X_R\alpha_{n-1},
\label{app:eq:W_projection}
\\
G_{yv}\alpha_n+G_{yw}\beta_n+G_{yy}\zeta_n
&=0.
\label{app:eq:Y_projection}
\end{align}
Assuming that $G_{yy}$ is invertible, the inactive amplitude is
\begin{equation}
\zeta_n
=
-G_{yy}^{-1}
\left(
G_{yv}\alpha_n+G_{yw}\beta_n
\right).
\label{app:eq:inactive_amplitude}
\end{equation}
Substitution into Eqs. \eqref{app:eq:V_projection} and
\eqref{app:eq:W_projection} produces the Schur-complement blocks
\begin{equation}
\overline{G}_{ab}
=
G_{ab}-G_{ay}G_{yy}^{-1}G_{yb},
\qquad
a,b\in\{v,w\},
\label{app:eq:schur_complement}
\end{equation}
and the reduced active-channel equations
\begin{align}
\overline{G}_{vv}\alpha_n+\overline{G}_{vw}\beta_n
&=X_L\beta_{n+1},
\label{app:eq:reduced_V_equation}
\\
\overline{G}_{wv}\alpha_n+\overline{G}_{ww}\beta_n
&=X_R\alpha_{n-1}.
\label{app:eq:reduced_W_equation}
\end{align}
Although $\zeta_n$ is no longer explicit, its coupling to the active
amplitudes is retained through the correction term in
Eq. \eqref{app:eq:schur_complement} \cite{ZhangSchur2005}.

Solving Eq. \eqref{app:eq:reduced_W_equation} for $\alpha_n$ gives
\begin{equation}
\alpha_n
=
-\overline{G}_{wv}^{-1}\overline{G}_{ww}\beta_n
+\overline{G}_{wv}^{-1}X_R\alpha_{n-1}.
\label{app:eq:alpha_solution}
\end{equation}
Substituting this into Eq. \eqref{app:eq:reduced_V_equation} yields
\begin{align}
\beta_{n+1}
={}&
X_L^{-1}
\left[
\overline{G}_{vw}
-\overline{G}_{vv}\overline{G}_{wv}^{-1}\overline{G}_{ww}
\right]\beta_n
\nonumber\\
&+
X_L^{-1}\overline{G}_{vv}
\overline{G}_{wv}^{-1}X_R\alpha_{n-1}.
\label{app:eq:beta_solution}
\end{align}
Therefore, for
\begin{equation}
\Phi_n=
\begin{pmatrix}
\beta_n\\
\alpha_{n-1}
\end{pmatrix},
\qquad
\Phi_{n+1}=M_n(E)\Phi_n,
\label{app:eq:transfer_state}
\end{equation}
the reduced transfer matrix is
\begin{equation}
M_n(E)=
\begin{pmatrix}
X_L^{-1}\!\left[
\overline{G}_{vw}
-\overline{G}_{vv}\overline{G}_{wv}^{-1}\overline{G}_{ww}
\right]
&
X_L^{-1}\overline{G}_{vv}\overline{G}_{wv}^{-1}X_R
\\[1.4ex]
-\overline{G}_{wv}^{-1}\overline{G}_{ww}
&
\overline{G}_{wv}^{-1}X_R
\end{pmatrix}.
\label{app:eq:general_transfer_matrix}
\end{equation}
This is the detailed derivation of Eq. \eqref{eq:generalSVDtransfer}. Its
dimension is $2\rho\times2\rho$, determined by the active hopping rank rather
than by the full supercell dimension $\fn$.

When the active $V$ and $W$ subspaces span the complete supercell Hilbert
space, the $Y$ sector is absent and
$\overline{G}_{ab}=G_{ab}$. In addition,
$G_{vw}=-K_{vw}$ and $G_{wv}=-K_{wv}$, so
Eq. \eqref{app:eq:general_transfer_matrix} reduces to
\begin{equation}
M_n=
\begin{pmatrix}
X_L^{-1}
\left[
G_{vv}K_{wv}^{-1}G_{ww}-K_{vw}
\right]
&
-X_L^{-1}G_{vv}K_{wv}^{-1}X_R
\\[1.4ex]
K_{wv}^{-1}G_{ww}
&
-K_{wv}^{-1}X_R
\end{pmatrix}.
\label{Transfer matrix of supercells}
\end{equation}

\subsection{Featureless leads and the scattering matrix}

We connect the finite superlattice to two semi-infinite featureless leads.
Following Refs. \cite{ghaemiPRA2021,ghaemiPRB2023}, we employ the wide-band
limit, in which the effective channel-space blocks satisfy
\begin{equation}
\overline{G}_{vv}
=
\overline{G}_{ww}
=0_{\rho},
\qquad
\overline{G}_{vw}
=
\overline{G}_{wv}
=-wI_{\rho},
\label{app:eq:wide_band_G}
\end{equation}
together with
\begin{equation}
X_L=X_R=wI_{\rho},
\qquad
w<0.
\label{app:eq:wide_band_X}
\end{equation}
These conditions are equivalent to the conventional wide-band choice
$K_{vv}=K_{ww}=EI_{\rho}$ and $K_{vw}=K_{wv}=wI_{\rho}$ because
$G_{vv}=EI_{\rho}-K_{vv}$ and, using $V^{\dagger}W=0$,
$G_{vw}=-K_{vw}$ and $G_{wv}=-K_{wv}$. When the inactive sector is absent,
$\overline{G}_{ab}=G_{ab}$.

With these conditions, the reduced equations give
$\beta_{n+1}=-\beta_{n}$ and $\alpha_{n}=-\alpha_{n-1}$ in the leads, so
the lead transfer matrix per supercell is $-I_{2\rho}$. This is the
propagation of plane waves with factors $\pm i$ per lattice site. Each
lead supercell consists of two lattice sites, with $\beta_{n}$ on site
$j=2n-1$ and $\alpha_{n}$ on site $j=2n$. Writing $\psi_{j}$ for the
amplitude on lattice site $j$, the lead solutions are
\begin{align}
	\psi_{j}&=\phi^{(+)}\,i^{\,j-j_{+}}+\phi^{(-)}\,(-i)^{\,j-j_{-}},
	\label{lead solution1}\\
	\beta_{n}&=\psi_{2n-1},\qquad \alpha_{n}=\psi_{2n},
	\label{lead solution2}
\end{align}
where
$\phi^{(\pm)}=(\phi^{(\pm)}_{1},\ldots,\phi^{(\pm)}_{\rho})^{T}$ are the
position-independent amplitudes of the right- and left-propagating
channels. The offsets $j_{\pm}$ fix the $U(1)$ phase convention. At each
interface we take $j_{+}$ to be the site carrying $\alpha_{n-1}$ and
$j_{-}$ the site carrying $\beta_{n}$, so that
$(\beta_{n},\alpha_{n-1})^{T}=\mathcal{Q}\,(\phi^{(+)},\phi^{(-)})^{T}$
with $\mathcal{Q}$ given by
\begin{equation}
	\mathcal{Q}
	=
	\begin{pmatrix}
		iI_{\rho} & I_{\rho}\\
		I_{\rho} & iI_{\rho}
	\end{pmatrix},
	\qquad
	\mathcal{Q}^{-1}
	=
	\frac{1}{2}
	\begin{pmatrix}
		-iI_{\rho} & I_{\rho}\\
		I_{\rho} & -iI_{\rho}
	\end{pmatrix}.
	\label{app:eq:Q_matrices}
\end{equation}
The total real-space transfer matrix is the ordered product
\begin{equation}
M=M_NM_{N-1}\cdots M_1.
\label{app:eq:total_transfer_matrix}
\end{equation}
Then, the propagating-wave transfer matrix is
\begin{equation}
\mathcal{M}
=
\mathcal{Q}^{-1}M\mathcal{Q}
\equiv
\begin{pmatrix}
\mathcal{M}_{11} & \mathcal{M}_{12}\\
\mathcal{M}_{21} & \mathcal{M}_{22}
\end{pmatrix},
\label{app:eq:modal_transfer_matrix}
\end{equation}
where each block is $\rho\times\rho$. Thus,
\begin{equation}
\begin{pmatrix}
\phi^{(+,R)}\\
\phi^{(-,R)}
\end{pmatrix}
=
\mathcal{M}
\begin{pmatrix}
\phi^{(+,L)}\\
\phi^{(-,L)}
\end{pmatrix}.
\label{app:eq:modal_transfer_relation}
\end{equation}
Here the superscripts $+$ and $-$ denote right- and left-moving
waves, respectively, while $L$ and $R$ identify the lead. Thus,
$\phi^{(+,L)}$ and $\phi^{(-,R)}$ are incident amplitudes, whereas
$\phi^{(-,L)}$ and $\phi^{(+,R)}$ are outgoing amplitudes. Each amplitude belongs to the active channel space
$\mathbb{C}^{\rho}$.\\ The scattering matrix is defined by
\begin{equation}
\begin{pmatrix}
\phi^{(-,L)}\\
\phi^{(+,R)}
\end{pmatrix}
=
S
\begin{pmatrix}
\phi^{(+,L)}\\
\phi^{(-,R)}
\end{pmatrix},
\qquad
S=
\begin{pmatrix}
r_L & t_{R\rightarrow L}\\
t_{L\rightarrow R} & r_R
\end{pmatrix}.
\label{app:eq:scattering_matrix}
\end{equation}
For invertible $\mathcal{M}_{22}$, solving the second row of
Eq. \eqref{app:eq:modal_transfer_relation} for $\phi^{(-,L)}$ gives
\begin{align}
r_L
&=-\mathcal{M}_{22}^{-1}\mathcal{M}_{21},
&
t_{L\rightarrow R}
&=
\mathcal{M}_{11}
-\mathcal{M}_{12}\mathcal{M}_{22}^{-1}\mathcal{M}_{21},
\nonumber\\
r_R
&=\mathcal{M}_{12}\mathcal{M}_{22}^{-1},
&
t_{R\rightarrow L}
&=\mathcal{M}_{22}^{-1}.
\label{app:eq:scattering_blocks}
\end{align}
\subsection{Transport boundary conditions}
\label{app:transport_boundary_conditions}
\label{sec:boundary_conditions}

The propagating-wave amplitudes introduced in the preceding
subsection belong to the active channel space
$\mathbb{C}^{\rho}$. To simplify the notation, we denote incoming
amplitudes by $\chi$ and outgoing amplitudes by $\eta$, with the
subscript identifying the lead in which the amplitude is defined.
The correspondence with the coefficients
$\phi^{(\pm,L/R)}$ is summarized in
Table \ref{tab:channel_amplitude_notation}.

\begin{table}[t]
    \caption{
    Relation between the channel amplitudes used below and the
    propagating-wave coefficients introduced in the preceding
    subsection. The subscripts $L$ and $R$ identify the lead.
    }
    \label{tab:channel_amplitude_notation}
    \begin{ruledtabular}
        \begin{tabular}{ccc}
            Channel & Lead amplitude & Physical role\\
            \hline
            $\chi_L$ & $\phi^{(+,L)}$ & incoming from left\\
            $\eta_L$ & $\phi^{(-,L)}$ & outgoing into left\\
            $\eta_R$ & $\phi^{(+,R)}$ & outgoing into right\\
            $\chi_R$ & $\phi^{(-,R)}$ & incoming from right
        \end{tabular}
    \end{ruledtabular}
\end{table}

With this notation, the propagating-wave transfer relation becomes
\begin{equation}
\begin{pmatrix}
\eta_R\\
\chi_R
\end{pmatrix}
=
\begin{pmatrix}
\mathcal{M}_{11}&\mathcal{M}_{12}\\
\mathcal{M}_{21}&\mathcal{M}_{22}
\end{pmatrix}
\begin{pmatrix}
\chi_L\\
\eta_L
\end{pmatrix}.
\label{app:eq:channel_transfer_relation}
\end{equation}
For the spinful model, $\rho=2$, and each channel amplitude is a
two-component spinor,
\begin{equation}
\chi_{\alpha}
=
\begin{pmatrix}
\chi_{\alpha,+}\\
\chi_{\alpha,-}
\end{pmatrix},
\qquad
\eta_{\alpha}
=
\begin{pmatrix}
\eta_{\alpha,+}\\
\eta_{\alpha,-}
\end{pmatrix},
\qquad
\alpha=L,R.
\end{equation}
The subscripts $+$ and $-$ on the components denote the two spin
states and should not be confused with the propagation labels on
$\phi^{(\pm,L/R)}$.

A boundary condition is called \emph{strong} when it holds as a
matrix identity over the full active-channel space and therefore
applies to every channel. More generally, the same effect may occur
only for a particular nonzero channel vector, defining a transport
eigenchannel.\\
For transparency and invisibility, $\tau_{0}$ denotes the transmission
amplitude through a segment of the featureless lead with the same number
of lattice sites as the scattering region. In the wide-band phase
convention used here, a right-propagating wave acquires a factor of $i$
per lattice site, so that $\tau_{0}=i^{N_{\mathrm{ref}}}$ for a reference
segment of $N_{\mathrm{ref}}$ sites. For the uniform $N$-site sample
considered in the main text, $\tau_{0}=i^{N}$.

Table \ref{tab:transport_BC} collects the closed-chain spectral
conditions and the strong transport boundary conditions in the
real-space and propagating-wave bases. We examine the corresponding
channel-resolved conditions below.

\begin{table*}[t]
    \caption{
    Boundary conditions expressed in terms of the real-space transfer
    matrix $\mathbf{M}$ and the propagating-wave transfer matrix
    $\mathcal{M}$. Matrix equalities represent strong multichannel
    conditions. The quasibound and scattering-zero conditions generally
    occur at complex energies; their real-energy realizations correspond
    to lasing and coherent perfect absorption (CPA), respectively.
    The quantity $\tau_0$ is the transmission amplitude of an equivalent
    segment of the featureless lead.
    }
    \label{tab:transport_BC}
    \begin{ruledtabular}
        \begin{tabular}{lll}
            Situation
            &
            Real-space transfer condition
            &
            Propagating-wave transfer condition
            \\
            \hline

            Open
            &
            $\mathbf{M}_{11}=0_{\rho}$
            &
            $\mathcal{M}_{11}+\mathcal{M}_{22}
            +i\mathcal{M}_{12}-i\mathcal{M}_{21}=0_{\rho}$
            \\[1mm]

            Periodic
            &
            $\det\!\left(\mathbf{M}-e^{ik}I_{2\rho}\right)=0$
            &
            $\det\!\left(\mathcal{M}-e^{ik}I_{2\rho}\right)=0$
            \\[1mm]

            Quasibound / lasing
            &
            $\mathbf{M}_{11}+\mathbf{M}_{22}
            +i\mathbf{M}_{12}-i\mathbf{M}_{21}=0_{\rho}$
            &
            $\mathcal{M}_{22}=0_{\rho}$
            \\[1mm]

            Scattering zero / CPA
            &
            $\mathbf{M}_{11}+\mathbf{M}_{22}
            -i\mathbf{M}_{12}+i\mathbf{M}_{21}=0_{\rho}$
            &
            $\mathcal{M}_{11}=0_{\rho}$
            \\[1mm]

            Left reflectionless
            &
            $i\mathbf{M}_{11}+\mathbf{M}_{12}
            +\mathbf{M}_{21}-i\mathbf{M}_{22}=0_{\rho}$
            &
            $\mathcal{M}_{21}=0_{\rho}$
            \\[1mm]

            Right reflectionless
            &
            $-i\mathbf{M}_{11}+\mathbf{M}_{12}
            +\mathbf{M}_{21}+i\mathbf{M}_{22}=0_{\rho}$
            &
            $\mathcal{M}_{12}=0_{\rho}$
            \\[1mm]

            Left transparent
            &
            $\displaystyle
            (\mathbf{M}^{-1})_{11}+(\mathbf{M}^{-1})_{22}
            -i(\mathbf{M}^{-1})_{12}+i(\mathbf{M}^{-1})_{21}
            =2\tau_0^{-1}I_{\rho}$
            &
            $(\mathcal{M}^{-1})_{11}=\tau_0^{-1}I_{\rho}$
            \\[2mm]

            Right transparent
            &
            $\mathbf{M}_{11}+\mathbf{M}_{22}
            +i\mathbf{M}_{12}-i\mathbf{M}_{21}
            =2\tau_0^{-1}I_{\rho}$
            &
            $\mathcal{M}_{22}=\tau_0^{-1}I_{\rho}$
            \\[1mm]

            Left invisible
            &
            $\begin{gathered}
                i\mathbf{M}_{11}+\mathbf{M}_{12}
                +\mathbf{M}_{21}-i\mathbf{M}_{22}=0_{\rho},\\
                \mathbf{M}_{11}+\mathbf{M}_{22}
                -i\mathbf{M}_{12}+i\mathbf{M}_{21}
                =2\tau_0 I_{\rho}
            \end{gathered}$
            &
            $\begin{gathered}
                \mathcal{M}_{21}=0_{\rho},\\
                \mathcal{M}_{11}=\tau_0 I_{\rho}
            \end{gathered}$
            \\[3mm]

            Right invisible
            &
            $\begin{gathered}
                -i\mathbf{M}_{11}+\mathbf{M}_{12}
                +\mathbf{M}_{21}+i\mathbf{M}_{22}=0_{\rho},\\
                \mathbf{M}_{11}+\mathbf{M}_{22}
                +i\mathbf{M}_{12}-i\mathbf{M}_{21}
                =2\tau_0^{-1}I_{\rho}
            \end{gathered}$
            &
            $\begin{gathered}
                \mathcal{M}_{12}=0_{\rho},\\
                \mathcal{M}_{22}=\tau_0^{-1}I_{\rho}
            \end{gathered}$
        \end{tabular}
    \end{ruledtabular}
\end{table*}

For open boundary conditions, let
$u_{\mathrm{OBC}}\in\mathbb{C}^{\rho}$ denote the nonzero boundary
vector in the real-space transfer state. This vector is not a lead
amplitude because the closed chain is not attached to leads. The
channel-resolved OBC condition is
\begin{equation}
\mathbf{M}_{11}(E)u_{\mathrm{OBC}}=0,
\qquad
u_{\mathrm{OBC}}\neq0,
\qquad
\det\mathbf{M}_{11}(E)=0.
\label{app:eq:open_boundary_condition}
\end{equation}
The corresponding strong relation is
$\mathbf{M}_{11}=0_{\rho}$.

For periodic boundary conditions, the transfer state reproduces
itself up to the boundary phase accumulated over the interval
represented by $\mathbf{M}$. Hence,
\begin{equation}
\det\!\left[
\mathbf{M}(E)-e^{ik}I_{2\rho}
\right]=0.
\label{eq:general_PBC}
\end{equation}
Because $\mathbf{M}$ and $\mathcal{M}$ are related by a similarity
transformation, the same condition holds in the propagating-wave
basis. If $\mathbf{M}$ represents one unit cell, $k$ is the Bloch
quasimomentum. If it represents an entire finite ring, $e^{ik}$
denotes the boundary phase across the ring, with ordinary PBC
corresponding to $e^{ik}=1$.

A quasibound state has no incident waves,
\begin{equation}
\chi_L=\chi_R=0,
\end{equation}
and at least one outgoing amplitude must be nonzero. Taking
$\eta_L\neq0$ as the independent outgoing amplitude,
Eq. \eqref{app:eq:channel_transfer_relation} gives
\begin{equation}
\mathcal{M}_{22}(E)\eta_L=0,
\qquad
\eta_R=\mathcal{M}_{12}(E)\eta_L.
\label{eq:quasibound_eigenchannel}
\end{equation}
A quasibound state therefore exists when
$\det\mathcal{M}_{22}(E)=0$. These solutions generally occur at
complex energies. When a pole reaches the real-energy axis, the
purely outgoing solution becomes a stationary lasing eigenchannel.
The corresponding strong condition is
$\mathcal{M}_{22}=0_{\rho}$.

A scattering zero has no outgoing waves,
\begin{equation}
\eta_L=\eta_R=0.
\end{equation}
Taking $\chi_L\neq0$ as the independent incoming amplitude,
Eq. \eqref{app:eq:channel_transfer_relation} gives
\begin{equation}
\mathcal{M}_{11}(E)\chi_L=0,
\qquad
\chi_R=\mathcal{M}_{21}(E)\chi_L.
\label{eq:scattering_zero_eigenchannel}
\end{equation}
Thus, $\det\mathcal{M}_{11}(E)=0$ determines a scattering zero. A
real-energy scattering zero is a CPA eigenchannel. The corresponding
strong scattering-zero or CPA condition is
$\mathcal{M}_{11}=0_{\rho}$.

Reflectionless transport is directional. For incidence from the
left, $\chi_L\neq0$ and there is no incident wave from the right,
so $\chi_R=0$. Left reflectionlessness requires the outgoing
amplitude in the left lead to vanish, $\eta_L=0$. The transfer
relation then gives
\begin{equation}
\mathcal{M}_{21}\chi_L=0,
\qquad
\eta_R=\mathcal{M}_{11}\chi_L.
\label{eq:left_reflectionless_eigenchannel}
\end{equation}
A left-reflectionless eigenchannel exists when
$\det\mathcal{M}_{21}=0$. Here $\chi_L$ is the incident spinor and
$\eta_R$ is the corresponding transmitted spinor.

For incidence from the right, $\chi_R\neq0$ and there is no incident
wave from the left, so $\chi_L=0$. Right reflectionlessness requires
the outgoing amplitude in the right lead to vanish, $\eta_R=0$.
Equation \eqref{app:eq:channel_transfer_relation} then gives
\begin{equation}
\mathcal{M}_{12}\eta_L=0,
\qquad
\chi_R=\mathcal{M}_{22}\eta_L.
\label{eq:right_reflectionless_eigenchannel}
\end{equation}
A right-reflectionless eigenchannel exists when
$\det\mathcal{M}_{12}=0$. Here $\eta_L$ is the spinor transmitted
into the left lead, while the physical incident spinor is obtained
from $\chi_R=\mathcal{M}_{22}\eta_L$.

The strong conditions
$\mathcal{M}_{21}=0_{\rho}$ and
$\mathcal{M}_{12}=0_{\rho}$ imply vanishing reflection for every
incident channel from the left and right, respectively. For example,
\begin{equation}
\mathcal{M}_{21}=0_{\rho}
\quad\Longrightarrow\quad
\det\mathcal{M}_{21}=0,
\label{app:eq:strong_to_eigenchannel}
\end{equation}
and analogously for $\mathcal{M}_{12}$. The converse does not
generally hold: a zero determinant guarantees only one
reflectionless eigenchannel.

A channel is transparent when it is transmitted with the same
amplitude and spin composition as through the reference lead
segment. For incidence from the left, the transmitted amplitude is
$\eta_R=t_{L\rightarrow R}\chi_L$, whereas for incidence from the
right it is $\eta_L=t_{R\rightarrow L}\chi_R$. The transparency
conditions are therefore
\begin{align}
t_{L\rightarrow R}\chi_L
&=
\eta_R
=
\tau_0\chi_L,
\label{eq:left_transparent_eigenchannel}
\\
t_{R\rightarrow L}\chi_R
&=
\eta_L
=
\tau_0\chi_R.
\label{eq:right_transparent_eigenchannel}
\end{align}
The corresponding determinant conditions are
\begin{equation}
\det\!\left(
t_{L\rightarrow R}-\tau_0I_{\rho}
\right)=0,
\qquad
\det\!\left(
t_{R\rightarrow L}-\tau_0I_{\rho}
\right)=0.
\label{eq:transparent_determinants}
\end{equation}
Transparency for every active channel gives the strong matrix
conditions listed in Table \ref{tab:transport_BC}.

Finally, invisibility requires reflectionlessness and transparency
for the same incident channel. For incidence from the left, the
amplitudes satisfy
\begin{equation}
\chi_R=0,
\qquad
\eta_L=0,
\qquad
\eta_R=\tau_0\chi_L.
\label{eq:left_invisible_amplitudes}
\end{equation}
Equation \eqref{app:eq:channel_transfer_relation} therefore gives
\begin{equation}
\mathcal{M}_{21}\chi_L=0,
\qquad
\left(
\mathcal{M}_{11}-\tau_0I_{\rho}
\right)\chi_L=0.
\label{eq:left_invisible_channel}
\end{equation}

For incidence from the right, the amplitudes satisfy
\begin{equation}
\chi_L=0,
\qquad
\eta_R=0,
\qquad
\eta_L=\tau_0\chi_R.
\label{eq:right_invisible_amplitudes}
\end{equation}
Using $\chi_R=\mathcal{M}_{22}\eta_L$, the right-invisibility
conditions become
\begin{equation}
\mathcal{M}_{12}\eta_L=0,
\qquad
\left(
\mathcal{M}_{22}-\tau_0^{-1}I_{\rho}
\right)\eta_L=0.
\label{eq:right_invisible_channel}
\end{equation}
At right invisibility,
\begin{equation}
\chi_R
=
\mathcal{M}_{22}\eta_L
=
\tau_0^{-1}\eta_L,
\qquad
\eta_L=\tau_0\chi_R.
\label{eq:right_invisible_chi_eta_relation}
\end{equation}
Thus, $\chi_R$ and $\eta_L$ are proportional and have the same
normalized spin polarization at an invisible point.

Separate zeros of the reflectionless and transparency determinants
are not sufficient to establish invisibility, because they may
select different channel vectors. The same nonzero vector must
satisfy both equations in Eq. \eqref{eq:left_invisible_channel} or
Eq. \eqref{eq:right_invisible_channel}. Replacing these
channel-resolved equations by matrix identities gives the strong
left- and right-invisibility conditions in
Table \ref{tab:transport_BC}.
\bibliography{refs}

\begin{thebibliography}{52}%
\makeatletter
\providecommand \@ifxundefined [1]{%
 \@ifx{#1\undefined}
}%
\providecommand \@ifnum [1]{%
 \ifnum #1\expandafter \@firstoftwo
 \else \expandafter \@secondoftwo
 \fi
}%
\providecommand \@ifx [1]{%
 \ifx #1\expandafter \@firstoftwo
 \else \expandafter \@secondoftwo
 \fi
}%
\providecommand \natexlab [1]{#1}%
\providecommand \enquote  [1]{``#1''}%
\providecommand \bibnamefont  [1]{#1}%
\providecommand \bibfnamefont [1]{#1}%
\providecommand \citenamefont [1]{#1}%
\providecommand \href@noop [0]{\@secondoftwo}%
\providecommand \href [0]{\begingroup \@sanitize@url \@href}%
\providecommand \@href[1]{\@@startlink{#1}\@@href}%
\providecommand \@@href[1]{\endgroup#1\@@endlink}%
\providecommand \@sanitize@url [0]{\catcode `\\12\catcode `\$12\catcode `\&12\catcode `\#12\catcode `\^12\catcode `\_12\catcode `\%12\relax}%
\providecommand \@@startlink[1]{}%
\providecommand \@@endlink[0]{}%
\providecommand \url  [0]{\begingroup\@sanitize@url \@url }%
\providecommand \@url [1]{\endgroup\@href {#1}{\urlprefix }}%
\providecommand \urlprefix  [0]{URL }%
\providecommand \Eprint [0]{\href }%
\providecommand \doibase [0]{https://doi.org/}%
\providecommand \selectlanguage [0]{\@gobble}%
\providecommand \bibinfo  [0]{\@secondoftwo}%
\providecommand \bibfield  [0]{\@secondoftwo}%
\providecommand \translation [1]{[#1]}%
\providecommand \BibitemOpen [0]{}%
\providecommand \bibitemStop [0]{}%
\providecommand \bibitemNoStop [0]{.\EOS\space}%
\providecommand \EOS [0]{\spacefactor3000\relax}%
\providecommand \BibitemShut  [1]{\csname bibitem#1\endcsname}%
\let\auto@bib@innerbib\@empty
\bibitem [{\citenamefont {Moiseyev}(2011)}]{Moiseyev2011}%
  \BibitemOpen
  \bibfield  {author} {\bibinfo {author} {\bibfnamefont {N.}~\bibnamefont {Moiseyev}},\ }\href {https://doi.org/10.1017/CBO9780511976186} {\emph {\bibinfo {title} {Non-Hermitian Quantum Mechanics}}}\ (\bibinfo  {publisher} {Cambridge University Press},\ \bibinfo {address} {Cambridge},\ \bibinfo {year} {2011})\BibitemShut {NoStop}%
\bibitem [{\citenamefont {El-Ganainy}\ \emph {et~al.}(2018)\citenamefont {El-Ganainy}, \citenamefont {Makris}, \citenamefont {Khajavikhan}, \citenamefont {Musslimani}, \citenamefont {Rotter},\ and\ \citenamefont {Christodoulides}}]{ElGanainy2018}%
  \BibitemOpen
  \bibfield  {author} {\bibinfo {author} {\bibfnamefont {R.}~\bibnamefont {El-Ganainy}}, \bibinfo {author} {\bibfnamefont {K.~G.}\ \bibnamefont {Makris}}, \bibinfo {author} {\bibfnamefont {M.}~\bibnamefont {Khajavikhan}}, \bibinfo {author} {\bibfnamefont {Z.~H.}\ \bibnamefont {Musslimani}}, \bibinfo {author} {\bibfnamefont {S.}~\bibnamefont {Rotter}},\ and\ \bibinfo {author} {\bibfnamefont {D.~N.}\ \bibnamefont {Christodoulides}},\ }\bibfield  {title} {\bibinfo {title} {Non-hermitian physics and {PT} symmetry},\ }\href {https://doi.org/10.1038/nphys4323} {\bibfield  {journal} {\bibinfo  {journal} {Nature Physics}\ }\textbf {\bibinfo {volume} {14}},\ \bibinfo {pages} {11} (\bibinfo {year} {2018})}\BibitemShut {NoStop}%
\bibitem [{\citenamefont {Ashida}\ \emph {et~al.}(2020)\citenamefont {Ashida}, \citenamefont {Gong},\ and\ \citenamefont {Ueda}}]{Ashida2020}%
  \BibitemOpen
  \bibfield  {author} {\bibinfo {author} {\bibfnamefont {Y.}~\bibnamefont {Ashida}}, \bibinfo {author} {\bibfnamefont {Z.}~\bibnamefont {Gong}},\ and\ \bibinfo {author} {\bibfnamefont {M.}~\bibnamefont {Ueda}},\ }\bibfield  {title} {\bibinfo {title} {Non-hermitian physics},\ }\href {https://doi.org/10.1080/00018732.2021.1876991} {\bibfield  {journal} {\bibinfo  {journal} {Advances in Physics}\ }\textbf {\bibinfo {volume} {69}},\ \bibinfo {pages} {249} (\bibinfo {year} {2020})}\BibitemShut {NoStop}%
\bibitem [{\citenamefont {Ding}\ \emph {et~al.}(2022)\citenamefont {Ding}, \citenamefont {Fang},\ and\ \citenamefont {Ma}}]{Ding2022}%
  \BibitemOpen
  \bibfield  {author} {\bibinfo {author} {\bibfnamefont {K.}~\bibnamefont {Ding}}, \bibinfo {author} {\bibfnamefont {C.}~\bibnamefont {Fang}},\ and\ \bibinfo {author} {\bibfnamefont {G.}~\bibnamefont {Ma}},\ }\bibfield  {title} {\bibinfo {title} {Non-hermitian topology and exceptional-point geometries},\ }\href {https://doi.org/10.1038/s42254-022-00516-5} {\bibfield  {journal} {\bibinfo  {journal} {Nature Reviews Physics}\ }\textbf {\bibinfo {volume} {4}},\ \bibinfo {pages} {745} (\bibinfo {year} {2022})}\BibitemShut {NoStop}%
\bibitem [{\citenamefont {Roccati}\ \emph {et~al.}(2022)\citenamefont {Roccati}, \citenamefont {Palma}, \citenamefont {Ciccarello},\ and\ \citenamefont {Bagarello}}]{Roccati2022}%
  \BibitemOpen
  \bibfield  {author} {\bibinfo {author} {\bibfnamefont {F.}~\bibnamefont {Roccati}}, \bibinfo {author} {\bibfnamefont {G.~M.}\ \bibnamefont {Palma}}, \bibinfo {author} {\bibfnamefont {F.}~\bibnamefont {Ciccarello}},\ and\ \bibinfo {author} {\bibfnamefont {F.}~\bibnamefont {Bagarello}},\ }\bibfield  {title} {\bibinfo {title} {Non-hermitian physics and master equations},\ }\href {https://doi.org/10.1142/S1230161222500044} {\bibfield  {journal} {\bibinfo  {journal} {Open Systems \& Information Dynamics}\ }\textbf {\bibinfo {volume} {29}},\ \bibinfo {pages} {2250004} (\bibinfo {year} {2022})}\BibitemShut {NoStop}%
\bibitem [{\citenamefont {Ghaemi-Dizicheh}\ and\ \citenamefont {Ramezani}(2023)}]{Ghaemi-Dizicheh:23}%
  \BibitemOpen
  \bibfield  {author} {\bibinfo {author} {\bibfnamefont {H.}~\bibnamefont {Ghaemi-Dizicheh}}\ and\ \bibinfo {author} {\bibfnamefont {H.}~\bibnamefont {Ramezani}},\ }\bibfield  {title} {\bibinfo {title} {Non-hermitian floquet-free analytically solvable time-dependent systems},\ }\href {https://doi.org/10.1364/OME.483188} {\bibfield  {journal} {\bibinfo  {journal} {Opt. Mater. Express}\ }\textbf {\bibinfo {volume} {13}},\ \bibinfo {pages} {678} (\bibinfo {year} {2023})}\BibitemShut {NoStop}%
\bibitem [{\citenamefont {Ghaemi-Dizicheh}(2024)}]{Ghaemi-Dizicheh_2024}%
  \BibitemOpen
  \bibfield  {author} {\bibinfo {author} {\bibfnamefont {H.}~\bibnamefont {Ghaemi-Dizicheh}},\ }\bibfield  {title} {\bibinfo {title} {A class of stable nonlinear non-hermitian skin modes},\ }\href {https://doi.org/10.1088/1402-4896/ad91f0} {\bibfield  {journal} {\bibinfo  {journal} {Physica Scripta}\ }\textbf {\bibinfo {volume} {99}},\ \bibinfo {pages} {125411} (\bibinfo {year} {2024})}\BibitemShut {NoStop}%
\bibitem [{\citenamefont {Ghaemidizicheh}\ \emph {et~al.}(2025)\citenamefont {Ghaemidizicheh}, \citenamefont {Dehdashti}, \citenamefont {Hanke}, \citenamefont {Touhami},\ and\ \citenamefont {N\"{o}tzel}}]{Ghaemidizicheh:25}%
  \BibitemOpen
  \bibfield  {author} {\bibinfo {author} {\bibfnamefont {H.}~\bibnamefont {Ghaemidizicheh}}, \bibinfo {author} {\bibfnamefont {S.}~\bibnamefont {Dehdashti}}, \bibinfo {author} {\bibfnamefont {A.}~\bibnamefont {Hanke}}, \bibinfo {author} {\bibfnamefont {A.}~\bibnamefont {Touhami}},\ and\ \bibinfo {author} {\bibfnamefont {J.}~\bibnamefont {N\"{o}tzel}},\ }\bibfield  {title} {\bibinfo {title} {Higher-order exceptional points in a non-reciprocal waveguide beam splitter},\ }\href {https://doi.org/10.1364/OE.563594} {\bibfield  {journal} {\bibinfo  {journal} {Opt. Express}\ }\textbf {\bibinfo {volume} {33}},\ \bibinfo {pages} {26329} (\bibinfo {year} {2025})}\BibitemShut {NoStop}%
\bibitem [{\citenamefont {Rodr\'{i}guez-Lara}\ \emph {et~al.}(2026)\citenamefont {Rodr\'{i}guez-Lara}, \citenamefont {Ghaemi-Dizicheh}, \citenamefont {Dehdashti}, \citenamefont {Hanke}, \citenamefont {Touhami},\ and\ \citenamefont {N\"{o}tzel}}]{Rodriguez-Lara:26}%
  \BibitemOpen
  \bibfield  {author} {\bibinfo {author} {\bibfnamefont {B.~M.}\ \bibnamefont {Rodr\'{i}guez-Lara}}, \bibinfo {author} {\bibfnamefont {H.}~\bibnamefont {Ghaemi-Dizicheh}}, \bibinfo {author} {\bibfnamefont {S.}~\bibnamefont {Dehdashti}}, \bibinfo {author} {\bibfnamefont {A.}~\bibnamefont {Hanke}}, \bibinfo {author} {\bibfnamefont {A.}~\bibnamefont {Touhami}},\ and\ \bibinfo {author} {\bibfnamefont {J.}~\bibnamefont {N\"{o}tzel}},\ }\bibfield  {title} {\bibinfo {title} {Non-hermitian sl(3,c) three-mode couplers},\ }\href {https://doi.org/10.1364/PRJ.583226} {\bibfield  {journal} {\bibinfo  {journal} {Photon. Res.}\ }\textbf {\bibinfo {volume} {14}},\ \bibinfo {pages} {1604} (\bibinfo {year} {2026})}\BibitemShut {NoStop}%
\bibitem [{\citenamefont {Rodriguez-Lara}\ \emph {et~al.}(2026)\citenamefont {Rodriguez-Lara}, \citenamefont {Ghaemi-Dizicheh}, \citenamefont {Dehdashti}, \citenamefont {Hanke}, \citenamefont {Touhami},\ and\ \citenamefont {N{\"o}tzel}}]{rodriguez2026grid}%
  \BibitemOpen
  \bibfield  {author} {\bibinfo {author} {\bibfnamefont {B.}~\bibnamefont {Rodriguez-Lara}}, \bibinfo {author} {\bibfnamefont {H.}~\bibnamefont {Ghaemi-Dizicheh}}, \bibinfo {author} {\bibfnamefont {S.}~\bibnamefont {Dehdashti}}, \bibinfo {author} {\bibfnamefont {A.}~\bibnamefont {Hanke}}, \bibinfo {author} {\bibfnamefont {A.}~\bibnamefont {Touhami}},\ and\ \bibinfo {author} {\bibfnamefont {J.}~\bibnamefont {N{\"o}tzel}},\ }\bibfield  {title} {\bibinfo {title} {Grid-state deformation in a no-jump non-hermitian bosonic dimer},\ }\href@noop {} {\bibfield  {journal} {\bibinfo  {journal} {arXiv preprint arXiv:2606.17036}\ } (\bibinfo {year} {2026})}\BibitemShut {NoStop}%
\bibitem [{\citenamefont {Huang}\ \emph {et~al.}(2024)\citenamefont {Huang}, \citenamefont {Huang}, \citenamefont {Shen}, \citenamefont {Yves}, \citenamefont {Pilipchuk}, \citenamefont {Ni}, \citenamefont {Kim}, \citenamefont {Chiang}, \citenamefont {Powell}, \citenamefont {Zhu}, \citenamefont {Cheng}, \citenamefont {Li}, \citenamefont {Sadreev}, \citenamefont {Al{\`u}},\ and\ \citenamefont {Miroshnichenko}}]{Huang2024Acoustic}%
  \BibitemOpen
  \bibfield  {author} {\bibinfo {author} {\bibfnamefont {L.}~\bibnamefont {Huang}}, \bibinfo {author} {\bibfnamefont {S.}~\bibnamefont {Huang}}, \bibinfo {author} {\bibfnamefont {C.}~\bibnamefont {Shen}}, \bibinfo {author} {\bibfnamefont {S.}~\bibnamefont {Yves}}, \bibinfo {author} {\bibfnamefont {A.~S.}\ \bibnamefont {Pilipchuk}}, \bibinfo {author} {\bibfnamefont {X.}~\bibnamefont {Ni}}, \bibinfo {author} {\bibfnamefont {S.}~\bibnamefont {Kim}}, \bibinfo {author} {\bibfnamefont {Y.~K.}\ \bibnamefont {Chiang}}, \bibinfo {author} {\bibfnamefont {D.~A.}\ \bibnamefont {Powell}}, \bibinfo {author} {\bibfnamefont {J.}~\bibnamefont {Zhu}}, \bibinfo {author} {\bibfnamefont {Y.}~\bibnamefont {Cheng}}, \bibinfo {author} {\bibfnamefont {Y.}~\bibnamefont {Li}}, \bibinfo {author} {\bibfnamefont {A.~F.}\ \bibnamefont {Sadreev}}, \bibinfo {author} {\bibfnamefont {A.}~\bibnamefont {Al{\`u}}},\ and\ \bibinfo {author} {\bibfnamefont {A.~E.}\ \bibnamefont {Miroshnichenko}},\ }\bibfield  {title} {\bibinfo {title} {Acoustic resonances in non-hermitian open systems},\ }\href {https://doi.org/10.1038/s42254-023-00659-z} {\bibfield  {journal} {\bibinfo  {journal} {Nature Reviews Physics}\ }\textbf {\bibinfo {volume} {6}},\ \bibinfo {pages} {11} (\bibinfo {year} {2024})}\BibitemShut {NoStop}%
\bibitem [{\citenamefont {Xiao}\ \emph {et~al.}(2025)\citenamefont {Xiao}, \citenamefont {Wang}, \citenamefont {Qu}, \citenamefont {Gao}, \citenamefont {Lin}, \citenamefont {Bian}, \citenamefont {Zhan},\ and\ \citenamefont {Xue}}]{Xiao2025Photonics}%
  \BibitemOpen
  \bibfield  {author} {\bibinfo {author} {\bibfnamefont {L.}~\bibnamefont {Xiao}}, \bibinfo {author} {\bibfnamefont {K.}~\bibnamefont {Wang}}, \bibinfo {author} {\bibfnamefont {D.}~\bibnamefont {Qu}}, \bibinfo {author} {\bibfnamefont {H.}~\bibnamefont {Gao}}, \bibinfo {author} {\bibfnamefont {Q.}~\bibnamefont {Lin}}, \bibinfo {author} {\bibfnamefont {Z.}~\bibnamefont {Bian}}, \bibinfo {author} {\bibfnamefont {X.}~\bibnamefont {Zhan}},\ and\ \bibinfo {author} {\bibfnamefont {P.}~\bibnamefont {Xue}},\ }\bibfield  {title} {\bibinfo {title} {Non-hermitian physics in photonic systems},\ }\href {https://doi.org/10.3788/PI.2025.R09} {\bibfield  {journal} {\bibinfo  {journal} {Photonics Insights}\ }\textbf {\bibinfo {volume} {4}},\ \bibinfo {pages} {R09} (\bibinfo {year} {2025})}\BibitemShut {NoStop}%
\bibitem [{\citenamefont {Bergholtz}\ \emph {et~al.}(2021)\citenamefont {Bergholtz}, \citenamefont {Budich},\ and\ \citenamefont {Kunst}}]{Bergholtz2021}%
  \BibitemOpen
  \bibfield  {author} {\bibinfo {author} {\bibfnamefont {E.~J.}\ \bibnamefont {Bergholtz}}, \bibinfo {author} {\bibfnamefont {J.~C.}\ \bibnamefont {Budich}},\ and\ \bibinfo {author} {\bibfnamefont {F.~K.}\ \bibnamefont {Kunst}},\ }\bibfield  {title} {\bibinfo {title} {Exceptional topology of non-hermitian systems},\ }\href {https://doi.org/10.1103/RevModPhys.93.015005} {\bibfield  {journal} {\bibinfo  {journal} {Rev. Mod. Phys.}\ }\textbf {\bibinfo {volume} {93}},\ \bibinfo {pages} {015005} (\bibinfo {year} {2021})}\BibitemShut {NoStop}%
\bibitem [{\citenamefont {Okuma}\ and\ \citenamefont {Sato}(2023)}]{OkumaSato2023}%
  \BibitemOpen
  \bibfield  {author} {\bibinfo {author} {\bibfnamefont {N.}~\bibnamefont {Okuma}}\ and\ \bibinfo {author} {\bibfnamefont {M.}~\bibnamefont {Sato}},\ }\bibfield  {title} {\bibinfo {title} {Non-hermitian topological phenomena: A review},\ }\href {https://doi.org/10.1146/annurev-conmatphys-040521-033133} {\bibfield  {journal} {\bibinfo  {journal} {Annual Review of Condensed Matter Physics}\ }\textbf {\bibinfo {volume} {14}},\ \bibinfo {pages} {83} (\bibinfo {year} {2023})}\BibitemShut {NoStop}%
\bibitem [{\citenamefont {Hatano}\ and\ \citenamefont {Nelson}(1996)}]{HatanoNelson1996}%
  \BibitemOpen
  \bibfield  {author} {\bibinfo {author} {\bibfnamefont {N.}~\bibnamefont {Hatano}}\ and\ \bibinfo {author} {\bibfnamefont {D.~R.}\ \bibnamefont {Nelson}},\ }\bibfield  {title} {\bibinfo {title} {Localization transitions in non-hermitian quantum mechanics},\ }\href {https://doi.org/10.1103/PhysRevLett.77.570} {\bibfield  {journal} {\bibinfo  {journal} {Phys. Rev. Lett.}\ }\textbf {\bibinfo {volume} {77}},\ \bibinfo {pages} {570} (\bibinfo {year} {1996})}\BibitemShut {NoStop}%
\bibitem [{\citenamefont {Yao}\ and\ \citenamefont {Wang}(2018)}]{YaoWang2018}%
  \BibitemOpen
  \bibfield  {author} {\bibinfo {author} {\bibfnamefont {S.}~\bibnamefont {Yao}}\ and\ \bibinfo {author} {\bibfnamefont {Z.}~\bibnamefont {Wang}},\ }\bibfield  {title} {\bibinfo {title} {Edge states and topological invariants of non-hermitian systems},\ }\href {https://doi.org/10.1103/PhysRevLett.121.086803} {\bibfield  {journal} {\bibinfo  {journal} {Phys. Rev. Lett.}\ }\textbf {\bibinfo {volume} {121}},\ \bibinfo {pages} {086803} (\bibinfo {year} {2018})}\BibitemShut {NoStop}%
\bibitem [{\citenamefont {Okuma}\ \emph {et~al.}(2020)\citenamefont {Okuma}, \citenamefont {Kawabata}, \citenamefont {Shiozaki},\ and\ \citenamefont {Sato}}]{Okuma2020}%
  \BibitemOpen
  \bibfield  {author} {\bibinfo {author} {\bibfnamefont {N.}~\bibnamefont {Okuma}}, \bibinfo {author} {\bibfnamefont {K.}~\bibnamefont {Kawabata}}, \bibinfo {author} {\bibfnamefont {K.}~\bibnamefont {Shiozaki}},\ and\ \bibinfo {author} {\bibfnamefont {M.}~\bibnamefont {Sato}},\ }\bibfield  {title} {\bibinfo {title} {Topological origin of non-hermitian skin effects},\ }\href {https://doi.org/10.1103/PhysRevLett.124.086801} {\bibfield  {journal} {\bibinfo  {journal} {Phys. Rev. Lett.}\ }\textbf {\bibinfo {volume} {124}},\ \bibinfo {pages} {086801} (\bibinfo {year} {2020})}\BibitemShut {NoStop}%
\bibitem [{\citenamefont {Lin}\ \emph {et~al.}(2023)\citenamefont {Lin}, \citenamefont {Tai}, \citenamefont {Li},\ and\ \citenamefont {Lee}}]{LinTaiLiLee2023}%
  \BibitemOpen
  \bibfield  {author} {\bibinfo {author} {\bibfnamefont {R.}~\bibnamefont {Lin}}, \bibinfo {author} {\bibfnamefont {T.}~\bibnamefont {Tai}}, \bibinfo {author} {\bibfnamefont {L.}~\bibnamefont {Li}},\ and\ \bibinfo {author} {\bibfnamefont {C.~H.}\ \bibnamefont {Lee}},\ }\bibfield  {title} {\bibinfo {title} {Topological non-hermitian skin effect},\ }\href {https://doi.org/10.1007/s11467-023-1309-z} {\bibfield  {journal} {\bibinfo  {journal} {Frontiers of Physics}\ }\textbf {\bibinfo {volume} {18}},\ \bibinfo {pages} {53605} (\bibinfo {year} {2023})}\BibitemShut {NoStop}%
\bibitem [{\citenamefont {Yokomizo}\ and\ \citenamefont {Murakami}(2019)}]{Yokomizo2019}%
  \BibitemOpen
  \bibfield  {author} {\bibinfo {author} {\bibfnamefont {K.}~\bibnamefont {Yokomizo}}\ and\ \bibinfo {author} {\bibfnamefont {S.}~\bibnamefont {Murakami}},\ }\bibfield  {title} {\bibinfo {title} {Non-bloch band theory of non-hermitian systems},\ }\href {https://doi.org/10.1103/PhysRevLett.123.066404} {\bibfield  {journal} {\bibinfo  {journal} {Phys. Rev. Lett.}\ }\textbf {\bibinfo {volume} {123}},\ \bibinfo {pages} {066404} (\bibinfo {year} {2019})}\BibitemShut {NoStop}%
\bibitem [{\citenamefont {Longhi}(2019)}]{Longhi2019}%
  \BibitemOpen
  \bibfield  {author} {\bibinfo {author} {\bibfnamefont {S.}~\bibnamefont {Longhi}},\ }\bibfield  {title} {\bibinfo {title} {Probing non-hermitian skin effect and non-bloch phase transitions},\ }\href {https://doi.org/10.1103/PhysRevResearch.1.023013} {\bibfield  {journal} {\bibinfo  {journal} {Physical Review Research}\ }\textbf {\bibinfo {volume} {1}},\ \bibinfo {pages} {023013} (\bibinfo {year} {2019})}\BibitemShut {NoStop}%
\bibitem [{\citenamefont {Mostafazadeh}(2009)}]{Mostafazadeh2009}%
  \BibitemOpen
  \bibfield  {author} {\bibinfo {author} {\bibfnamefont {A.}~\bibnamefont {Mostafazadeh}},\ }\bibfield  {title} {\bibinfo {title} {Spectral singularities of complex scattering potentials and infinite reflection and transmission coefficients at real energies},\ }\href {https://doi.org/10.1103/PhysRevLett.102.220402} {\bibfield  {journal} {\bibinfo  {journal} {Physical Review Letters}\ }\textbf {\bibinfo {volume} {102}},\ \bibinfo {pages} {220402} (\bibinfo {year} {2009})}\BibitemShut {NoStop}%
\bibitem [{\citenamefont {Chong}\ \emph {et~al.}(2010)\citenamefont {Chong}, \citenamefont {Ge}, \citenamefont {Cao},\ and\ \citenamefont {Stone}}]{Chong2010}%
  \BibitemOpen
  \bibfield  {author} {\bibinfo {author} {\bibfnamefont {Y.~D.}\ \bibnamefont {Chong}}, \bibinfo {author} {\bibfnamefont {L.}~\bibnamefont {Ge}}, \bibinfo {author} {\bibfnamefont {H.}~\bibnamefont {Cao}},\ and\ \bibinfo {author} {\bibfnamefont {A.~D.}\ \bibnamefont {Stone}},\ }\bibfield  {title} {\bibinfo {title} {Coherent perfect absorbers: Time-reversed lasers},\ }\href {https://doi.org/10.1103/PhysRevLett.105.053901} {\bibfield  {journal} {\bibinfo  {journal} {Phys. Rev. Lett.}\ }\textbf {\bibinfo {volume} {105}},\ \bibinfo {pages} {053901} (\bibinfo {year} {2010})}\BibitemShut {NoStop}%
\bibitem [{\citenamefont {Lin}\ \emph {et~al.}(2011)\citenamefont {Lin}, \citenamefont {Ramezani}, \citenamefont {Eichelkraut}, \citenamefont {Kottos}, \citenamefont {Cao},\ and\ \citenamefont {Christodoulides}}]{Lin2011}%
  \BibitemOpen
  \bibfield  {author} {\bibinfo {author} {\bibfnamefont {Z.}~\bibnamefont {Lin}}, \bibinfo {author} {\bibfnamefont {H.}~\bibnamefont {Ramezani}}, \bibinfo {author} {\bibfnamefont {T.}~\bibnamefont {Eichelkraut}}, \bibinfo {author} {\bibfnamefont {T.}~\bibnamefont {Kottos}}, \bibinfo {author} {\bibfnamefont {H.}~\bibnamefont {Cao}},\ and\ \bibinfo {author} {\bibfnamefont {D.~N.}\ \bibnamefont {Christodoulides}},\ }\bibfield  {title} {\bibinfo {title} {Unidirectional invisibility induced by {PT}-symmetric periodic structures},\ }\href {https://doi.org/10.1103/PhysRevLett.106.213901} {\bibfield  {journal} {\bibinfo  {journal} {Phys. Rev. Lett.}\ }\textbf {\bibinfo {volume} {106}},\ \bibinfo {pages} {213901} (\bibinfo {year} {2011})}\BibitemShut {NoStop}%
\bibitem [{\citenamefont {Longhi}\ \emph {et~al.}(2015)\citenamefont {Longhi}, \citenamefont {Gatti},\ and\ \citenamefont {Della~Valle}}]{Longhi2015}%
  \BibitemOpen
  \bibfield  {author} {\bibinfo {author} {\bibfnamefont {S.}~\bibnamefont {Longhi}}, \bibinfo {author} {\bibfnamefont {D.}~\bibnamefont {Gatti}},\ and\ \bibinfo {author} {\bibfnamefont {G.}~\bibnamefont {Della~Valle}},\ }\bibfield  {title} {\bibinfo {title} {Non-hermitian transparency and one-way transport in low-dimensional lattices by an imaginary gauge field},\ }\href {https://doi.org/10.1103/PhysRevB.92.094204} {\bibfield  {journal} {\bibinfo  {journal} {Phys. Rev. B}\ }\textbf {\bibinfo {volume} {92}},\ \bibinfo {pages} {094204} (\bibinfo {year} {2015})}\BibitemShut {NoStop}%
\bibitem [{\citenamefont {Mostafazadeh}(2020)}]{Mostafazadeh2020}%
  \BibitemOpen
  \bibfield  {author} {\bibinfo {author} {\bibfnamefont {A.}~\bibnamefont {Mostafazadeh}},\ }\bibfield  {title} {\bibinfo {title} {Transfer matrix in scattering theory: A survey of basic properties and recent developments},\ }\href@noop {} {\bibfield  {journal} {\bibinfo  {journal} {Turk. J. Phys.}\ }\textbf {\bibinfo {volume} {44}},\ \bibinfo {pages} {472} (\bibinfo {year} {2020})}\BibitemShut {NoStop}%
\bibitem [{\citenamefont {Dwivedi}\ and\ \citenamefont {Chua}(2016)}]{Dwivedi2016}%
  \BibitemOpen
  \bibfield  {author} {\bibinfo {author} {\bibfnamefont {V.}~\bibnamefont {Dwivedi}}\ and\ \bibinfo {author} {\bibfnamefont {V.}~\bibnamefont {Chua}},\ }\bibfield  {title} {\bibinfo {title} {Of bulk and boundaries: Generalized transfer matrices for tight-binding models},\ }\href {https://doi.org/10.1103/PhysRevB.93.134304} {\bibfield  {journal} {\bibinfo  {journal} {Phys. Rev. B}\ }\textbf {\bibinfo {volume} {93}},\ \bibinfo {pages} {134304} (\bibinfo {year} {2016})}\BibitemShut {NoStop}%
\bibitem [{\citenamefont {Ghaemi-Dizicheh}\ and\ \citenamefont {Schomerus}(2021)}]{ghaemiPRA2021}%
  \BibitemOpen
  \bibfield  {author} {\bibinfo {author} {\bibfnamefont {H.}~\bibnamefont {Ghaemi-Dizicheh}}\ and\ \bibinfo {author} {\bibfnamefont {H.}~\bibnamefont {Schomerus}},\ }\bibfield  {title} {\bibinfo {title} {Compatibility of transport effects in non-hermitian nonreciprocal systems},\ }\href {https://doi.org/10.1103/PhysRevA.104.023515} {\bibfield  {journal} {\bibinfo  {journal} {Phys. Rev. A}\ }\textbf {\bibinfo {volume} {104}},\ \bibinfo {pages} {023515} (\bibinfo {year} {2021})}\BibitemShut {NoStop}%
\bibitem [{\citenamefont {Ghaemi-Dizicheh}(2023)}]{ghaemiPRB2023}%
  \BibitemOpen
  \bibfield  {author} {\bibinfo {author} {\bibfnamefont {H.}~\bibnamefont {Ghaemi-Dizicheh}},\ }\bibfield  {title} {\bibinfo {title} {Transport effects in non-hermitian nonreciprocal systems: General approach},\ }\href {https://doi.org/10.1103/PhysRevB.107.125155} {\bibfield  {journal} {\bibinfo  {journal} {Phys. Rev. B}\ }\textbf {\bibinfo {volume} {107}},\ \bibinfo {pages} {125155} (\bibinfo {year} {2023})}\BibitemShut {NoStop}%
\bibitem [{\citenamefont {Kunst}\ and\ \citenamefont {Dwivedi}(2019)}]{KunstDwivedi2019}%
  \BibitemOpen
  \bibfield  {author} {\bibinfo {author} {\bibfnamefont {F.~K.}\ \bibnamefont {Kunst}}\ and\ \bibinfo {author} {\bibfnamefont {V.}~\bibnamefont {Dwivedi}},\ }\bibfield  {title} {\bibinfo {title} {Non-hermitian systems and topology: A transfer-matrix perspective},\ }\href {https://doi.org/10.1103/PhysRevB.99.245116} {\bibfield  {journal} {\bibinfo  {journal} {Phys. Rev. B}\ }\textbf {\bibinfo {volume} {99}},\ \bibinfo {pages} {245116} (\bibinfo {year} {2019})}\BibitemShut {NoStop}%
\bibitem [{\citenamefont {Chen}\ and\ \citenamefont {Zhang}(2025)}]{ChenZhang2025}%
  \BibitemOpen
  \bibfield  {author} {\bibinfo {author} {\bibfnamefont {H.}~\bibnamefont {Chen}}\ and\ \bibinfo {author} {\bibfnamefont {Y.}~\bibnamefont {Zhang}},\ }\bibfield  {title} {\bibinfo {title} {Recurrence method in non-hermitian systems},\ }\href {https://doi.org/10.1103/PhysRevB.111.165118} {\bibfield  {journal} {\bibinfo  {journal} {Physical Review B}\ }\textbf {\bibinfo {volume} {111}},\ \bibinfo {pages} {165118} (\bibinfo {year} {2025})}\BibitemShut {NoStop}%
\bibitem [{\citenamefont {Osterloh}\ \emph {et~al.}(2005)\citenamefont {Osterloh}, \citenamefont {Baig}, \citenamefont {Santos}, \citenamefont {Zoller},\ and\ \citenamefont {Lewenstein}}]{Osterloh2005}%
  \BibitemOpen
  \bibfield  {author} {\bibinfo {author} {\bibfnamefont {K.}~\bibnamefont {Osterloh}}, \bibinfo {author} {\bibfnamefont {M.}~\bibnamefont {Baig}}, \bibinfo {author} {\bibfnamefont {L.}~\bibnamefont {Santos}}, \bibinfo {author} {\bibfnamefont {P.}~\bibnamefont {Zoller}},\ and\ \bibinfo {author} {\bibfnamefont {M.}~\bibnamefont {Lewenstein}},\ }\bibfield  {title} {\bibinfo {title} {Cold atoms in non-abelian gauge potentials: From the hofstadter ``moth'' to lattice gauge theory},\ }\href {https://doi.org/10.1103/PhysRevLett.95.010403} {\bibfield  {journal} {\bibinfo  {journal} {Physical Review Letters}\ }\textbf {\bibinfo {volume} {95}},\ \bibinfo {pages} {010403} (\bibinfo {year} {2005})}\BibitemShut {NoStop}%
\bibitem [{\citenamefont {Ruseckas}\ \emph {et~al.}(2005)\citenamefont {Ruseckas}, \citenamefont {Juzeliunas}, \citenamefont {Ohberg},\ and\ \citenamefont {Fleischhauer}}]{Ruseckas2005}%
  \BibitemOpen
  \bibfield  {author} {\bibinfo {author} {\bibfnamefont {J.}~\bibnamefont {Ruseckas}}, \bibinfo {author} {\bibfnamefont {G.}~\bibnamefont {Juzeliunas}}, \bibinfo {author} {\bibfnamefont {P.}~\bibnamefont {Ohberg}},\ and\ \bibinfo {author} {\bibfnamefont {M.}~\bibnamefont {Fleischhauer}},\ }\bibfield  {title} {\bibinfo {title} {Non-abelian gauge potentials for ultracold atoms with degenerate dark states},\ }\href {https://doi.org/10.1103/PhysRevLett.95.010404} {\bibfield  {journal} {\bibinfo  {journal} {Physical Review Letters}\ }\textbf {\bibinfo {volume} {95}},\ \bibinfo {pages} {010404} (\bibinfo {year} {2005})}\BibitemShut {NoStop}%
\bibitem [{\citenamefont {Goldman}\ \emph {et~al.}(2014)\citenamefont {Goldman}, \citenamefont {Juzeliunas}, \citenamefont {Ohberg},\ and\ \citenamefont {Spielman}}]{Goldman2014}%
  \BibitemOpen
  \bibfield  {author} {\bibinfo {author} {\bibfnamefont {N.}~\bibnamefont {Goldman}}, \bibinfo {author} {\bibfnamefont {G.}~\bibnamefont {Juzeliunas}}, \bibinfo {author} {\bibfnamefont {P.}~\bibnamefont {Ohberg}},\ and\ \bibinfo {author} {\bibfnamefont {I.~B.}\ \bibnamefont {Spielman}},\ }\bibfield  {title} {\bibinfo {title} {Light-induced gauge fields for ultracold atoms},\ }\href {https://doi.org/10.1088/0034-4885/77/12/126401} {\bibfield  {journal} {\bibinfo  {journal} {Reports on Progress in Physics}\ }\textbf {\bibinfo {volume} {77}},\ \bibinfo {pages} {126401} (\bibinfo {year} {2014})}\BibitemShut {NoStop}%
\bibitem [{\citenamefont {Ozawa}\ \emph {et~al.}(2019)\citenamefont {Ozawa}, \citenamefont {Price}, \citenamefont {Amo}, \citenamefont {Goldman}, \citenamefont {Hafezi}, \citenamefont {Lu}, \citenamefont {Rechtsman}, \citenamefont {Schuster}, \citenamefont {Simon}, \citenamefont {Zilberberg},\ and\ \citenamefont {Carusotto}}]{Ozawa2019}%
  \BibitemOpen
  \bibfield  {author} {\bibinfo {author} {\bibfnamefont {T.}~\bibnamefont {Ozawa}}, \bibinfo {author} {\bibfnamefont {H.~M.}\ \bibnamefont {Price}}, \bibinfo {author} {\bibfnamefont {A.}~\bibnamefont {Amo}}, \bibinfo {author} {\bibfnamefont {N.}~\bibnamefont {Goldman}}, \bibinfo {author} {\bibfnamefont {M.}~\bibnamefont {Hafezi}}, \bibinfo {author} {\bibfnamefont {L.}~\bibnamefont {Lu}}, \bibinfo {author} {\bibfnamefont {M.~C.}\ \bibnamefont {Rechtsman}}, \bibinfo {author} {\bibfnamefont {D.}~\bibnamefont {Schuster}}, \bibinfo {author} {\bibfnamefont {J.}~\bibnamefont {Simon}}, \bibinfo {author} {\bibfnamefont {O.}~\bibnamefont {Zilberberg}},\ and\ \bibinfo {author} {\bibfnamefont {I.}~\bibnamefont {Carusotto}},\ }\bibfield  {title} {\bibinfo {title} {Topological photonics},\ }\href {https://doi.org/10.1103/RevModPhys.91.015006} {\bibfield  {journal} {\bibinfo  {journal} {Reviews of Modern Physics}\ }\textbf {\bibinfo {volume} {91}},\ \bibinfo {pages} {015006} (\bibinfo {year} {2019})}\BibitemShut {NoStop}%
\bibitem [{\citenamefont {Pang}\ \emph {et~al.}(2024)\citenamefont {Pang}, \citenamefont {Wong}, \citenamefont {Hu},\ and\ \citenamefont {Yang}}]{Pang2024}%
  \BibitemOpen
  \bibfield  {author} {\bibinfo {author} {\bibfnamefont {Z.}~\bibnamefont {Pang}}, \bibinfo {author} {\bibfnamefont {B.~T.~T.}\ \bibnamefont {Wong}}, \bibinfo {author} {\bibfnamefont {J.}~\bibnamefont {Hu}},\ and\ \bibinfo {author} {\bibfnamefont {Y.}~\bibnamefont {Yang}},\ }\bibfield  {title} {\bibinfo {title} {Synthetic non-abelian gauge fields for non-hermitian systems},\ }\href {https://doi.org/10.1103/PhysRevLett.132.043804} {\bibfield  {journal} {\bibinfo  {journal} {Phys. Rev. Lett.}\ }\textbf {\bibinfo {volume} {132}},\ \bibinfo {pages} {043804} (\bibinfo {year} {2024})}\BibitemShut {NoStop}%
\bibitem [{\citenamefont {Zhao}\ \emph {et~al.}(2025)\citenamefont {Zhao}, \citenamefont {Miao}, \citenamefont {Xing}, \citenamefont {Qiu}, \citenamefont {Ma},\ and\ \citenamefont {Zhao}}]{Zhao2025NonAbelian}%
  \BibitemOpen
  \bibfield  {author} {\bibinfo {author} {\bibfnamefont {Y.}~\bibnamefont {Zhao}}, \bibinfo {author} {\bibfnamefont {Y.}~\bibnamefont {Miao}}, \bibinfo {author} {\bibfnamefont {Y.}~\bibnamefont {Xing}}, \bibinfo {author} {\bibfnamefont {T.}~\bibnamefont {Qiu}}, \bibinfo {author} {\bibfnamefont {H.}~\bibnamefont {Ma}},\ and\ \bibinfo {author} {\bibfnamefont {X.}~\bibnamefont {Zhao}},\ }\bibfield  {title} {\bibinfo {title} {Non-{Abelian} gauge effect for the non-{Hermitian} {Hatano--Nelson} model in cylindrical geometry},\ }\href {https://doi.org/10.1103/xmlw-b6b9} {\bibfield  {journal} {\bibinfo  {journal} {Phys. Rev. B}\ }\textbf {\bibinfo {volume} {112}},\ \bibinfo {pages} {174317} (\bibinfo {year} {2025})}\BibitemShut {NoStop}%
\bibitem [{\citenamefont {Tang}\ \emph {et~al.}(2026)\citenamefont {Tang}, \citenamefont {Zhang}, \citenamefont {Wang}, \citenamefont {Tang}, \citenamefont {Song}, \citenamefont {Xu}, \citenamefont {Zhang}, \citenamefont {Buljan}, \citenamefont {Zhang},\ and\ \citenamefont {Chen}}]{Tang2026Z2}%
  \BibitemOpen
  \bibfield  {author} {\bibinfo {author} {\bibfnamefont {H.}~\bibnamefont {Tang}}, \bibinfo {author} {\bibfnamefont {Y.}~\bibnamefont {Zhang}}, \bibinfo {author} {\bibfnamefont {Z.}~\bibnamefont {Wang}}, \bibinfo {author} {\bibfnamefont {L.}~\bibnamefont {Tang}}, \bibinfo {author} {\bibfnamefont {D.}~\bibnamefont {Song}}, \bibinfo {author} {\bibfnamefont {J.}~\bibnamefont {Xu}}, \bibinfo {author} {\bibfnamefont {W.}~\bibnamefont {Zhang}}, \bibinfo {author} {\bibfnamefont {H.}~\bibnamefont {Buljan}}, \bibinfo {author} {\bibfnamefont {X.}~\bibnamefont {Zhang}},\ and\ \bibinfo {author} {\bibfnamefont {Z.}~\bibnamefont {Chen}},\ }\bibfield  {title} {\bibinfo {title} {Non-{Abelian} route to {$\mathbb{Z}_2$} non-{Hermitian} skin effects},\ }\href {https://doi.org/10.1103/b728-gh5v} {\bibfield  {journal} {\bibinfo  {journal} {Phys. Rev. Lett.}\ }\textbf {\bibinfo {volume} {137}},\ \bibinfo {pages} {126605} (\bibinfo {year} {2026})}\BibitemShut {NoStop}%
\bibitem [{\citenamefont {Sanahal}\ \emph {et~al.}(2025)\citenamefont {Sanahal}, \citenamefont {Panda},\ and\ \citenamefont {Nandy}}]{Sanahal2025}%
  \BibitemOpen
  \bibfield  {author} {\bibinfo {author} {\bibfnamefont {M.}~\bibnamefont {Sanahal}}, \bibinfo {author} {\bibfnamefont {S.}~\bibnamefont {Panda}},\ and\ \bibinfo {author} {\bibfnamefont {S.}~\bibnamefont {Nandy}},\ }\bibfield  {title} {\bibinfo {title} {Gauge field induced skin effect in spinful non-hermitian systems},\ }\href {https://doi.org/10.1103/vhz9-xwf4} {\bibfield  {journal} {\bibinfo  {journal} {Phys. Rev. B}\ }\textbf {\bibinfo {volume} {112}},\ \bibinfo {pages} {125149} (\bibinfo {year} {2025})}\BibitemShut {NoStop}%
\bibitem [{\citenamefont {Chen}\ \emph {et~al.}(2026)\citenamefont {Chen}, \citenamefont {Wu}, \citenamefont {Chen}, \citenamefont {Pu}, \citenamefont {Hu}, \citenamefont {Lu}, \citenamefont {Ke}, \citenamefont {Deng},\ and\ \citenamefont {Liu}}]{ChenNonAbelian2026}%
  \BibitemOpen
  \bibfield  {author} {\bibinfo {author} {\bibfnamefont {X.}~\bibnamefont {Chen}}, \bibinfo {author} {\bibfnamefont {J.}~\bibnamefont {Wu}}, \bibinfo {author} {\bibfnamefont {X.}~\bibnamefont {Chen}}, \bibinfo {author} {\bibfnamefont {Z.}~\bibnamefont {Pu}}, \bibinfo {author} {\bibfnamefont {Y.}~\bibnamefont {Hu}}, \bibinfo {author} {\bibfnamefont {J.}~\bibnamefont {Lu}}, \bibinfo {author} {\bibfnamefont {M.}~\bibnamefont {Ke}}, \bibinfo {author} {\bibfnamefont {W.}~\bibnamefont {Deng}},\ and\ \bibinfo {author} {\bibfnamefont {Z.}~\bibnamefont {Liu}},\ }\bibfield  {title} {\bibinfo {title} {Implementing non-abelian hatano-nelson model in electric circuits},\ }\href {https://doi.org/10.1103/48wx-5gmj} {\bibfield  {journal} {\bibinfo  {journal} {Physical Review Letters}\ }\textbf {\bibinfo {volume} {136}},\ \bibinfo {pages} {106604} (\bibinfo {year} {2026})}\BibitemShut {NoStop}%
\bibitem [{\citenamefont {Wu}\ \emph {et~al.}(2026)\citenamefont {Wu}, \citenamefont {Zhang}, \citenamefont {Cui}, \citenamefont {Yan},\ and\ \citenamefont {Chen}}]{WuNonAbelian2026}%
  \BibitemOpen
  \bibfield  {author} {\bibinfo {author} {\bibfnamefont {C.}~\bibnamefont {Wu}}, \bibinfo {author} {\bibfnamefont {X.}~\bibnamefont {Zhang}}, \bibinfo {author} {\bibfnamefont {Z.}~\bibnamefont {Cui}}, \bibinfo {author} {\bibfnamefont {M.}~\bibnamefont {Yan}},\ and\ \bibinfo {author} {\bibfnamefont {G.}~\bibnamefont {Chen}},\ }\bibfield  {title} {\bibinfo {title} {Non-hermitian non-abelian lattice gauge fields in electrical circuits},\ }\href {https://doi.org/10.1007/s11433-026-2972-0} {\bibfield  {journal} {\bibinfo  {journal} {Science China Physics, Mechanics \& Astronomy}\ }\textbf {\bibinfo {volume} {69}},\ \bibinfo {pages} {277011} (\bibinfo {year} {2026})}\BibitemShut {NoStop}%
\bibitem [{\citenamefont {Velkovsky}\ \emph {et~al.}(2026)\citenamefont {Velkovsky}, \citenamefont {Camacho}, \citenamefont {Ozawa}, \citenamefont {Price},\ and\ \citenamefont {Gadway}}]{Velkovsky2026non}%
  \BibitemOpen
  \bibfield  {author} {\bibinfo {author} {\bibfnamefont {I.}~\bibnamefont {Velkovsky}}, \bibinfo {author} {\bibfnamefont {C.}~\bibnamefont {Camacho}}, \bibinfo {author} {\bibfnamefont {T.}~\bibnamefont {Ozawa}}, \bibinfo {author} {\bibfnamefont {H.}~\bibnamefont {Price}},\ and\ \bibinfo {author} {\bibfnamefont {B.}~\bibnamefont {Gadway}},\ }\bibfield  {title} {\bibinfo {title} {Non-abelian gauge field mechanics},\ }\href@noop {} {\bibfield  {journal} {\bibinfo  {journal} {arXiv preprint arXiv:2607.18215}\ } (\bibinfo {year} {2026})}\BibitemShut {NoStop}%
\bibitem [{\citenamefont {Zhang}\ \emph {et~al.}(2025)\citenamefont {Zhang}, \citenamefont {Hu}, \citenamefont {Zhang}, \citenamefont {Liu}, \citenamefont {Shen}, \citenamefont {Veronis}, \citenamefont {Al{\`u}}, \citenamefont {Huang},\ and\ \citenamefont {Luo}}]{ZhangSU22025}%
  \BibitemOpen
  \bibfield  {author} {\bibinfo {author} {\bibfnamefont {W.}~\bibnamefont {Zhang}}, \bibinfo {author} {\bibfnamefont {Y.}~\bibnamefont {Hu}}, \bibinfo {author} {\bibfnamefont {H.}~\bibnamefont {Zhang}}, \bibinfo {author} {\bibfnamefont {X.}~\bibnamefont {Liu}}, \bibinfo {author} {\bibfnamefont {Y.}~\bibnamefont {Shen}}, \bibinfo {author} {\bibfnamefont {G.}~\bibnamefont {Veronis}}, \bibinfo {author} {\bibfnamefont {A.}~\bibnamefont {Al{\`u}}}, \bibinfo {author} {\bibfnamefont {Y.}~\bibnamefont {Huang}},\ and\ \bibinfo {author} {\bibfnamefont {W.}~\bibnamefont {Luo}},\ }\bibfield  {title} {\bibinfo {title} {Skin effect in non-hermitian systems with {SU(2)} gauge fields},\ }\href {https://doi.org/10.1103/cqmg-qzt4} {\bibfield  {journal} {\bibinfo  {journal} {Physical Review B}\ }\textbf {\bibinfo {volume} {112}},\ \bibinfo {pages} {125164} (\bibinfo {year} {2025})}\BibitemShut {NoStop}%
\bibitem [{\citenamefont {Zhang}(2005)}]{ZhangSchur2005}%
  \BibitemOpen
  \bibinfo {editor} {\bibfnamefont {F.}~\bibnamefont {Zhang}},\ ed.,\ \href {https://doi.org/10.1007/b105056} {\emph {\bibinfo {title} {The Schur Complement and Its Applications}}}\ (\bibinfo  {publisher} {Springer},\ \bibinfo {year} {2005})\BibitemShut {NoStop}%
\bibitem [{Note1()}]{Note1}%
  \BibitemOpen
  \bibinfo {note} {General site-dependent onsite terms, including spatially modulated $\protect \mathcal {PT}$-symmetric gain--loss profiles, are described by the ordered product of the local transfer matrices. Periodic profiles can be treated by enlarging the unit cell.}\BibitemShut {Stop}%
\bibitem [{\citenamefont {Chow}\ \emph {et~al.}(2004)\citenamefont {Chow}, \citenamefont {Grot}, \citenamefont {Mirkarimi}, \citenamefont {Sigalas},\ and\ \citenamefont {Girolami}}]{Chow2004RefractiveIndex}%
  \BibitemOpen
  \bibfield  {author} {\bibinfo {author} {\bibfnamefont {E.}~\bibnamefont {Chow}}, \bibinfo {author} {\bibfnamefont {A.}~\bibnamefont {Grot}}, \bibinfo {author} {\bibfnamefont {L.~W.}\ \bibnamefont {Mirkarimi}}, \bibinfo {author} {\bibfnamefont {M.}~\bibnamefont {Sigalas}},\ and\ \bibinfo {author} {\bibfnamefont {G.}~\bibnamefont {Girolami}},\ }\bibfield  {title} {\bibinfo {title} {Ultracompact biochemical sensor built with two-dimensional photonic crystal microcavity},\ }\href {https://doi.org/10.1364/OL.29.001093} {\bibfield  {journal} {\bibinfo  {journal} {Optics Letters}\ }\textbf {\bibinfo {volume} {29}},\ \bibinfo {pages} {1093} (\bibinfo {year} {2004})}\BibitemShut {NoStop}%
\bibitem [{\citenamefont {Huang}\ and\ \citenamefont {Lehmann}(2008)}]{Huang2008Birefringence}%
  \BibitemOpen
  \bibfield  {author} {\bibinfo {author} {\bibfnamefont {H.}~\bibnamefont {Huang}}\ and\ \bibinfo {author} {\bibfnamefont {K.~K.}\ \bibnamefont {Lehmann}},\ }\bibfield  {title} {\bibinfo {title} {Effects of linear birefringence and polarization-dependent loss of supermirrors in cavity ring-down spectroscopy},\ }\href {https://doi.org/10.1364/AO.47.003817} {\bibfield  {journal} {\bibinfo  {journal} {Applied Optics}\ }\textbf {\bibinfo {volume} {47}},\ \bibinfo {pages} {3817} (\bibinfo {year} {2008})}\BibitemShut {NoStop}%
\bibitem [{\citenamefont {Ruan}\ \emph {et~al.}(2025)\citenamefont {Ruan}, \citenamefont {Tang}, \citenamefont {Li}, \citenamefont {Wu}, \citenamefont {Zhou}, \citenamefont {Xiao}, \citenamefont {Chen}, \citenamefont {Ge}, \citenamefont {Hu}, \citenamefont {Zhang}, \citenamefont {Qiu}, \citenamefont {Liu}, \citenamefont {Jing}, \citenamefont {Lu},\ and\ \citenamefont {Xia}}]{Ruan2025MagnetoOptical}%
  \BibitemOpen
  \bibfield  {author} {\bibinfo {author} {\bibfnamefont {Y.-P.}\ \bibnamefont {Ruan}}, \bibinfo {author} {\bibfnamefont {J.-S.}\ \bibnamefont {Tang}}, \bibinfo {author} {\bibfnamefont {Z.}~\bibnamefont {Li}}, \bibinfo {author} {\bibfnamefont {H.}~\bibnamefont {Wu}}, \bibinfo {author} {\bibfnamefont {W.}~\bibnamefont {Zhou}}, \bibinfo {author} {\bibfnamefont {L.}~\bibnamefont {Xiao}}, \bibinfo {author} {\bibfnamefont {J.}~\bibnamefont {Chen}}, \bibinfo {author} {\bibfnamefont {S.-J.}\ \bibnamefont {Ge}}, \bibinfo {author} {\bibfnamefont {W.}~\bibnamefont {Hu}}, \bibinfo {author} {\bibfnamefont {H.}~\bibnamefont {Zhang}}, \bibinfo {author} {\bibfnamefont {C.-W.}\ \bibnamefont {Qiu}}, \bibinfo {author} {\bibfnamefont {W.}~\bibnamefont {Liu}}, \bibinfo {author} {\bibfnamefont {H.}~\bibnamefont {Jing}}, \bibinfo {author} {\bibfnamefont {Y.-Q.}\ \bibnamefont {Lu}},\ and\ \bibinfo {author} {\bibfnamefont {K.}~\bibnamefont {Xia}},\ }\bibfield  {title} {\bibinfo {title} {Observation of loss-enhanced magneto-optical effect},\ }\href {https://doi.org/10.1038/s41566-024-01592-y} {\bibfield  {journal} {\bibinfo  {journal} {Nature Photonics}\ }\textbf {\bibinfo {volume} {19}},\ \bibinfo {pages} {109} (\bibinfo {year} {2025})}\BibitemShut {NoStop}%
\bibitem [{\citenamefont {Zhu}\ \emph {et~al.}(2010)\citenamefont {Zhu}, \citenamefont {{\"O}zdemir}, \citenamefont {Xiao}, \citenamefont {Li}, \citenamefont {He}, \citenamefont {Chen},\ and\ \citenamefont {Yang}}]{Zhu2010ModeSplitting}%
  \BibitemOpen
  \bibfield  {author} {\bibinfo {author} {\bibfnamefont {J.}~\bibnamefont {Zhu}}, \bibinfo {author} {\bibfnamefont {{\c{S}}.~K.}\ \bibnamefont {{\"O}zdemir}}, \bibinfo {author} {\bibfnamefont {Y.-F.}\ \bibnamefont {Xiao}}, \bibinfo {author} {\bibfnamefont {L.}~\bibnamefont {Li}}, \bibinfo {author} {\bibfnamefont {L.}~\bibnamefont {He}}, \bibinfo {author} {\bibfnamefont {D.-R.}\ \bibnamefont {Chen}},\ and\ \bibinfo {author} {\bibfnamefont {L.}~\bibnamefont {Yang}},\ }\bibfield  {title} {\bibinfo {title} {On-chip single nanoparticle detection and sizing by mode splitting in an ultrahigh-$q$ microresonator},\ }\href {https://doi.org/10.1038/nphoton.2009.237} {\bibfield  {journal} {\bibinfo  {journal} {Nature Photonics}\ }\textbf {\bibinfo {volume} {4}},\ \bibinfo {pages} {46} (\bibinfo {year} {2010})}\BibitemShut {NoStop}%
\bibitem [{\citenamefont {He}\ \emph {et~al.}(2011)\citenamefont {He}, \citenamefont {{\"O}zdemir}, \citenamefont {Zhu}, \citenamefont {Kim},\ and\ \citenamefont {Yang}}]{He2011}%
  \BibitemOpen
  \bibfield  {author} {\bibinfo {author} {\bibfnamefont {L.}~\bibnamefont {He}}, \bibinfo {author} {\bibfnamefont {{\c{S}}.~K.}\ \bibnamefont {{\"O}zdemir}}, \bibinfo {author} {\bibfnamefont {J.}~\bibnamefont {Zhu}}, \bibinfo {author} {\bibfnamefont {W.}~\bibnamefont {Kim}},\ and\ \bibinfo {author} {\bibfnamefont {L.}~\bibnamefont {Yang}},\ }\bibfield  {title} {\bibinfo {title} {Detecting single viruses and nanoparticles using whispering gallery microlasers},\ }\href {https://doi.org/10.1038/nnano.2011.99} {\bibfield  {journal} {\bibinfo  {journal} {Nat. Nanotechnol.}\ }\textbf {\bibinfo {volume} {6}},\ \bibinfo {pages} {428} (\bibinfo {year} {2011})}\BibitemShut {NoStop}%
\bibitem [{\citenamefont {Lau}\ and\ \citenamefont {Clerk}(2018)}]{LauClerk2018}%
  \BibitemOpen
  \bibfield  {author} {\bibinfo {author} {\bibfnamefont {H.-K.}\ \bibnamefont {Lau}}\ and\ \bibinfo {author} {\bibfnamefont {A.~A.}\ \bibnamefont {Clerk}},\ }\bibfield  {title} {\bibinfo {title} {Fundamental limits and non-reciprocal approaches in non-hermitian quantum sensing},\ }\href {https://doi.org/10.1038/s41467-018-06477-7} {\bibfield  {journal} {\bibinfo  {journal} {Nature Communications}\ }\textbf {\bibinfo {volume} {9}},\ \bibinfo {pages} {4320} (\bibinfo {year} {2018})}\BibitemShut {NoStop}%
\bibitem [{\citenamefont {McDonald}\ and\ \citenamefont {Clerk}(2020)}]{McDonald2020}%
  \BibitemOpen
  \bibfield  {author} {\bibinfo {author} {\bibfnamefont {A.}~\bibnamefont {McDonald}}\ and\ \bibinfo {author} {\bibfnamefont {A.~A.}\ \bibnamefont {Clerk}},\ }\bibfield  {title} {\bibinfo {title} {Exponentially-enhanced quantum sensing with non-{H}ermitian lattice dynamics},\ }\href {https://doi.org/10.1038/s41467-020-19090-4} {\bibfield  {journal} {\bibinfo  {journal} {Nat. Commun.}\ }\textbf {\bibinfo {volume} {11}},\ \bibinfo {pages} {5382} (\bibinfo {year} {2020})}\BibitemShut {NoStop}%
\bibitem [{\citenamefont {Aidelsburger}\ \emph {et~al.}(2018)\citenamefont {Aidelsburger}, \citenamefont {Nascimbene},\ and\ \citenamefont {Goldman}}]{Aidelsburger2018}%
  \BibitemOpen
  \bibfield  {author} {\bibinfo {author} {\bibfnamefont {M.}~\bibnamefont {Aidelsburger}}, \bibinfo {author} {\bibfnamefont {S.}~\bibnamefont {Nascimbene}},\ and\ \bibinfo {author} {\bibfnamefont {N.}~\bibnamefont {Goldman}},\ }\bibfield  {title} {\bibinfo {title} {Artificial gauge fields in materials and engineered systems},\ }\href {https://doi.org/10.1016/j.crhy.2018.03.002} {\bibfield  {journal} {\bibinfo  {journal} {C. R. Phys.}\ }\textbf {\bibinfo {volume} {19}},\ \bibinfo {pages} {394} (\bibinfo {year} {2018})}\BibitemShut {NoStop}%
\end{thebibliography}%
	
\end{document}